\documentclass[11pt,twoside,reqno]{atmp}

\usepackage{amsmath,amssymb,amsfonts,amsthm}
\usepackage{bm,bbm}
\usepackage{graphicx}
\usepackage[T1]{fontenc}
\usepackage{esint}
\usepackage{color}
\usepackage[linktocpage,breaklinks]{hyperref}

\newcommand{\be}{\begin{equation}}
\newcommand{\ee}{\end{equation}}
\newcommand{\ba}{\begin{eqnarray}}
\newcommand{\ea}{\end{eqnarray}}
\def\bs{\begin{subequations}}
\def\es{\end{subequations}}
\def\a{\alpha}
\def\b{\beta}
\def\de{\delta}
\def\g{\gamma}
\def\la{\lambda}

\def\e{\epsilon}
\def\ve{\varepsilon}
\def\Om{\Omega}
\def\om{\omega}

\def\s{\sigma}
\def\vr{\varrho}
\def\vp{\varphi}
\def\N{\nabla}
\def\cA{\mathcal{A}}
\def\cB{\mathcal{B}}
\def\cC{\mathcal{C}}
\def\cD{\mathcal{D}}
\def\cE{\mathcal{E}}
\def\cF{\mathcal{F}}

\def\cJ{\mathcal{J}}
\def\cK{\mathcal{K}}
\def\cL{\mathcal{L}}
\def\cM{\mathcal{M}}

\def\cO{\mathcal{O}}
\def\cP{\mathcal{P}}
\def\cR{\mathcal{R}}
\def\cS{\mathcal{S}}

\def\cV{\mathcal{V}}
\def\ds{d_{\rm S}}
\def\dh{d_{\rm H}}

\def\dw{d_{\rm W}}
\def\p{\partial}
\def\bp{\bar{\partial}}
\def\B{\Box}
\newcommand{\Eq}[1]{(\ref{#1})}
\def\com{\color{magenta}}
\def\cob{\color{blue}}
\newcommand{\oarX}[1]{\href{http://arxiv.org/abs/#1}{{\ttfamily\com arXiv:#1}}}
\newcommand{\arX}[1]{\href{http://arxiv.org/abs/#1}{{\ttfamily\com arXiv:#1}}}
\newcommand{\doin}[5]{\href{http://dx.doi.org/#1}{\cob #2 {\bf #3} (#5), #4}}
\newcommand{\doij}[5]{\href{http://dx.doi.org/#1}{\cob #2 {\bf #3} (#5), #4}}
\newcommand{\ndoin}[5]{\href{#1}{\cob #2 {\bf #3} (#5), #4}}
\newcommand{\tia}[1]{\textit{#1}, }
\newcommand{\boxd}[1]{\boxed{\phantom{\Biggl(}#1\phantom{\Biggl)}}}

\def\rme{\text{e}}
\def\rmi{\text{i}}
\def\rmd{d}
\def\bd{\mathbbm{d}}
\def\bD{\mathbbm{D}}
\def\x{q}
\def\bx{\bar{q}}
\def\tI{\tilde{I}}
\def\y{{q'}}

\numberwithin{equation}{section}

\begin{document}

\begin{center}
\href{http://www.intlpress.com/ATMP/ATMP-issue_16_2.php}{Adv.\ Theor.\ Math.\ Phys.\ 16 (2012) 549-644}\\ \medskip

June 28, 2011 \hspace{2cm} AEI-2011-029 \hspace{2cm} \arX{1106.5787}
\end{center}

\barefootnote{Present address: Instituto de Estructura de la Materia, CSIC, Serrano 121, 28006 Madrid, Spain}

\title{Geometry of fractional spaces}

\author{Gianluca Calcagni}
\address{Max Planck Institute for Gravitational Physics (Albert Einstein Institute)\\
Am M\"uhlenberg 1, D-14476 Golm, Germany}
\addressemail{calcagni@iem.cfmac.csic.es}


\begin{abstract}
We introduce fractional flat space, described by a continuous geometry with constant non-integer Hausdorff and spectral dimensions. This is the analogue of Euclidean space, but with anomalous scaling and diffusion properties. The basic tool is fractional calculus, which is cast in a way convenient for the definition of the differential structure, distances, volumes, and symmetries. By an extensive use of concepts and techniques of fractal geometry, we clarify the relation between fractional calculus and fractals, showing that fractional spaces can be regarded as fractals when the ratio of their Hausdorff and spectral dimension is greater than one. All the results are analytic and constitute the foundation for field theories living on multi-fractal spacetimes, which are presented in a companion paper.
\end{abstract}



\keywords{Models of Quantum Gravity, Field Theories in Lower Dimensions, Fractal Geometry}

\maketitle

\setcounter{tocdepth}{2}
\tableofcontents


\section{Introduction}


\subsection{Motivation}

The quest for a quantum theory of gravity has reached such a level of sophistication that several independent approaches to this problem have been developed by now. Each proposes its own physical picture, but they are often related one to the other \cite{Ori09}. Apart from string theory, we mention in particular group field theory \cite{Ori06}, loop quantum gravity (LQG) \cite{Rov06,Thi07}, spin foam models \cite{Ori01,Per03}, asymptotically safe gravity (or quantum Einstein gravity, QEG) \cite{Nie06}--
\cite{CPR}, and simplicial quantum gravity \cite{Wil1}--
\cite{Wil3}, the latter having a particular incarnation in causal dynamical triangulations (CDT) \cite{lol08}. All these approaches, starting from different assumptions and using different techniques, have achieved important results concerning both the kinematical description of quantum space and its dynamics. However, many formal aspects of such dynamics are still to be understood and we still lack a conclusive proof that, within any of these models, the proposed quantum dynamics of space as a discrete entity leads to a continuum description of the same, and to the general relativistic dynamics at large scales and in the classical limit. 

Another important issue is the extraction of effective models of quantum gravity dynamics for both spacetime and matter. These effective descriptions should be used to predict new phenomena and quantum gravity corrections to known particle and astroparticle physics, as well as to large-scale cosmological scenarios. An example of the type of phenomenology one might deal with is \emph{dimensional flow}. It has been noticed that independent models such as CDT \cite{AJL4,BeH}, QEG \cite{LaR5}, and Ho\v{r}ava--Lifshitz gravity \cite{Hor3,SVW1} all exhibit a running of the spectral dimension of spacetime such that at short scales the physics is effectively two-dimensional (see also \cite{MoN}). This number is not accidental and plays an important role in the construction and renormalizability properties of quantum gravity \cite{Car09}--
\cite{fra1}. The change of dimensionality at different scales and its acquiring non-integer values is typical of multi-fractals, so it is customary to describe dimensional flow as a ``fractal property'' of spacetime. The idea that spacetime is ``fractal'' in extreme regimes has been hovering for a while \cite{Bar83}, especially in relation with the classical mixmaster behaviour of the BKL big bang singularity \cite{Bar81}--
\cite{Mot03} and with the notion that renormalization properties of gravity improve on a microscopic ``spacetime foam'' \cite{CrS1,CrS2}. The concept of fractal spacetime has been often shrouded in a halo of vagueness, devoid of any clear-cut definition, and it has begun to be realized concretely in quantum models only in recent times. Aside from the above cases, there are hints of fractal spacetime behaviour at high energies also in effective $\kappa$-Minkowski non-commutative field theories \cite{Ben08}. These findings prompted similar investigations in the context of LQG spin foam models, with some preliminary results obtained using certain approximations to the full spin foam dynamics \cite{Mod08}--
\cite{MPM}. 

The scenario is that of a fundamental dynamics where the usual notions of space, time and geometry emerge only in specific regimes and approximations of the theory. At high energies/small scales, the effective dimension of spacetime is two, while at lower energies the dimension should instead run to four, and the dynamics be well-described by general relativity. This suggests the intriguing possibility that Nature admits a multi-fractal formulation with good ultraviolet (UV) behaviour.

In the attempt to encode dimensional flow in a physically intuitive framework, in \cite{fra1,fra2,fra3} a field theory living in a fractal spacetime was proposed. The key point is to replace the standard Lebesgue measure in the action with a Lebesgue--Stieltjes measure with anomalous scaling,
\be\label{mear}
\rmd^D x\to \rmd\vr(x)\,,\qquad [\vr]=-D\a\geq -D\,,
\ee
where $D$ is the number of topological dimensions, $\vr$ is a (possibly very irregular) distribution, square brackets denote the engineering dimension in momentum units, and $0<\a\leq 1$ is a real parameter interpreted as running with the scale. In the ultraviolet, $\a$ achieves the critical value
\be\label{ast}
\a\to \a_*:=\frac{2}{D}\,,
\ee
and the measure becomes effectively two-dimensional. Several characteristic features were highlighted for the generic measure \Eq{mear} \cite{fra1,fra3}, but the Lebesgue--Stieltjes formalism is too general to be manipulated. Therefore, the special case of an absolutely continuous measure was considered, where $\rmd\vr$ can be written as
\be\label{vrv}
\rmd\vr(x)=\rmd^D x\,v(x)\,,
\ee
for some scalar $v$. In particular, the Poincar\'e algebra of these models is deformed \cite{fra2} and at sufficiently small scales an observer should see deviations from Lorentz invariance. At the quantum level, \Eq{ast} helps field theories with Lebesgue--Stieltjes measure to be power-counting renormalizable, gravity included. The reason is that the dimension of the couplings changes with the scale and goes to zero in the ultraviolet. Dimensional flow also affects cosmology, leading to possible applications to inflation and the cosmological constant problem \cite{fra2}. 

Seen as effective models of quantum gravity, fractal field theories can provide a useful tool to describe the physics at scales large enough to smoothen any discrete structure into a continuous spacetime, but small enough to retain some properties (such as dimensional flow) of the putative full theory. 

Although Ho\v{r}ava--Lifshitz gravity is presently facing some criticism, one of its major merits has been to advance the perspective that treating gravity as an ordinary field theory has nothing wrong in itself, but it requires some rethinking of the geometric structure of spacetime. In this spirit, the framework of \cite{fra1,fra2,fra3} can be regarded simply as an alternative to any of the more complicated quantum gravity scenarios mentioned at the beginning, and the interest shifts from a comparison with the continuum limit of discrete models to a detailed study of properties such as renormalization and the breaking of ordinary Poincar\'e symmetries. Also, there is a conspicuous hiatus between the rigorous mathematical literature on fractals and generic ideas of ``fractal'' structures in field theory and quantum gravity. Letting aside dimensional flow, in what sense is spacetime ``fractal''? What is the geometric meaning of a fractal in Lorentzian signature \cite{fra2}? 

It is the purpose of this paper to begin a study of these problems, triggered by the qualitative treatment of \cite{fra2,fra3}, in a more rigorous framework. Fractional calculus is the natural candidate, since it is known to provide a reliable continuum description of certain properties of fractals. Fractional measures have the desired characteristics of having anomalous scaling and inducing power-counting renormalizability. Their study with the techniques of fractal geometry will fulfill the initial expectations and unravel a number of new, and perhaps surprising, phenomena.

The relation between matter and spacetime and the concept of spacetime itself are radically different when comparing field theories in Minkowski and general relativity. Although realistic models of Nature should include gravity, the introduction of the present formalism of fractional dynamics in a flat non-dynamical space will help in clarifying how to construct a covariant notion of spacetime with non-integer dimension. This will eventually result in a model of multi-fractal geometry with non-Euclidean signature \cite{fra4,frc2}.

We spell out again the physical motivations for studying fractional spaces. We can distinguish between direct and indirect applications of these objects to quantum gravity approaches. Indirect applications concern the description of certain features of quantum gravity such as dimensional flow and the transition from a discrete to a continuum spacetime texture. Fractional space is the building block of a more sofisticated environment, multi-fractional spacetime \cite{fra4,frc2}. In turn, the geometry of multi-fractional spacetime is an explicit realization of the above-mentioned features. The present paper is the starting point of this programme.

Direct applications of fractional spaces are possible whenever they naturally emerge as the limit of some quantum spacetime model. Presently, we are aware only of two such examples. The first is non-commutative geometry. In \cite{ACOS}, we establish a connection between a certain class of non-commutative geometries and fractional spacetimes. The case of $\kappa$-Minkowski is special, inasmuch as it is obtained as the asymptotic regime of multi-fractional geometry at the lowest possible scale. (In multi-fractional spacetimes, there exists a hierarchy of scales starting from a fundamental length we can identify with the Planck length.) This mapping is realized as an equivalence of spacetime measures in both theories. On the non-commutative side, these measures preserve the property of cyclicity in the action. Thus, certain non-commutative geometries, including the best known one ($\kappa$-Minkowski), correspond to spacetimes with well-defined and precise fractal properties. These spacetimes are nothing but fractional spacetimes. A second example, asymptotic safety, will be discussed elsewhere \cite{fra7}.

Finally, with no reference to other approaches, our programme aims at the construction of a UV-finite perturbative quantum field theory. As a field theory, one can apply well-known techniques (modulo some modifications) to check the desired property of renormalizability. This is illustrated with the example of a scalar field theory in \cite{frc2}. The final goal is to show that perturbative quantum gravity is well defined in a multi-fractional setting.


\subsection{Strategy and plan of the paper}\label{strate}

Our programme follows the logic of gradually introducing all the necessary concepts to formalize the idea of a general multi-fractional spacetime:
\begin{enumerate}
\item[1.] First, the fractional analogue of Euclidean space is defined via an embedding abstract space (ordinary Euclidean space) and a choice of fractional calculus of fixed real order $\a$. In this arena, we can understand elementary notions of geometry such as distances, volumes and dimension.
\item[2.] Fractional Euclidean space is a particular model of fractal spacetime. It is very important to clarify this concept using tools of fractal geometry which are seldom or only partially employed in the literature of quantum gravity. The relations between fractional measures and deterministic and random fractals, and between different rigorous definitions of dimension, are discussed. The Hausdorff ($\dh$) and spectral ($\ds$) dimensions of space are non-integer and constant at all scales. In general, as expected in fractals, $\dh\neq \ds$, although in the simplest cases $\dh=\ds$.
\end{enumerate}
These two points are the subject of the present paper. There is still little physics in this setting for many reasons: the dimension of space is fixed and non-integer (in particular, different from 4), the signature is Euclidean, and gravity and matter fields are not mentioned. To continue our agenda:
\begin{enumerate}
\item[3.] Fractional Euclidean spaces are generalized to fractional Minkowski spacetimes with Lorentzian signature. Their symmetry structure is thoroughly analyzed.
\item[4.] A scalar field is introduced as matter, in order to test renormalization properties of field theories in fractional spacetimes.
\item[5.] The Hausdorff and spectral dimensions are made scale-dependent via multi-fractal geometry techniques. At large scales spacetime is four-dimensional, while at small scales it is two-dimensional. Contrary to other approaches where phenomenology is the main guiding principle, both these numbers are constrained by geometric arguments.
\item[6.] A major advancement in the description of fractal spacetimes is to allow the order of fractional operators to be complex. This entails a radical change of the physical picture at ultra-microscopic scales, and opens up remarkable connections with non-commutative spacetimes and discrete quantum gravity models. Discrete symmetries are fundamental, and are progressively averaged at larger scales to finally give way to effective continuous symmetries, up to the point where ordinary Lorentz invariance is achieved.
\end{enumerate}
These features were sketched in \cite{fra4} and full details are presented in the companion paper \cite{frc2}. Finally, as a future work,
\begin{enumerate}
\item[7.] One should introduce the notion of fractional manifold in order to realize a dynamical fractal spacetime and gravity. Consequences for the cosmology of the early universe are expected.
\end{enumerate}
The strategy, plan, and results of this paper are the following.
\begin{itemize}
\item {\bf Section \ref{fc}.} We begin by reviewing the basic tools of fractional calculus in one dimension. Our target reader is the physicist unfamiliar with this area of mathematics and for this reason we will not attempt a rigorous presentation of the latter, preferring intuitive descriptions of its properties. This section is not a mere compilation of results taken from the literature. In fact, it also includes several proofs of properties of the Caputo derivative, which are rarely found in textbooks (Section \ref{core}), and a novel discussion on mixed operators (Section \ref{mix}). Fractional integral measures are a particular case of Lebesgue--Stieltjes measure, whose mathematical and physical interpretation is given in Section \ref{phit}. The reader acquainted with fractional calculus can skip this section.
\item {\bf Section \ref{fes}.} The first step towards the description of the physical world as a field theory on a multi-fractional manifold is to consider a much simpler object; namely, a fractional generalization of empty Euclidean space with fixed dimension, denoted as $\cE_\a^D$. This is the ideal playground whereon to understand the properties of fractional geometry (Section \ref{fgra}). The notation of fractional calculus is not always suitable for this task and it is convenient to introduce a new notation (Section \ref{gn}), which makes explicit the geometric properties of $\cE_\a^D$. After providing hopefully fresh insights on the relations between different fractional derivatives (Section \ref{which}), we define the distance between two points (Section \ref{dis}). The volume calculations of Section \ref{volume} are instructive for several reasons. First, they provide examples of multiple fractional integration on non-rectangular domains. Second, they highlight how different choices of fractional integration affect volume measurements, both in strongly fractional theories and in spaces with almost integer dimension. Last but not least, they show how the Hausdorff dimension of fractional space is related to the fractional order of the measure (Section \ref{dim}).
\item {\bf Section \ref{fractals}.} We discuss at length in which sense fractional Euclidean space is a fractal. On one hand, by considering in detail each of the most characteristic features of fractal sets, we will show that $\cE_\a^D$ can be regarded as a genuine but peculiar fractal, with a continuum structure and anomalous scaling properties  (Sections \ref{finir}--\ref{othe}). These are hybrid properties interpolating between the trivial case of smooth space $\mathbb{R}^D$ (technically a fractal, but with a most boring structure) and self-similar sets. By this analysis, we identify the symmetries of $\cE_\a^D$ as affine transformations of geometric coordinates, equivalent to non-linear transformations in the embedding coordinates (Section \ref{sesi}). On the other hand, fractional calculus can be also seen as an approximation of discrete/disconnected fractals under certain assumptions (Section \ref{fmaf}). We specify the relation between fractional theories and genuine fractal models in this section, where the state of the art of this often confusing subject is reviewed and clarified.
\item {\bf Section \ref{spe}.} The rigorous definition of the spectral dimension $\ds$ on self-similar fractals is reviewed. In smooth spaces, it can be found via an operational procedure involving a diffusion process. This is the most widely used tool to probe ``fractal'' properties of spacetime in quantum gravity theories, and we revisit it at length in Section \ref{diffusion}. Examining every single ingredient of the recipe will allow us to find the spectral dimension of fractional space in Section \ref{dsfras}. As a byproduct, fractional momentum space and the transform thereon, replacing the ordinary Fourier transform, are constructed \cite{frc3}.
\end{itemize}


\subsection{Original and review material}

Due to the length of the paper, it may be useful to tell apart original from review material in detail. Fractional space $\cE_\a^D$ is a novel object and all related statements are presented here for the first time. Many mathematical concepts composing it had been already explored by a number of authors, but they had not been collected together into a unified physical proposal. We employ the well-established mathematical tools of fractional calculus and fractal geometry, and all statements about these two sectors (including definitions and theorems) are, modulo some exceptions, taken from the literature, which is quoted in the text. Attempting a somewhat artificial discrimination, Section \ref{fc} is a review, except Section \ref{core} (commutation relations are usually reported only for the Riemann--Liouville derivative, not for the Caputo derivative) and the text discussion in Section \ref{mix}. Section \ref{fes} is original except Section \ref{fgra}, \Eq{geomi} to \Eq{linel}, and the Definition \Eq{defhau} of Hausdorff dimension. Section \ref{fractals} is original except general definitions and statements about self-similarity (\Eq{simd}--\Eq{Rdsim} and \Eq{ssim}--\Eq{mfssm2}) and Section \ref{fmaf} from \Eq{convo} to \Eq{net}. Sections \ref{dsdef} and \ref{diffusion} are reviews, while Section \ref{dsfras} is new, including the discussion of certain conjectures advanced in fractal geometry at large.

Most of previous ``fractal'' field theoretical proposals assumed spacetime to have a non-integer but fixed, non-dynamical dimensionality, thus drawing the attention to $4-\e$ dimensions with $0<\e\ll1$. On the other hand, here the dimension of space is fixed, but in \cite{frc2} we shall enforce a non-trivial dimensional flow via a superposition of fractional measures. Therefore, we shall not be concerned with the almost-integer regime except as the infrared limiting case of the multi-fractional theory. A global comparison with the early literature on spacetime models in non-integer dimension will be done in \cite{frc2}.


\section{Fractional calculus in one dimension}\label{fc}

We begin by reviewing some basic formul\ae\ of fractional calculus. The material in this section can be found in the dedicated literature. Among the many excellent textbooks on fractional integrals and derivatives, we mention \cite{Pod99}--
\cite{MR} (for an historical account, see also \cite{Ros75}). For reasons which will become clear later, here we make extensive use of Caputo derivatives. Multi-dimensional vector calculus based on these derivatives has been developed in \cite{Tar12} (see references therein for works adopting other types of integro-differential operators).

Unfortunately, this branch of mathematics may sometimes produce a feeling of estrangement in the reader with field theory/high energy background, thus giving her/him the impression of a bizarre topic. This may be due to several reasons: (i) Contrary to ordinary calculus, there is no unique definition of derivative (Riemann--Liouville, Caputo, Erd\'elyi--Kober, Gr\"unwald--Letnikov, Hadamard, Nishimoto, Riesz, Weyl); (ii) The output of fractional operators is almost always quite different from that of ordinary calculus and, in this respect, counterintuitive; (iii) Fractional operators are very seldom employed in particle physics, cosmology, and physics beyond the Standard Model. 

A deeper inspection, however, shows that: (i) All definitions of fractional derivatives are related to one another in a precise way (eventually, differences amount both to the convergence properties of the functional space on which these operators act and to boundary terms in the formul\ae), so there is no mutual contradiction between different operators; (ii) For any choice of fractional operators, the calculus can be defined self-consistently, whatever the look of the formul\ae; (iii) Fractional operators are regularly employed in a number of fields in physics and mathematics, such as statistics, diffusing or dissipative processes with residual memory like weather and stochastic financial models \cite{Mis08}, and system modeling and control in engineering \cite{Das08}. Moreover, in our context the choice of the Caputo operators over the others will be motivated by independent arguments, thus further neutralizing issue (i).

Although by now some particular symbols are widely employed, for the sake of simplicity, and in order not to overburden the unfamiliar reader, we will reserve a simplified notation for the operators we use the most.


\subsection{Integrals and derivatives}

Let
\be
x\in [x_0,x_1]
\ee
be a real coordinate variable defined on an interval with constant extrema $x_0$ and $x_1$, which may be taken to infinity if desired. We define a space of functions $f(x)$ on this interval, such that all the following integro-differential operators will be well defined. A space of particular interest is $AC^n[x_0,x_1]$, that of functions which are absolutely continuous on $[x_0,x_1]$ up to their $n-1$ derivative. This is equivalent to the space of Lebesgue-summable functions with summable derivatives $\p^j$, $j=0,\dots,n$, almost everywhere in the interval. Another space we shall use more or less implicitly is $L_p(x_0,x_1)$, that of Lebesgue-measurable functions $f$ on $[x_0,x_1]$ with finite $p$-norm $\left\|f\right\|_p$. From now on, we assume that for every mathematical statement the functional space is suitably chosen. One can find precise theorem declarations in \cite{Pod99,KST}.

Let $f\in L_p(x_0,x_1)$ and let $\theta$ be the Heaviside distribution:
\be
\theta(x)=\left\{ \begin{matrix} 1\,,&\quad x> 0\\
                                 0\,,&\quad x<0\end{matrix}\right. \,.
\ee
We introduce the \emph{left fractional integral of order $\a$} as
\be\label{I}
(I^\a f)(x) :=\frac{1}{\Gamma(\a)}\int_{x_0}^{x_1} \frac{\rmd x'}{(x-x')^{1-\a}} \theta(x-x')f(x')\,.
\ee
Here $\a\in\mathbb{C}$ is a complex constant parameter, which we shall restrict to be real for our purposes. This formula is naturally suggested as a generalization to non-integer $n$ of the Cauchy formula for the $n$-time repeated integration:
\ba
(I^n f)(x) &=& \int_{x_0}^{x} \rmd y_1\int_{x_0}^{y_1}\rmd y_2\cdots \int_{x_0}^{y_{n-1}}\rmd y_n f(y_n)\nonumber\\
&=&\frac{1}{(n-1)!}\int_{x_0}^{x} \rmd x'\,(x-x')^{n-1} f(x')\,.\nonumber
\ea
One can also define the \emph{right fractional integral of order $\a$}
\be\label{baI}
(\bar{I}^\a f)(x) :=\frac{1}{\Gamma(\a)}\int_{x_0}^{x_1} \frac{\rmd x'}{(x'-x)^{1-\a}} \theta(x'-x)f(x')\,,
\ee
where integration now is from $x$ to the end of the interval.\footnote{The left fractional integral is variedly indicated in the literature with the symbols ${}_{x_0}I^\a_x$, ${}_{x_0}D^{-\a}_x$, ${}_{x_0}d^{-\a}_x$, $I^\a_{x_0-}$. The right integral, with ${}_{x}I^\a_{x_1}$, ${}_{x}D^{-\a}_{x_1}$, ${}_{x}d^{-\a}_{x_1}$, $I^\a_{x_1+}$.} Because of the $x$ dependence in the step function, the output of fractional integrals is a function of $x$. These operators are bounded if $f\in L_p(x_0,x_1)$, for every $1\leq p\leq +\infty$ \cite[(2.1.23)]{KST}. Sometimes we will need to specify the integration domain explicitly and we shall denote the left and right integrals with $I^\a =I^\a_{x_0,x}$ and $\bar{I}^\a =\bar{I}^\a_{x,x_1}$.

In parallel, for any $f\in AC^n[x_0,x_1]$ the \emph{left} and \emph{right Caputo derivatives of order $\a$} \cite{cap1,cap2} exist almost everywhere in $[x_0,x_1]$ \cite[Theorem 2.1]{KST}:
\ba
(\p^\a f)(x) &:=& (I^{n-\a}\p^n f)(x)\,,\qquad n-1\leq \a<n\,,\label{lcd}\\
&=&\frac{1}{\Gamma(n-\a)}\int_{x_0}^{x_1} \frac{\rmd x'}{(x-x')^{\a+1-n}} \theta(x-x')\p^n_{x'}f(x')\,,\label{pan}\\
(\bp^\a f)(x) &:=& (\bar{I}^{n-\a}\p^n f)(x)\,,\qquad n-1\leq \a<n\,,\\
&=&\frac{(-1)^n}{\Gamma(n-\a)}\int_{x_0}^{x_1} \frac{\rmd x'}{(x'-x)^{\a+1-n}} \theta(x'-x)\p^n_{x'}f(x')\,,\label{bpan}
\ea
where $\p$ is the ordinary first-order partial derivative and $n\geq 1$ is a natural number.\footnote{The left fractional derivative is variedly indicated in the literature with the symbols ${}_{x_0}D^\a_x$, ${}_{x_0}d^\a_x$, $D^\a_{x_0-}$. The right derivative, with ${}_{x}D^\a_{x_1}$, ${}_{x}d^{\a}_{x_1}$, $D^\a_{x_1+}$. All these symbols are further decorated with tilde's, in bold font, or with the superscript $C$ when they denote the Caputo derivative. Fractional operators are often called differintegrals, since one can analytically continue derivative expressions to $\a<0$ and vice versa.} We shall be interested in the particular case
\be
0\leq \a<1\qquad (n=1)\,,
\ee
which simplifies the above expression for the left derivative as
\be
(\p^\a f)(x) = \frac{1}{\Gamma(1-\a)}\int_{x_0}^{x_1} \frac{\rmd x'}{(x-x')^{\a}} \theta(x-x')\p_{x'}f(x')\,,\qquad 0\leq \a<1\,;
\ee
a similar statement holds for the right derivative. An elementary but typically ignored fact is that, under the transformation
\ba\label{pari1}
x\to x_0+x_1-x\,,
\ea
left operators are mapped into right operators:
\be\label{pari2}
(\bar{I}^\a f)(x)=(I^\a F)(x_0+x_1-x)\,,\qquad (\bp^\a f)(x)=(\p^\a F)(x_0+x_1-x)\,,
\ee
where $F(x):=f(x_0+x_1-x)$. Therefore, it is sufficient to study the properties of, say, left operators, and infer their right counterparts by using \Eq{pari2}. For this reason, in the following we mainly concentrate on left operators.

An alternative, inequivalent definition of fractional derivation is obtained by exchanging the order of integration and derivation. This corresponds to the left and right Riemann--Liouville derivatives ($n-1\leq \a<n$)
\ba
({}_\textsc{rl}\p^\a f)(x) &:=& (\p^n I^{n-\a}f)(x)\nonumber\\
&=&\frac{1}{\Gamma(n-\a)}\p^n_x\int_{x_0}^{x_1} \frac{\rmd x'}{(x-x')^{\a+1-n}} \theta(x-x')f(x')\,,\qquad\\
({}_\textsc{rl}\bp^\a f)(x) &:=& (\p^n \bar{I}^{n-\a}f)(x)\nonumber\\
&=&\frac{1}{\Gamma(n-\a)}\p^n_x\int_{x_0}^{x_1} \frac{\rmd x'}{(x'-x)^{\a+1-n}} \theta(x'-x)f(x')\,.\qquad
\ea
There is a precise relation between Caputo and Riemann--Liouville derivatives. If $\p_x^jf$ is continuous on $[x_0,x]$ for $j=1,\dots,n$, one has
\be\label{caprl}
(\p^\a f)(x)=({}_\textsc{rl}\p^\a f)(x)-\sum_{j=0}^{n-1} \frac{(x-x_0)^{j-\a}}{\Gamma(1+j-\a)} (\p^j f)(x_0)\,,\qquad n-1\leq\a<n\,.
\ee
In particular, for $n=1$
\be\label{crl}
(\p^\a f)(x)=({}_\textsc{rl}\p^\a f)(x)-\frac{(x-x_0)^{-\a}}{\Gamma(1-\a)}f(x_0)\,,\qquad 0\leq\a<1\,,
\ee
and the two derivatives are the same if $f(x_0)=0$. In general, the two types of derivatives differ in the boundary conditions. While $(\p^\a f)(x_0)=0$ by definition, because of \Eq{caprl} 
\be\label{rlus}
({}_\textsc{rl}\p^\a f)(x_0)=0\qquad \Leftrightarrow \qquad (\p^j f)(x_0)=0\,,\qquad j=0,\dots,n-1\,.
\ee
So, the Riemann--Liouville derivative of a constant is not zero. Right derivatives obey similar relations.


\subsection{Examples of fractional derivatives}

Let us see some examples of calculation with Caputo left fractional derivatives with $n-1<\a<n$. A simple function to consider is a power law,
\be
f(x)=(x-x_*)^\b\,,\qquad x_*\in [x_0,x_1]\,,\qquad \b\in\mathbb{R}\,.
\ee
If $\b=m= 0,1,\dots,n-1$, then both the left and right Caputo derivatives vanish for every $x_*$,
\be\label{psm}
\p^\a (x-x_*)^m = 0 = \bp^\a (x-x_*)^m\,,\qquad m= 0,1,\dots,n-1\,.
\ee
Barring these cases, and assuming $x_0\neq -\infty$, one has
\ba
\p^\a (x-x_*)^\b &=& \frac{1}{\Gamma(n-\a)}\int_{x_0}^{x} \frac{\rmd x'}{(x-x')^{\a+1-n}} \p^n_{x'}(x'-x_*)^\b \nonumber\\
&=&\frac{\Gamma(\b+1)}{\Gamma(n-\a)\Gamma(\b-n+1)}\int_{x_0}^{x} \rmd x'\,\frac{(x'-x_*)^{\b-n}}{(x-x')^{\a+1-n}}\nonumber\\
&\ \stackrel{y=x'-x_0}{=}\ & \frac{\Gamma(\b+1)}{\Gamma(n-\a)\Gamma(\b-n+1)}\int_{0}^{x-x_0}\rmd y\, \frac{[y+(x_0-x_*)]^{\b-n}}{[(x-x_0)-y]^{\a+1-n}} \nonumber\\
&=& \frac{\Gamma(\b+1)}{\Gamma(n-\a+1)\Gamma(\b-n+1)} (x_0-x_*)^{\b-n}(x-x_0)^{n-\a}\nonumber\\
&&\times{}_2F_1\left(1,n-\b;n+1-\a,\frac{x-x_0}{x_*-x_0}\right)\,,\label{speci0}
\ea
where we used \cite[3.196.1]{GR} and ${}_2F_1$ is the hypergeometric function. In the special case $x_*=x_0$, we get \Eq{psm} and \cite[3.191.1]{GR}
\be
\p^\a (x-x_0)^\b =\frac{\Gamma(\b+1)}{\Gamma(\b-\a+1)}(x-x_0)^{\b-\a}\,,\qquad \b\neq 0,1,\dots,n-1\,.\label{speci}
\ee
When the lower extremum is $x_0=-\infty$, one can employ 3.196.2 of \cite{GR}:
\ba
{}_{\infty}\p^\a (x-x_*)^\b &=&\frac{\Gamma(\b+1)}{\Gamma(n-\a)\Gamma(\b-n+1)}\int_{-\infty}^{x} \rmd x'\,\frac{(x'-x_*)^{\b-n}}{(x-x')^{\a+1-n}}\nonumber\\
&\ \stackrel{y=-x'}{=}\ & \frac{(-1)^{\b-n}\Gamma(\b+1)}{\Gamma(n-\a)\Gamma(\b-n+1)}\int_{-x}^{+\infty}\rmd y\, \frac{(y+x_*)^{\b-n}}{(y+x)^{\a+1-n}} \nonumber\\
&=& \frac{(-1)^{\b-n}\Gamma(\b+1)\Gamma(\a-\b)}{\Gamma(1-n+\b)\Gamma(n-\b)} (x_*-x)^{\b-\a}\nonumber\\
&=& (-1)^{\a}\frac{\Gamma(\b+1)}{\Gamma(\b+1-\a)}\frac{\sin(\pi\b)}{\sin[\pi(\b-\a)]} (x-x_*)^{\b-\a},\nonumber\\\label{218}
\ea
where we used the property $\Gamma(z)\Gamma(1-z)=\pi/\sin(\pi z)$. This expression is ill defined for $\b=\a$. Otherwise, it is real under certain conditions on the values of $\a$ and $\b$ and the sign of $x-x_*$; consistently, it vanishes for $\b=m$. The Riemann--Liouville, Caputo and Gr\"unwald--Letnikov derivatives all collapse to the same operator when $x_0=-\infty$, the \emph{Liouville fractional derivative} ${}_{\infty}\p^\a$ (in particular, ${}_{\infty}\p^\a 1=0$). When regarded as an approximation in the limit $t\gg t_0$, this operator is employed in mechanics to describe ``steady state'' systems, that is, systems which evolved well after the initial transient phase at time $t_0$. The right derivative (or integral) with $x_1=+\infty$ is called \emph{Weyl derivative} (or integral). To get the Weyl differintegral from the Liouville differintegral, it is sufficient to set $x_0=-x_1$ in \Eq{pari2} and then take the limit $x_0\to-\infty$:
\be\label{pari3}
({}_{\infty}\bar{I}^\a f)(x)=({}_{\infty}I^\a F)(-x)\,,\qquad ({}_{\infty}\bp^\a f)(x)=({}_{\infty}\p^\a F)(-x)\,,
\ee
where $F(x):=f(-x)$.

The $\a$-th order integral of $(x-x_*)^\b$ is given by the analytic continuation $\a\to-\a$ of the above formul\ae\ for any $\b$. In particular, to prove some commutation theorems we will need the integral \cite[(2.1.16)]{KST}
\be
I^\a (x-x_0)^\b =\frac{\Gamma(\b+1)}{\Gamma(\b+\a+1)}(x-x_0)^{\b+\a}\,.\label{speci2}
\ee
The extension of \Eq{218} to negative $\a$ is
\ba
{}_{\infty}I^\a (x-x_*)^\b &=&\frac{1}{\Gamma(\a)}\int_{-\infty}^{x} \rmd x'\,\frac{(x'-x_*)^{\b}}{(x-x')^{1-\a}}\nonumber\\
&\ \stackrel{y=-x'}{=}\ & \frac{(-1)^{\b}}{\Gamma(\a)}\int_{-x}^{+\infty}\rmd y\, \frac{(y+x_*)^{\b}}{(y+x)^{1-\a}} \nonumber\\
&=& \frac{(-1)^{\b}\Gamma(-\a-\b)}{\Gamma(-\b)} (x_*-x)^{\b+\a}\nonumber\\
&=& \frac{(-1)^{-\a}\Gamma(-\a-\b)}{\Gamma(-\b)}(x-x_*)^{\b+\a}\,.\label{218I}
\ea

Another example is the derivative of the exponential function,
\be
f(x) = \rme^{\la x}\,,\qquad 0\neq\la\in\mathbb{R}\,.
\ee
For finite $x_0$, we have
\ba
\p^\a \rme^{\la x} &=& \frac{\la^n}{\Gamma(n-\a)}\int_{x_0}^{x}\rmd x'\, \frac{\rme^{\la x'}}{(x-x')^{\a+1-n}}\nonumber\\
&\stackrel{y=x'-x_0}{=} \ & \frac{\la^n \rme^{\la x_0}}{\Gamma(n-\a)}\int_{0}^{x-x_0}\rmd y\, \frac{\rme^{\la y}}{[(x-x_0)-y]^{\a+1-n}}\nonumber\\
&=& \frac{\la^\a}{\Gamma(n-\a)}\,\rme^{\la x}\gamma[n-\a,\la(x-x_0)]\,,\label{227}
\ea
where we used \cite[3.382.1]{GR} and $\gamma$ is the incomplete gamma function. Comparing its series representation
\be\nonumber
\gamma(b-1,z)=\Gamma(b-1) z^{b-1} \rme^{-z}\sum_{k=0}^{+\infty}\frac{z^k}{\Gamma(k+b)}\,,
\ee
with that of the two-parameter Mittag-Leffler function \cite{HMS},
\be\label{mil}
E_{a,b}(z):=\sum_{k=0}^{+\infty}\frac{z^k}{\Gamma(ak+b)}\,,\qquad a>0\,,
\ee
one can also express \Eq{227} as
\be
\p^\a \rme^{\la x} = \la^n\rme^{\la x_0}(x-x_0)^{n-\a}E_{1,n+1-\a}[\la(x-x_0)]\,.
\ee

Since $\gamma(z,+\infty)=\Gamma(z)$, the exponential is an eigenfunction of the Liouville derivative ($x_0=-\infty$), with eigenvalues $\la^\a$:
\be\label{espo}
{}_{\infty}\p^\a \rme^{\la x} = \la^\a\rme^{\la x}\,.
\ee
This can be obtained also from \cite[3.382.2]{GR}. For finite $x_0$, the eigenfunction of the Caputo derivative is the one-parameter Mittag-Leffler function $E_\a(z):= E_{\a,1}(z)$:
\be\label{mile}
\p^\a E_\a[\la(x-x_0)^\a] = \la E_\a[\la(x-x_0)^\a]\,,
\ee
which stems from differentiating \Eq{mil} term by term via \Eq{speci} (and remembering that the constant term $k=0$ gives zero). Thus, the Mittag-Leffler function $E_\a[\la(x-x_0)^\a]$ can be considered as the fractional generalization of the exponential $\rme^z=E_1(z)$. With the same procedure, one can get other expressions such as
\be\label{gemile}
\p^\a \{(x-x_0)^\b E_{a,\b+1}[\la(x-x_0)^a]\} = \la (x-x_0)^{\b-\a}E_{a,\b-\a+1}[\la(x-x_0)^a]\,.
\ee
Notice, however, that one should exercise care in inferring some results by analytic continuation of others. For instance, one does not recover \Eq{mile} from \Eq{gemile} in the limit $a\to\a$, $\b\to 0$, because $E_{\a,1-\a}(z)=[\Gamma(1-\a)]^{-1}+z E_\a(z)$. The first term would give an extra contribution $(x-x_0)^{-\a}/\Gamma(1-\a)$ to \Eq{mile}. The reason is that, contrary to the case of the Riemann--Liouville derivative, the limit $\b\to 0$ of \Eq{speci} does not give the correct result \Eq{psm}. In fact, when differentiating the left-hand side of \Eq{gemile} term by term one picks up also the $k=0$ contribution, which is zero if $\b$ is set to zero from the beginning.

Analogous formul\ae\ can be obtained for the right derivative by making use of \Eq{pari2} and \Eq{pari3}. For example, the right version of \Eq{speci}, \Eq{218} and \Eq{espo} are simply
\ba
\bp^\a (x_1-x)^\b &=&\frac{\Gamma(\b+1)}{\Gamma(\b-\a+1)}(x_1-x)^{\b-\a}\,,\quad \b\neq 0,\dots,n-1\,,\qquad\label{specir}\\
{}_{\infty}\bp^\a (x-x_*)^\b &=&\frac{\Gamma(\b+1)}{\Gamma(\b+1-\a)}\frac{\sin(\pi\b)}{\sin[\pi(\b-\a)]} (x-x_*)^{\b-\a}\,,\label{218r}\\
{}_{\infty}\bp^\a \rme^{\la x} &=& (-\la)^\a\rme^{\la x}\,.\label{espo2}
\ea


\subsection{Properties of fractional operators}

We list some properties of fractional operators, focussing on left Caputo differintegrals but quoting some results also for Riemann--Liouville operators. Here, $n-1<\a< n$ and $m-1<\b<m$.

We detail the derivation of most of the results for two reasons. One is that some of the theorems below on Caputo derivatives cannot be found in \cite{Pod99,KST}. Another is to give the unacquainted reader a few examples of the subtleties of fractional calculus. In order to prove some of the statements, we will invoke the first and second theorems of fundamental calculus on the interval $[x_0,x_1]$:
\ba
(\p^n I^n f)(x) &=& f(x)\,,\qquad n\in \mathbb{N}\,,\label{ftc1n}\\
(I^n\p^n f)(x)  &=& f(x)-\sum_{j=1}^{n-1} \frac{1}{j!}(x-x_0)^{j} (\p^j f)(x_0)\,.\label{ftc2n}
\ea

\subsubsection{Limit to ordinary calculus and linearity}

When $\a=n\in\mathbb{N}$, one recovers ordinary calculus of integer order $n$ \cite[(2.4.14)]{Pod99}:
\be
\lim_{\a\to n}\p^\a = \p^n\,,\qquad \lim_{\a\to n}\bp^\a = (-1)^n\p^n\,.
\ee
In particular, when $\a=0$ derivatives and integrals collapse to the identity operator.

Fractional operators are linear: for $\cO^\a=\p^\a,\bp^\a,I^{\a},\bar{I}^{\a}$,
\be
\cO^\a[c_1 f(x)+c_2 g(x)]=c_1 (\cO^\a f)(x)+c_2(\cO^\a g)(x)\,.
\ee

\subsubsection{Commutation relations}\label{core}

The commutation relations
\be\label{commu}
\cO^\a \cO^\b=\cO^\b \cO^\a\,,\qquad \cO^{\a>0}=\p^\a\,,\qquad \cO^{\a<0} =I^{\a}
\ee
are valid under one of the following conditions.
\begin{itemize}
\item $I^\a I^\b \stackrel{?}{=} I^\b I^\a$.

If $\a,\b<0$, at almost every point $x\in [x_0,x_1]$, fractional integrals obey the semi-group property (\cite[(2.100)]{Pod99}; \cite[(2.1.30)]{KST})
\be\label{sgpI}
I^\a I^\b=I^\b I^\a = I^{\a+\b}\,,\qquad \forall~ \a,\b>0\,.
\ee
This property holds everywhere in $[x_0,x_1]$ if $\a+\b>1$.
\item $\p^\a I^\b \stackrel{?}{=} I^\b\p^\a$.

If $\b\geq \a$,
\ba
\p^\a I^\b &\ \stackrel{\Eq{sgpI}}{=}\ & \p^\a I^\a I^{\b-\a}\nonumber\\
           &=& I^{\b-\a}\,,\label{paib}
\ea
while if $\b<\a$ we have, letting $k-1\leq \a-\b <k$ (so that $k=n-m\leq n$),
\ba
(\p^\a I^\b f)(x)&  \stackrel{\Eq{ftc2n}}{=} & (\p^\a I^\b I^k\p^k f)(x)+\sum_{j=1}^{k-1} \frac{\p^\a I^\b(x-x_0)^{j}}{j!} (\p^j f)(x_0)\nonumber\\
&  \stackrel{\Eq{speci2}}{=} & (\p^\a I^{k+\b-\a+\a}\p^k f)(x)+\sum_{j=1}^{k-1} \frac{\p^\a (x-x_0)^{j+\b}}{\Gamma(1+j+\b)} (\p^j f)(x_0)\nonumber\\
&  \stackrel{\Eq{speci}}{=} & [\p^\a I^\a I^{k-(\a-\b)}\p^k f](x)+\sum_{j=1}^{k-1} \frac{(x-x_0)^{j+\b-\a}}{\Gamma(1+j+\b-\a)} (\p^j f)(x_0)\nonumber\\
&  \stackrel{\Eq{ftc1n}}{=} & (\p^{\a-\b} f)(x)+\sum_{j=1}^{k-1} \frac{(x-x_0)^{j+\b-\a}}{\Gamma(1+j+\b-\a)} (\p^j f)(x_0)\,.
\ea
On the other hand, when $\b\geq \a$,
\ba
(I^\b\p^\a f)(x) & \stackrel{\Eq{lcd}}{=} & (I^\b I^{n-\a}\p^n f)(x)\nonumber\\
                 &=& I^{\b-\a}I^n\p^n f(x)\nonumber\\
          & \stackrel{\Eq{ftc2n}}{=} & I^{\b-\a}f(x)-\sum_{j=0}^{n-1}\frac{(x-x_0)^{j+\b-\a}}{\Gamma(1+j+\b-\a)}(\p^j f)(x_0)\,,\quad\label{uti0}
\ea
while for $\b<\a$
\ba
(I^\b \p^\a f)(x) &=& (I^\b I^{n-\a} \p^n f)(x)\nonumber\\
                 &\ \stackrel{\Eq{lcd}}{=}\ & [I^{n-(\a-\b)}\p^n f](x)\nonumber\\
                 &\ \stackrel{\Eq{sgpI}}{=}\ & [I^{k-(\a-\b)}I^{n-k}\p^n f](x)\nonumber\\
                 &\ \stackrel{\Eq{ftc1n}}{=}\ & [I^{k-(\a-\b)}\p^k I^k I^{n-k}\p^n f](x)\nonumber\\
                 &=& (\p^{\a-\b}I^n\p^n f)(x)\nonumber\\
                 &\ \stackrel{\Eq{ftc2n}}{=}&(\p^{\a-\b}f)(x)-\sum_{j=0}^{n-1} \frac{(x-x_0)^{j+\b-\a}}{\Gamma(j+1+\b-\a)} (\p^j f)(x_0).\qquad\label{util}
\ea
Therefore, \Eq{util} is valid for all $\a,\b>0$ and
\be
\p^\a I^\b = I^\b\p^\a,\quad \a,\b\geq 0\quad \Leftrightarrow\quad (\p^j f)(x_0)=0\,,\quad j=0,\dots,n-1\,.
\ee
In comparison, for $\a,\b>0$ the Riemann--Liouville relations are ${}_\textsc{rl}\p^\a I^\b=\p^{\a-\b}$ (for any $\a$ and $\b$, not just $\a\geq \b$) and
\be\nonumber
(I^\b{}_\textsc{rl} \p^\a f)(x) = ({}_\textsc{rl}\p^{\a-\b}f)(x)-\sum_{j=1}^{n-1} \frac{(x-x_0)^{\b-j}}{\Gamma(1-j+\b)} ({}_\textsc{rl}\p^{\a-j} f)(x_0)\,,
\ee
and, via \Eq{rlus}, one has ${}_\textsc{rl} \p^\a I^\b=I^\b{}_\textsc{rl} \p^\a$ if $(\p^j f)(x_0)=0$, $j=0,1,\dots,n-1$.

\item $\p^\a \p^m \stackrel{?}{=} \p^m\p^\a$.

If $\b=m\in\mathbb{N}^+$, \Eq{commu} holds if
\be\label{con}
(\p^j f)(x_0)=0\,,\qquad j=n,n+1,\dots,n+m-1\,.
\ee
In fact, on one hand we have 
\ba
\p^\a\p^m &=& I^{n-\a}\p^n\p^m\nonumber\\
          &=&I^{(n+m)-(\a+m)}\p^{n+m}\nonumber\\
          &=&\p^{\a+m}\,.\label{capam}
\ea
On the other hand, by using \Eq{caprl} twice and $\p^m{}_\textsc{rl}\p^\a={}_\textsc{rl}\p^{\a+m}$, one gets
\be\label{ma}
(\p^m\p^\a f)(x) = (\p^{\a+m}f)(x)+\sum_{j=n}^{n+m-1} \frac{(x-x_0)^{j-m-\a}}{\Gamma(1+j-m-\a)} (\p^j f)(x_0)\,,
\ee
hence the result. For the Riemann--Liouville counterpart of \Eq{commu}, plug \Eq{caprl} into \Eq{capam}: then,
\be\nonumber
({}_\textsc{rl}\p^\a\p^m f)(x) = ({}_\textsc{rl}\p^{\a+m}f)(x)-\sum_{j=0}^{m-1} \frac{(x-x_0)^{j-m-\a}}{\Gamma(1+j-m-\a)} (\p^j f)(x_0)\,,
\ee
and ${}_\textsc{rl}\p^\a\p^m=\p^m\,{}_\textsc{rl}\p^\a$ when $(\p^j f)(x_0)$ for $j=0,\dots,m-1$. 
\item $\p^\a \p^\b \stackrel{?}{=} \p^\b\p^\a$.

Another case of interest is the commutation relation between two Caputo derivatives. When $\a,\b>0$, one has
\ba
(\p^\a\p^\b f)(x) &\ \stackrel{\Eq{lcd}}{=}\ & (I^{n-\a}\p^n\p^\b f)(x)\nonumber\\
                &\ \stackrel{\Eq{ma}}{=}\ & (I^{n-\a}\p^{\b+n}f)(x)+\sum_{j=m}^{n+m-1} \frac{I^{n-\a}(x-x_0)^{j-n-\b}}{\Gamma(1+j-n-\b)} (\p^j f)(x_0)\nonumber\\
                &\ \stackrel{\Eq{speci2}}{=}\ &(I^{n-\a}\p^{\b+n}f)(x)+\sum_{j=m}^{n+m-1} \frac{(x-x_0)^{j-\a-\b}}{\Gamma(1+j-\a-\b)} (\p^j f)(x_0)\nonumber\\
                &\ \stackrel{\Eq{util}}{=}\ & (\p^{\a+\b}f)(x)-\sum_{j=0}^{m-1} \frac{(x-x_0)^{j-\a-\b}}{\Gamma(1+j-\a-\b)} (\p^j f)(x_0)\,.\label{abe}
\ea
(Incidentally, notice that one cannot analytically continue \Eq{abe} to the cases $\b=m$ and $\a=n$. For $\b=m$, there is an obstruction in the step from the first to the second line, while for $\a=n$ there is an obstruction from the third to the fourth line. The correct expressions \Eq{capam} and \Eq{ma} are obtained from the first and third line, respectively.) Switching $(\a,n)$ and $(\b,m)$ and comparing the expressions, one finds that
\be
\p^\a\p^\b=\p^\b\p^\a
\ee
if either
\be
n=m\,,
\ee
or
\be\label{abn}
\a+\b=n+m-1\,,
\ee
or
\be
(\p^j f)(x_0)=0\,,\qquad j=\underbar{$r$},\underbar{$r$}+1,\dots,\bar r-1\,,
\ee
where $\underbar{$r$}={\rm min}(n,m)$ and $\bar r={\rm max}(n,m)$. Interestingly, fractional derivatives do commute if $\a$ and $\b$ have same integer part. The reason is that \Eq{abe} is symmetric in $\a$ and $\b$. Notice also that $\p^\a\p^\b\neq \p^{\a+\b}$ unless $(\p^j f)(x_0)=0$ for $j=0,\dots,\bar r-1$ or \Eq{abn} holds. In the latter case there fall the values $\a=\b=1/2$, and one has $\p^{\frac12}\p^{\frac12} =\p$.

The commutation relation for the Riemann--Liouville derivatives is (\cite[(2.126)]{Pod99}; \cite[(2.1.42)]{KST})
\be\nonumber
({}_\textsc{rl}\p^\a\,{}_\textsc{rl}\p^\b f)(x) = ({}_\textsc{rl}\p^{\a+\b}f)(x)-\sum_{j=1}^{m} \frac{(x-x_0)^{-j-\a}}{\Gamma(1-j-\a)} ({}_\textsc{rl}\p^{\b-j} f)(x_0)\,.
\ee
Switching $(\a,n)$ and $(\b,m)$, the relation
\be\nonumber
{}_\textsc{rl}\p^\a\,{}_\textsc{rl}\p^\b={}_\textsc{rl}\p^\b\,{}_\textsc{rl}\p^\a
\ee
holds if, simultaneously, $({}_\textsc{rl}\p^{\b-j} f)(x_0)=0$ for $j=1,\dots,m$ and $({}_\textsc{rl}\p^{\a-j} f)(x_0)=0$ for $j=1,\dots,n$\,. Using \Eq{rlus}, the combined condition is
\be\nonumber
(\p^j f)(x_0)=0\,,\qquad j=0,1,\dots,\bar r-1\,.
\ee
\end{itemize}
Summarizing for $0<\a,\b<1$ and $m\in\mathbb{N}^+$,
\bs\ba
I^\a I^\b &=& I^\b I^\a\,,\qquad\forall~ \a,\b\,,\\
\p^\a I^\b &=& I^\b\p^\a\qquad \Leftrightarrow\qquad f(x_0)=0\,,\\
\p^\a \p^m &=& \p^m\p^\a\qquad \Leftrightarrow\qquad (\p f)(x_0)=0\,,\\
\p^\a\p^\b &=& \p^\b\p^\a\,,\qquad\forall~ \a,\b
\ea\es
for the Caputo left derivative, while
\bs\ba
{}_\textsc{rl}\p^\a I^\b &=& I^\b\,{}_\textsc{rl}\p^\a\qquad \Leftrightarrow\qquad f(x_0)=0\,,\\
{}_\textsc{rl}\p^\a \p^m &=& \p^m\,{}_\textsc{rl}\p^\a\qquad \Leftrightarrow\qquad f(x_0)=0\,,\\
{}_\textsc{rl}\p^\a\,{}_\textsc{rl}\p^\b &=& {}_\textsc{rl}\p^\b\,{}_\textsc{rl}\p^\a\qquad \Leftrightarrow\qquad f(x_0)=0
\ea\es
for the Riemann--Liouville left derivative. Finally, the Liouville differintegral ${}_{\infty}\p^\a$ always commutes, for sufficiently good functions $f$ (i.e., continuous with continuous derivatives and which fall to zero with their derivatives sufficiently fast for $x_0\to -\infty$) \cite{MR}:
\bs\ba
{}_{\infty}I^\a\,{}_{\infty}I^\b &=& {}_{\infty}I^\b\,{}_{\infty}I^\a={}_{\infty}I^{\a+\b}\,,\qquad \forall~ \a,\b>0\,,\\
{}_{\infty}\p^\a\,{}_{\infty}I^\b &=& {}_{\infty}I^\b\,{}_{\infty}\p^\a={}_{\infty}\p^{\a-\b}\,,\qquad \forall~ \a,\b>0\,,\\
{}_{\infty}\p^\a\,{}_{\infty}\p^\b &=& {}_{\infty}\p^\b\,{}_{\infty}\p^\a={}_{\infty}\p^{\a+\b}\,,\qquad \forall~ \a,\b>0\,.
\ea\es

\subsubsection{Fundamental theorems of calculus}\label{ftcs}

The fractional derivative is the left inverse of the integral. Setting $\a=\b$ in \Eq{paib},
\be\label{ftc1}
(\p^\a I^\a f)(x)=f(x)\,,\qquad \a>0.
\ee
This equation holds also for the Riemann--Liouville derivatives (\cite[(2.106)]{Pod99}; \cite[(2.1.31)]{KST}).

The extra term in \Eq{caprl} is responsible for the following, important difference between Riemann--Liouville and Caputo derivatives. One of the points where fractional calculus may show its worst trickiness is upon generalization of the theorems of calculus, such as the Newton--Leibniz formula
\be\nonumber
\int_{x_0}^{x} \rmd x'\, (\p f)(x') = f(x)-f(x_0)\,.
\ee
The same formula is not valid for the Riemann--Liouville derivative. In fact, one can show that (\cite[(2.108)]{Pod99}; \cite[(2.1.39)]{KST})
\be\nonumber
(I^\a\, {}_\textsc{rl}\p^\a f)(x) = f(x)-\sum_{j=1}^n \frac{(x-x_0)^{\a-j}}{\Gamma(\a-j+1)} (\p^{n-j}I^{n-\a} f)(x_0)\,,\quad n-1\leq\a<n\,,
\ee
so that for $n=1$
\be\nonumber
(I^\a\,{}_\textsc{rl}\p^\a f)(x) = f(x)-\frac{(x-x_0)^{\a-1}}{\Gamma(\a)} (I^{1-\a} f)(x_0)\,,\qquad 0\leq\a<1\,.
\ee
On the other hand, the Caputo derivative is the only fractional derivative obeying a simple Newton--Leibniz formula without imposing particular boundary conditions on the functional space. Thanks to \Eq{sgpI}, one has
\be\nonumber
(I^\a\p^\a f)(x)=(I^\a I^{n-\a}\p^n f)(x) = (I^n\p^n f)(x)\,,
\ee
so that we get
\be
(I^\a\p^\a f)(x)=f(x)-\sum_{j=0}^{n-1}\frac{1}{j!}(x-x_0)^j(\p^j f)(x_0)\,,\qquad n-1<\a\leq n\,,\quad\label{ftc2}
\ee
which could have been obtained from \Eq{util} with $\a=\b$. For $n=1$,
\be
(I^\a\p^\a f)(x)=f(x)-f(x_0)\,,\qquad 0<\a\leq 1\,.\label{nlf}
\ee
Therefore, both fundamental theorems of calculus are satisfied by the Caputo derivative. If the terminal points are not equal in $I^\a$ and $\p^\a$, composition laws become more complicated; we do not consider this case, since these operators are always thought of as defined on the same domain.

\subsubsection{Leibniz rule, non-locality, composite functions}

Unfortunately, the Leibniz rule of derivation for a product of functions $f$ and $g$ is complicated whatever the choice of derivative, since it contains an infinite number of terms \cite[(2.202)]{Pod99}. If $f,g\in C^\infty$ in $[x_0,x]$, then
\be\label{leru}
{}_\textsc{rl}\p^\a(fg)=\sum_{j=0}^{+\infty}\binom{\a}{j} (\p^j f)({}_\textsc{rl}\p^{\a-j}g)\,,\qquad \binom{\a}{j}=\frac{\Gamma(1+\a)}{\Gamma(\a-j+1)\Gamma(j+1)}\,,
\ee
where $\p^{\a-j}=I^{j-\a}$ are actually integrations for $j\geq 1$. 
If $f$ and $g$ are, respectively, analytic and continuous in $[x_0,x]$, \Eq{leru} is valid also  for $\a<0$ (i.e., for fractional derivatives replaced by fractional integrals) and for Liouville/Weyl operators \cite{MR}. 

Equation \Eq{leru} shows the non-local nature of fractional operators: fractional integration by parts or derivation gives rise to an infinite number of terms. Setting $g=1$ or $g=\theta(x-x_0)$, one obtains an expression of the Caputo derivative $\p^\a$ as an infinite series of ordinary derivatives. For $n=1$,
\ba
(\p^\a f)(x)
&=& \frac{1}{\Gamma(1-\a)}\frac{f(x)-f(x_0)}{(x-x_0)^{\a}}\nonumber\\
&&+\sum_{j=1}^{+\infty} \frac{\sin[\pi(j-\a)]}{\pi(j-\a)}\frac{\Gamma(1+\a)}{\Gamma(1+j)}(x-x_0)^{j-\a} (\p^j f)(x)\,,\qquad\label{leru2}
\ea
where we used \Eq{crl} and \Eq{speci2}. 

Equation \Eq{leru2} is useful for writing down the fractional derivative (or integral) of a composite function $f[g(x)]$. In fact, the simple rule 
\be\label{leru0}
\p_x[f(g)]=\frac{\p f}{\p g}\, \p_x g
\ee
no longer holds. The $j$-th derivative $\p_x^j[f(g)]$ can be further expanded as a series, using the Arbogast--Fa\`a di Bruno formula \cite[Section 2.7.3]{Pod99}.

\subsubsection{Integration by parts}\label{ibps}

For an ordinary integral where the integrand contains fractional derivatives, one has \cite[(2.1.50)]{KST} 
\be\label{ibp}
\int_{x_0}^{x_1}\rmd x\, f\,\p^\a g = \int_{x_0}^{x_1}\rmd x\, g\,\bp^\a f\,,
\ee
and the same formula, under suitable conditions on the functions, holds for $\p^\a$ and $\bp^\a$ replaced by $I^\a$ and $\bar{I}^\a$, respectively. One can also consider the opposite situation, i.e., a fractional integral where the integrand has only ordinary derivatives:
\ba
I^\a_{x_0,x_1}\left\{g \p f\right\} &=& \frac{1}{\Gamma(\a)}\int_{x_0}^{x_1} \frac{\rmd x}{(x_1-x)^{1-\a}} g \p f\nonumber\\
									 &=& \int_{x_0}^{x_1} \rmd x\, G \p f\qquad \left(G = \frac{1}{\Gamma(\a)}\frac{g}{(x_1-x)^{1-\a}}\right)\nonumber\\
									 &=& -\int_{x_0}^{x_1} \rmd x\, f\p G\nonumber\\
									 &=& -\frac{1}{\Gamma(\a)}\int_{x_0}^{x_1} \frac{\rmd x}{(x_1-x)^{1-\a}} f \left(\p g+\frac{1-\a}{x_1-x}g\right)\nonumber\\
									 &=& -I^\a_{x_0,x_1} \left\{f \p g+\frac{1-\a}{x_1-x}fg\right\}\,.\label{ibpo}
\ea
The integration by parts of fractional integrals of fractional integrands can be inferred by combining these two cases. As an example, we take a fractional integral over the positive semi-axis. Denoting with a left subscript the lower (upper) extremum of integration in left (right) derivatives,
\ba
\bar I^\a_{0,\infty}\left\{g \p^\b f\right\} &=& \frac{1}{\Gamma(\a)}\int_{0}^{+\infty} \rmd x\,x^{\a-1}\, g \,{}_{0}\p^\b f\nonumber\\
                      &\ \stackrel {\Eq{ibp}}{=}\ & \frac{1}{\Gamma(\a)}\int_{0}^{+\infty} \rmd x\, f(x)\,{}_{\infty}\bp^\b [x^{\a-1}g(x)]\,,
\ea
assuming that the functions $f,g$ are good enough for all steps to be well defined.


\subsection{Exterior derivative}

Fractional differentials were early proposed in \cite{BeA1}--
\cite{BeA4} for the Nishimoto derivative, in \cite{CSN1}--
\cite{CYZ} for the Riemann--Liouville derivative, and in \cite{Tar12,Tar9,Tar10} with Caputo derivatives; early applications of fractional differential forms to mechanical systems can be found in \cite{Tar10}--
\cite{Tar6}. We shall mainly follow the results of \cite{Tar12,CSN1}, with adaptations.

The fractional exterior derivative is defined via the left derivative as
\be\label{diff}
\rmd^\a := (\rmd x)^\a \p^\a\,\,,
\ee
and a right definition also exists: $\bar{\rmd}^\a := (\rmd x)^\a \bp^\a$. The reader may wonder, on one hand, whether this definition is natural and, on the other hand, about the meaning of the writing $(\rmd x)^\a$, the ``$\a$-th power of $\rmd x$.'' These two questions are actually interrelated. We shall postpone the answer to the first in Section \ref{gn}, where we will see how the second fundamental theorem of fractional calculus \Eq{nlf} immediately suggests \Eq{diff} as the obvious candidate for the fractional differential of a function. As for the second question, the object $(\rmd x)^\a$ is a compact rewriting of the fractional differential of a certain function of $x$, which will be later recognized as the natural coordinate in fractional space. By \Eq{speci}, one sees that $\p^\a (x-x_0)^\a =\Gamma(1+\a)$ and
\be\label{usef}
\rmd^\a(x-x_0)^\a = \Gamma(1+\a) (\rmd x)^\a\,,
\ee
and $(\rmd x)^\a$ is shown to be, up to a constant, the fractional differential of
\be\label{fraco}
\boxd{\x := \frac{(x-x_0)^\a}{\Gamma(1+\a)} \,,\qquad [\x]=-\a\,.}
\ee
Therefore, we can recast \Eq{diff} as
\be\label{diff2}
\rmd^\a = \rmd^\a\x\, \p^\a\,.
\ee
For integer $\a$, the fractional differential behaves as the ordinary one. Taking again \Eq{speci}, one has $\rmd^0(x-x_0)^\b=(x-x_0)^\b$, $\rmd^1(x-x_0)^\b= \b (x-x_0)^{\b-1}\rmd x$, $\rmd^2(x-x_0)^\b= \b(\b-1)(x-x_0)^{\b-2}(\rmd x)^2$, and so on.

Notice that the left-hand side of \Eq{usef} seems to be ill defined for $x=x_0$ or $x_0=-\infty$ at any $x$, but the right-hand side just shows that these are artifacts of the presentation in fractional coordinates. The internal structure of $\rmd^\a$ conspires with that of the fractional coordinate to give a finite, well-defined result.

An exact fractional 1-form is the differential of a scalar function, $\rmd^\a f$. A generic fractional 1-form is $\omega=(\rmd x)^\a f(x)$, for some function $f$. The exterior derivative of $\omega$ yields a 2-form, which requires an extension of the coordinate space to many dimensions. In that context we shall describe a geometric interpretation of fractional differentials \cite{BeA3,BeA4}.


\subsection{Mixed operators}\label{mix}

After choosing to work with the left or right sector, one can consistently define all the elements of differential calculus within the same sector. However, in a fractional calculus of variations we expect to have a mixing of the sectors because of \Eq{ibp}. This may be a first reason to also consider versions of fractional calculus with mixed operators.

As far as derivatives are concerned, Cresson defined the complex linear combination \cite{Cre07}\footnote{This operator was actually defined for Riemann--Liouville derivatives of mixed fractional order.}
\be\label{cres}
\cD_\la^\a := \frac{\rmi\la+1}{2}\p^\a+\frac{\rmi\la-1}{2}\bp^\a\,,
\ee
such that integration by parts becomes
\be
\int_{x_0}^{x_1}\rmd x\, f\,\cD_\la^\a g = -\int_{x_0}^{x_1}\rmd x\, g\,\cD_{-\la}^\a f\,.
\ee
When $\la=-\rmi$ and $\la=+\rmi$, one recovers $\p^\a$ and $-\bp^\a$, respectively. For $\la=0$, the same derivative operator would appear on both sides of the equation. However, the operators $\p^\a$ and $\bp^\a$ have complementary domains and it is not possible to define a generalized fractional coordinate associated with the operator $\cD^\a_\la$. In the interval $[x_0,x)$, the natural fractional coordinate is \Eq{fraco}, while in the interval $(x,x_1]$ it is
\be\label{fraco2}
\bx := \frac{(x_1-x)^\a}{\Gamma(1+\a)}\,.
\ee
One of the extrema in differintegral operators varies and, very roughly speaking, one cannot envisage a functional which is constant in the first \emph{and} second interval for any $x$ (compare \Eq{speci} and \Eq{specir} with $\b=\a$). There seems to be no natural geometric interpretation of a theory defined with $\cD_\la$, for any $\la$.

Another motivation to construct mixed operators is in the way fractional integration is carried out. In fact, there are different prescriptions for generalizing the definite Lebesgue integral
\be\nonumber
\int_{x_0}^{x_1}\rmd x' f(x')
\ee
to a fractional integral. The left operator $I^\a$, \Eq{I} with $x=x_1$, carries a measure weight $(x_1-x')^{\a-1}$, such that the major contribution to the integral comes from the area under the curve with $x\sim x_1$. On the other hand, the right operator $\bar{I}^\a$, \Eq{baI} with $x=x_0$, carries a measure weight $(x'-x_0)^{\a-1}$ dominating near the lower extremum, $x\sim x_0$: the output value of the two integrals is not the same. This will not result in different physical qualitative properties, but will eventually lead to different physical measurements, as we shall see in the next section. Therefore, it may be interesting to explore other possibilities. A mixed-type generalization of the definite integral draws inspiration from the splitting
\be\nonumber
\int_{x_0}^{x_1}\rmd x' f(x') = \int_{x_0}^{x_*}\rmd x' f(x')+\int_{x_*}^{x_1}\rmd x' f(x')\,,\qquad \forall~x_*\in[x_0,x_1]\,,
\ee
so that one can define
\be
\tI^\a_{x_*} f : = [(I^\a+\bar{I}^\a) f](x_*)\,,
\ee
or, more explicitly,
\be\label{tildeI}
\tI^\a_{x_0,x_*,x_1} f : = I^\a_{x_0,x_*} f+\bar{I}^\a_{x_*,x_1} f\,.
\ee
Actually, and contrary to the Lebesgue case, due to the fractional weights the splitting is no longer arbitrary and $\tI^\a_{x_*}$ is a class of inequivalent integrals parametrized by $x_*\in[x_0,x_1]$. 

Notably, $\tI^\a$ can be written in terms of just one sector \cite{Tar1,Tar2,Tar3b,Tar4}. After obvious coordinate transformations, one has
\be\nonumber
\tI^\a_{x_*} f = \frac{1}{\Gamma(\a)}\int_0^{x_*-x_0}\frac{\rmd x'}{{x'}^{1-\a}}\,f(x_*-x')+\frac{1}{\Gamma(\a)}\int_0^{x_1-x_*}\frac{\rmd x}{x^{1-\a}}\,f(x_*+x)\,.
\ee
Upon changing integration variable in the first term as $x'= [(x_*-x_0)/(x_1-x_*)]x$, we obtain
\ba
\bar{I}^\a_{x_*,x_1} f &=& \bar{I}^\a_{0,x_1-x_*} f_+\,,\\
I^\a_{x_0,x_*} f  &=& \left(\frac{x_*-x_0}{x_1-x_*}\right)^{\a}\bar{I}^\a_{0,x_1-x_*} f_-\,,
\ea
where
\be
f_+(x):=f(x_*+x)\,,\qquad f_-(x):= f\left(x_*-\frac{x_*-x_0}{x_1-x_*}x\right)\,.
\ee
Thus,
\ba
\tI^\a_{x_*} f &=& \frac{2}{\Gamma(\a)}\int_0^{x_1-x_*}\frac{\rmd x}{x^{1-\a}}\,\tilde f(x)\nonumber\\
&=& 2\bar{I}^\a_{0,x_1-x_*}\{\tilde f\}\,,\label{tI2}
\ea
where
\be
\tilde f(x) :=\frac12\left[f_+(x)+\left(\frac{x_*-x_0}{x_1-x_*}\right)^\a f_-(x)\right]\,. 
\ee
Therefore, the mixed integral $\tI^\a$ is equivalent to a right integral on the positive semi-axis (the weight dominating near the origin), acting on a modified function space. Expression \Eq{tI2} simplifies under some conditions. First, if the definite integral is symmetric ($x_0=-x_1=-R$), we have
\bs\label{symin}\ba
&&\tI^\a_{-R,x_*,R} f = \frac{2}{\Gamma(\a)}\int_0^{R-x_*}\frac{\rmd x}{x^{1-\a}}\,\tilde f(x)\,,\\
&&\tilde f(x) =\frac12\left[f(x_*+x)+\left(\frac{R+x_*}{R-x_*}\right)^\a f\left(x_*-\frac{R+x_*}{R-x_*}x\right)\right].\qquad
\ea\es
In particular, the symmetric choice $x_*=0$ yields
\bs\label{tosymin}\ba
\tI^\a_{-R,0,R} f &=& \frac{2}{\Gamma(\a)}\int_0^{R}\frac{\rmd x}{x^{1-\a}}\,\tilde f(x)\,,\\
\tilde f(x) &=&\frac12\left[f(x)+ f(-x)\right]\,.
\ea\es
When $f$ is even, $\tilde f(x)=f(x)$. We will use this property later to calculate multiple volume integrals. Also, sending $R\to+\infty$ in \Eq{symin}, the mixed fractional integral can be presented as
\bs\label{tIR}\ba
\tI^\a_{\mathbb{R},x_*} f &=& \frac{2}{\Gamma(\a)}\int_0^{+\infty}\frac{\rmd x}{x^{1-\a}}\,\tilde f(x)\,,\\
\tilde f(x) &=&\frac12\left[f(x_*+x)+ f(x_*-x)\right]\,,
\ea\es
which can be further specialized to $x_*=0$, if desired. Notice that this is a generalization of the ordinary integral  over the whole real axis, despite the fact that the presentation \Eq{tIR} is on the positive semi-axis. Taking the absolute value $|x|$ in the measure, one can formally extend this presentation to the whole axis, but we prefer to keep \Eq{tIR} because it makes explicit the existence of a boundary (the special point $x=0$). This will be the object of much discussion when defining the symmetries of fractional spacetime \cite{frc2}.

Unfortunately, also the mixed integral has unattractive properties. Integration by parts follows from the results of Section \ref{ibps}. When the integrand has only ordinary derivatives, one has
\ba
\tI^\a_{x_*}\left\{g \p f\right\} &\ \stackrel{\Eq{ibpo}}{=}\ & -\tI^\a_{x_*} \left\{f \p g+\frac{1-\a}{x_*-x}fg\right\}\nonumber\\
&&+\frac{(-1)^\a-1}{\Gamma(\a)}\lim_{x\to x_*}\frac{f(x)g(x)}{(x_*-x)^{1-\a}}\,.\label{ibpo2}
\ea
The last term must vanish, thus imposing a function space such that for any $f$ and $g$, $(fg)(x)\sim (x_*-x)^{1-\a+\e}$ near $x\sim x_*$, where $\e>0$. The physical meaning of this constraint is not clear \emph{a priori}. Finally, the fundamental theorems of calculus do not hold, since there is no left inverse of the mixed integral (a linear combination of left and right derivatives would produce cross-terms which, in general, do not cancel).



\subsection[Lebesgue--Stieltjes measure and interpretation of fractional integrals]{Lebesgue--Stieltjes measure and interpretation \\ of fractional integrals}\label{phit}

Equation \Eq{fraco} can be also regarded as the Lebesgue--Stieltjes measure associated with left fractional integrals. Indeed, the distribution $\x(x)$ defines a measure $\vr_\a$ over the interval $[x_0,x]$. This measure is the Carath\'eodory extension of $m_\vr((x_*,x]):=\x(x)-\x(x_*)$, $m_\vr(\{x_0\}):=0$, for any $x_*\in [x_0,x]$. In fact, $\vr_\a$ is monotonic, non-decreasing and right-continuous. Furthermore, the properties of measures are satisfied. We can see this intuitively by considering $m_\vr$: (i) $m_\vr(\emptyset)=0$, (ii) $m_\vr(A)\leq m_\vr(B)$ if $A\in B\notin [x_0,x]$, (iii) given a countable or finite union of sets, $m_\vr(\bigcup_i A_i)\leq \sum_i m_\vr(A_i)$, where the equality holds if the $A_i$ are disjoint Borel sets; in particular, $m_\vr(A\setminus B)=m_\vr(A)-m_\vr(B)$. Therefore, the Riemann--Liouville integral \Eq{I} can be regarded as a \emph{Lebesgue--Stieltjes} (or \emph{Radon}) \emph{integral} \cite{RN}--
\cite{deb}:
\be\label{iavr}
I^\a f = \int_{x_0}^x \rmd\vr_\a(x')\,f(x')\,,\qquad \vr_\a(x')=-\frac{(x-x')^\a}{\Gamma(1+\a)}\,,
\ee
where we made a slight abuse of notation and identified $\vr_\a(x)$ with $m_\vr((x_0,x])=\x(x)$. An important feature of the measure is the scaling property, inherited from $\x$,
\be\label{scavr}
\vr_\a(\la x) = \la^{\a}\vr_\a(x)\,,\qquad \la>0\,;
\ee
namely, the measure of the set obtained by a rescaling $x\to \la x$ is the same as the original set, multiplied by a factor $\la^{\a}$.

\begin{figure}
\includegraphics[width=\textwidth]{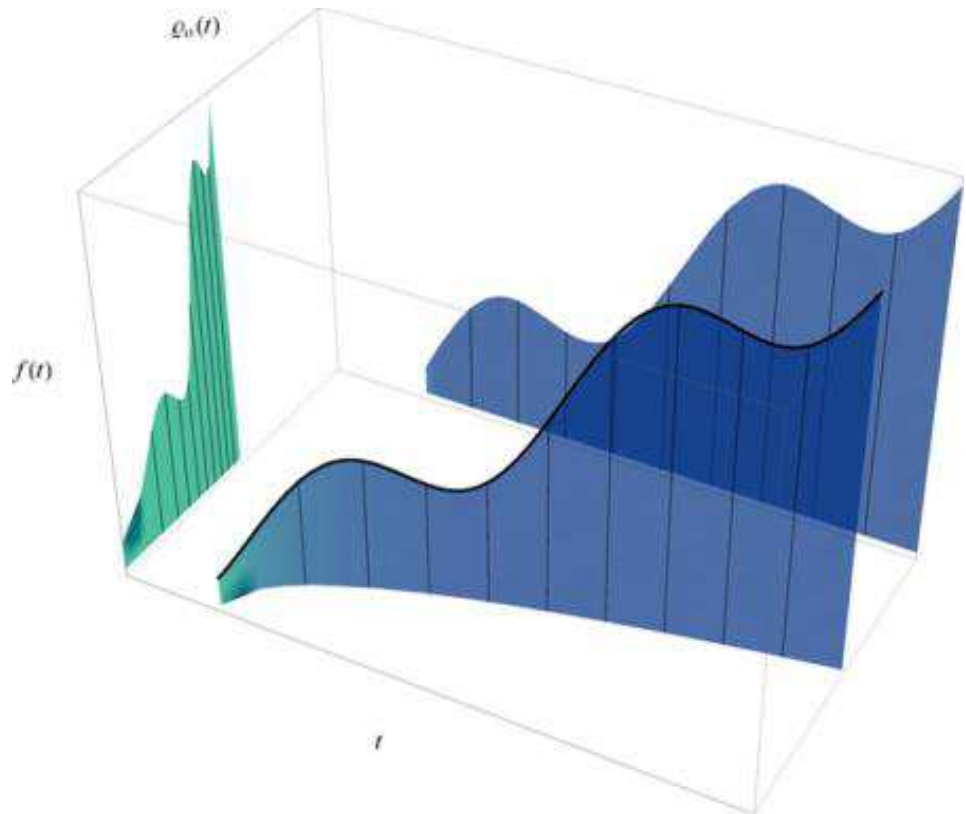}
\caption{Geometric interpretation of Lebesgue--Stieltjes integrals as ``shadows'' of a ``fence''. $f(t)$ is a generic smooth function and the measure $\vr_\a(t)$ is given in \Eq{iavr} (specifically, in the figure $\a=1/2$).}
\label{fig1}
\end{figure}

Fractional integrals admit neat geometrical \cite{Bul88,Pod02} and physical \cite{RYS,MH} interpretations. Consider a function $f(t)$ and the time integral
\ba
(I^\a f)(t_1) &=&\int_{t_0}^{t_1}\rmd t\,\frac{(t_1-t)^{\a-1}}{\Gamma(\a)} f(t)=:\int_{t_0}^{t_1}\rmd t\,v_\a(t_1-t) f(t)\nonumber\\
&=& \int_{t_0}^{t_1} \rmd\vr_\a(t)\,f(t)\,.\label{It}
\ea
The geometric meaning of the left fractional integral \Eq{It} with $\a\neq 1$ fixed is shown in figure \ref{fig1}. The continuous curve in the box is given parametrically by the set of points $\cC=\{(t,\vr_\a(t),f(t))\}$, where $f$ is some smooth function. Projection of $\cC$ onto the $t$-$f$ plane ($\vr_\a={\rm const.}$) gives $f(t)$, while projection onto the $t$-$\vr_\a$  plane ($f={\rm const.}$) yields $\vr_\a(t)$. Now, build a vertical ``fence'' under the curve $\cC$, and project it onto both planes. On the $t$-$f$ plane, the shadow of the fence is the ordinary integral,
\be\label{orin}
(I^1 f)(t_1) =\int_{t_0}^{t_1} \rmd t\, f(t)\,.
\ee
On the $\vr_\a$-$f$ plane, the shadow corresponds to the fractional integral \Eq{It}, the area under the projection of $\cC$ on such plane.

The behaviour of the measure weight $v_\a(t_1-t)$ leads to different physical scenarios at the extreme values of the interval $0\leq \a\leq 1$. We can regard this weight as a memory function and the fractional integral as a memory flux. If $\a=1$, the fence lies in the $t$-$f$ plane and the memory function $v_1=1$ equally weighs all the points from the initial time $t_0$ to the final time $t$, \Eq{orin}. Processes described by integer integrals retain all the memory of the past history. The integral has the usual meaning of ``area under the curve $f(t)$ in the interval $[t_0,t_1]$.'' Also, if $f(t)$ is the speed of a point particle, $I^1 f$ is the operational definition of the distance covered in the time interval $\Delta t=t_1-t_0$.

When $\a=0$, the limit of the memory function in the sense of distributions is a delta, $v_{\a\to 0}(t_1-t)=\delta(t_1-t)$, the fractional integral becomes the identity operator, and the integral of a function from $t_0$ to $t_1$ equals the function itself, evaluated at the final point $t_1$:
\be
\lim_{\a\to 0}(I^\a f)(t_1) =f(t_1)\,.
\ee
The past history is completely forgotten. Systems with no memory retention are called Markovian, and are well described by fractional calculus in the limit $\a\to 0$. Thinking of information as carried by ``states'', $\a$ roughly corresponds to the fraction of states preserved at a given time $t$. In turn, loss of information corresponds to a loss of energy, at a rate $1-\a$. Hence, fractional systems are dissipative \cite{Rie1,Rie2}. Examples are percolation clusters, porous media, collision systems, and Brownian motion. The one-sidedness of fractional operators, in fact, is responsible for the irreversibility of time \cite{Jum06}.


\section{Fractional Euclidean space}\label{fes}

The extension to $D$ topological dimensions (i.e., to $D$ coordinates, where $D\in\mathbb{N}_+$) is straightforward. Each direction is associated with a fractional ``charge'' $\a_{\mu}$.  The corresponding Lebesgue--Stieltjes measure is
\be\label{cross}
\vr_\a(x)=\bigotimes_{\mu =1}^D \vr_{\a_{\mu}}(x^\mu)\,,
\ee
with rescaling
\be\label{scavr2}
\boxd{\vr_\a(\la x) = \la^{\sum_\mu\a_{\mu}}\vr_\a(x)\,,\qquad \la>0\,.}
\ee
In general, the $D$ parameters $\a_{\mu}$ can be different from one another. To make the presentation as simple as possible, we shall make an ``isotropy'' assumption: namely, 
\be\
\a_{\mu}=\a\,,\qquad \forall~ \mu=1,\dots, D\,.
\ee
This restricts the analysis to fractional manifolds (Section \ref{gn}) where the fractional charge is equally distributed among the directions, and all of them are treated on an equal footing. Anisotropic configurations are possible and different choices of the set $\{\a_{\mu}\}$ (modulo permutations) correspond to inequivalent geometries. Some anisotropic measures were given in \cite{fra2}.

Before discussing fractional Euclidean space, we need the extension of fractional differentials to many dimensions. The partial fractional derivative along the $\mu$ direction is
\be
\p^\a_\mu :=\p^\a_{x^\mu}\,,\qquad [\p^\a_\mu]=\a\,. 
\ee
A simple summation over coordinates yields
\be\label{damu}
\rmd^\a := (\rmd x^\mu)^\a \p^\a_{\mu}\,,\qquad [d^\a]=0\,.
\ee
The Einstein convention of summing over repeated upper-lower indices is employed. Arbitrary fractional $n$-forms can be constructed \cite{Tar12,CSN1}. For instance, the exterior derivative of the 1-form 
\be\label{1form}
\omega=(\rmd x^\mu)^\a f_\mu(x)
\ee
is $\rmd^\a\omega({\bm x}) = (\p_\mu^\a f_\nu)({\bm x})\,(\rmd x^\mu)^\a\wedge (\rmd x^\nu)^\a$.


\subsection{Interpretation of fractional gradients}\label{fgra}

\begin{figure}
\includegraphics[width=\textwidth]{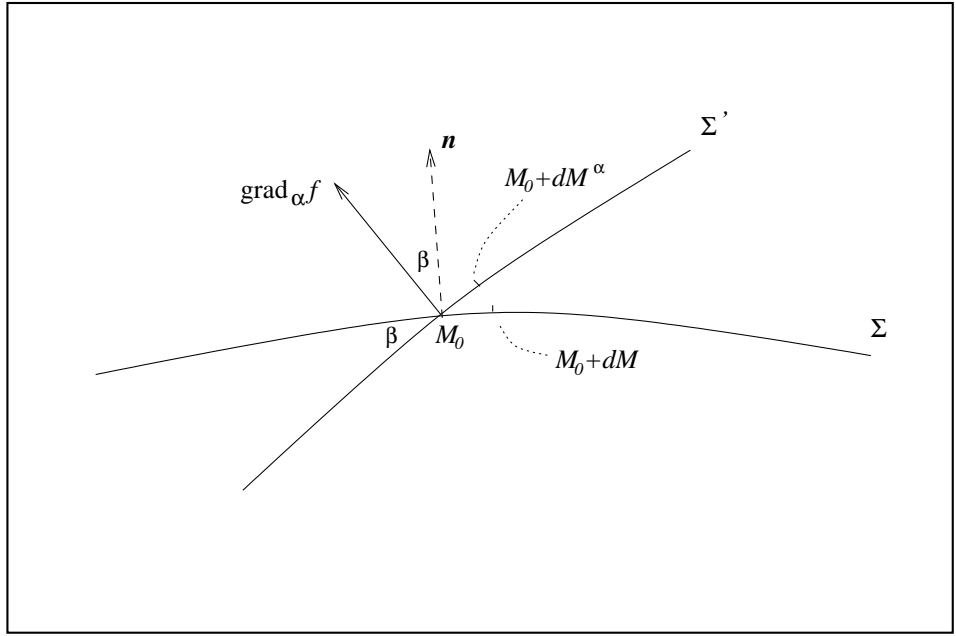}
\caption{Geometric interpretation of fractional gradient. The symbols are explained in the text.}
\label{fig2}
\end{figure}

Fractional gradients admit a geometric interpretation \cite{BeA3,BeA4}, illustrated in figure \ref{fig2}. Already in one dimension, the fractional derivative of an $\a$-differentiable function $f$ on $\mathbb{R}$ can be expressed as
\be
(\p^\a f)(x)=\lim_{y\to x^+} \frac{(T_\a f)(y,x)-(T_\a f)(x,x)}{(y-x)^\a}\,,
\ee
where $T_\a$ is a mapping suitably defined (see \cite{BeA2,BeA4} for details in the case of the Nishimoto derivative). In particular,
\be\label{hoh}
(T_\a f)(x+h,x)= f(x)+\rmd^\a f_{x}(h)+h^\a\ve(h)\,,
\ee
where $h>0$, $\rmd^\a f_{x}(h)=h^\a (\p^\a f)(x) $, and $\lim_{h\to 0}\ve(h)=0$. (Related to this equation or variations on the same theme, one can develop a fractional Taylor expansion \cite{Jum06}--
\cite{Ju05a}.) In many dimensions, one can consider a directional fractional derivative and a bilinear mapping $(T_\a f)({\bm y},{\bm x})$ acting on $\mathbb{R}^D\otimes \mathbb{R}^D$. Let $({\rm grad}_\a f)_\mu:=\p^\a_\mu f$ be the $\mu$-th component of the fractional gradient of a function $f$. Indicating a vector ${\bm x}$ as a point $M$ in $\mathbb{R}^D$, let the gradient exist at the point $M_0\in\mathbb{R}^D$. If $\rmd M$ is an infinitesimal vector displacement of the point, with coordinates $(\rmd x^1,\dots,\rmd x^D)$, one can identify a fractional displacement $\rmd M^\a$ as the vector $((\rmd x^1)^\a,\dots, (\rmd x^D)^\a)$. Therefore, the differential $\rmd^\a f$ is given by
\be
\rmd^\a f = {\rm grad}_\a f \cdot \rmd M^\a\,.
\ee
The set $\Sigma$ of points $M$ satisfying $(T_\a f)(M,M_0)=c$, where $c$ is a constant, is called a level surface passing through $M_0$. The point $M_0+\rmd M$ in a neighborhood of $M_0$ belongs to $\Sigma$, but the point $M_0+\rmd M^\a$ does not. In fact, the latter determines another level surface $\Sigma'$, $(T_\a f)(M,M_0)=c'$, at an angle $\b$ with $\Sigma$ determined by $\cos\b=|c-c'|/\|\rmd M^\a\|$, where $\|\cdot\|$ is the norm equipping the vector space. Since $\rmd^\a f$ vanishes on $\Sigma$, the vectors ${\rm grad}_\a f$ and $\rmd M^\a$ are orthogonal, so that ${\rm grad}_\a f$ is not orthogonal to (tangent vectors on) $\Sigma$ at $M_0$ for $0<\a<1$. The projection of ${\rm grad}_\a f$ on the unit vector ${\bm n}$ normal to $\Sigma$ at $M_0$ has modulus ${\rm grad}_\a f\cdot {\bm n}=|{\rm grad}_\a f|\sin\b$. When $\a=1$, the two level surfaces coincide and $c=c'$, $\b=0$.


\subsection{Which calculus?}\label{which}

We define \emph{fractional Euclidean space} $\mathcal{E}_\a^D$ of order $\a$ as Euclidean space $\mathbb{R}^D$ endowed with a set of rules ${\rm Calc}^\a=\{\p^\a, I^\a,\dots \}$ of integro-differential calculus, a measure $\vr_\a$ with a given support, a natural norm $\| \cdot \|$, and a Laplacian $\cK$:
\be\label{fres}
\boxd{\mathcal{E}_\a^D = (\mathbb{R}^D,\,{\rm Calc}^\a,\,\vr_\a,\,\| \cdot \|,\,\cK)\,.}
\ee
Different sets of fractional operators in ${\rm Calc}^\a$ can correspond to inequivalent fractional spaces. We should now make a commitment on the type of derivative and integration operators acting in $\mathcal{E}_\a^D$, and choose between left and/or right integrals, left and/or right Caputo or Riemann--Liouville derivatives, mixed operators, the range of $\a$, and so on. In fact, there are many more fractional derivatives we could have listed here, so the choice is actually larger. For the purposes of this section, the only ingredient we need to specify is calculus. In particular, for finite-volume calculations we pick the left fractional integral with $0<\a\leq 1$, while we elect the left Caputo derivative $\p^\a$ with $0<\a\leq 1$ as the building block of the differential structure of $\cE^D_\a$. We postpone the choice of the Laplacian $\cK$ to Section \ref{dsfras}, after the discussion of the so-called harmonic structure of fractal sets. Furthermore, we anticipate that the measure of $\cE_\a^D$ is, for each direction, the Weyl measure $\vr_\a(x)=x^\a/\Gamma(\a+1)$, with support on the positive real axis. Here we are interested in local properties of fractional Euclidean space, so we do not spell the reason why the support of $\vr_\a$ can be chosen as $[0,+\infty)$ and not something else. That is given in \cite{frc2}.

Let us justify in detail the calculus.
\begin{itemize}
\item \emph{Fractional versus non-fractional calculus}: Fractional calculus is among the most studied and best manageable frameworks generalizing ordinary calculus and, to the best of our knowledge, there are almost no other examples of calculi mimicking fractal behaviours in physics and statistics such as anomalous scaling of dimensionality and discrete scale invariance \cite{fra4}. Mathematical properties of fractals, however, can be analyzed by Connes quantized calculus \cite{Con94}, which we shall not consider here.
\item \emph{Left versus right operators}: While left operators involve integration from an initial point $x_0$ up to the arbitrary point $x$, right operators integrate over the complementary sub-interval $[x,x_1]$. In fractional mechanics, where $x=t$ is time and one studies dissipative classical systems, the output of these operators should depend on the past (rather than future) evolution of the system, and left derivatives seem more natural. However, on one hand in our context we do not have this type of physical interpretation and, on the other hand, both types of operators appear in fractional Lagrangian systems because of \Eq{ibp}. Fortunately, ``left versus right'' is a non-issue, since the two classes of operators are actually the same under coordinate reflection (\Eq{pari1}, \Eq{pari2} and \Eq{pari3}). Further discussion of this point can be found in \cite{frc2}.
\item $0<\a\leq 1$: This choice is empirical. One wishes to obtain a model of space whose dimension is smaller than the topological dimension $D$ of the embedding space. For a natural isotropic distribution of fractional charge over the $D$ coordinates, this is achieved precisely for this range of the order parameter. The range of $\a$ will be further restricted in Section \ref{dis}.
\item \emph{Caputo versus Riemann--Liouville (and others)}: The Caputo derivative carries several advantages over the Riemann--Liouville operator. 
	\begin{enumerate}
	\item[(i)] First, the Caputo derivative of a constant is zero for any $x_0$, while for the Riemann--Liouville derivative it is so only when $x_0=-\infty$. This helps in rendering fractional calculus more akin to the ordinary one without affecting other properties. We have seen, for instance, that the fundamental theorems of calculus are generalized in a simple way only for the Caputo derivative. Another difference, fully appreciated only when doing tensor calculus on curved manifolds, is that the frame metric $\eta_{\mu\nu}$ is indeed a constant matrix for the Caputo differentiation; hence, tensor calculus keeps many of its usual rules.\footnote{This fact
was recognized in studies of manifolds with non-holonomic structure \cite{vac4}--
\cite{BV2}, a framework employed to integrate non-linear dynamical equations such as Einstein's \cite{VSGG,vac3}.} In \cite{frc2}, we shall see that it is possible to define a simple fractional generalization of Poincar\'e symmetries precisely because of this property of the Caputo derivative.
  \item[(ii)] A popular reason to prefer Caputo over Riemann--Liouville operators is the existence, in the former case, of a standard Cauchy problem, where one needs to specify only the first $n$ ordinary derivatives \cite{Pod99}. On the other hand, the Riemann--Liouville operator requires the specification of $n$ initial conditions of the form $\lim_{x\to x_0} (I^{k-\a}f)$ $(x)$, $k=1,\dots, n$, which have no clear physical interpretation. Related considerations hold for general non-local theories, where the Cauchy problem must be reinterpreted; such a reinterpretation is known only for special non-local operators (e.g., \cite{MZ,cuta7}). 
  \item[(iii)] There is, moreover, a simple but not very well known argument setting the Caputo and Riemann--Liouville derivatives on a different footing \cite{Pod99}, and favouring the former as the most plausible operator to appear in dynamical equations. Taking the $\a\to n$ limit of \Eq{caprl}, one obtains
\be\label{caprln}
(\p^n f)(x)=({}_\textsc{rl}\p^n f)(x)-\sum_{j=0}^{n-1} \p^{n-1-j}\delta(x-x_0) (\p^j f)(x_0)\,.
\ee
This is nothing but the relation between the integer classical derivative, on the left-hand side, and the derivative in the sense of distributions on the right-hand side. Therefore, one can consider the Caputo derivative as the fractional generalization of classical differentiation, and the Riemann--Liouville derivative as the fractional generalization of functional differentiation. In this respect, the Caputo operator is a much more natural choice for the fractional derivative in actions defined on $\cE_\a^D$. 
  \item[(iv)] Another difference between Caputo and Riemann--Liouville derivatives emerges within initialized fractional calculus \cite{Das08,LH00}--
\cite{HL09}. We mention it for the sake of completeness, although it has no impact in the present discussion. Heuristically, while repeated use of ordinary integration and differentiation generates arbitrary integration constants which are fixed by the boundary conditions of the problem, fractional integration is a continuous operation leading to a non-trivial entire function, called \emph{complementary} (or complimentary). This function is determined by additional input on the boundary conditions. More precisely, in classical fractional mechanics one is interested in observing a system starting from a time $t_*$ later than the initial time $t_0$ when the system began to evolve; call $f(t)$ the observed history, a continuous function on $[t_*,t]$. The effect of past history is incorporated in the complementary function. It turns out that the inferred history for the Riemann--Liouville derivative is continuous throughout the evolution period, and it is described just by the function $f(t)$, analytically continued to the whole interval $[t_0,t]$. On the other hand, if the Caputo derivative is required to be equal to the initialized Riemann--Liouville derivative, the analytic continuation of $f(t)$ is the constant $f(t)=f(t_*)$, for $t<t_*$ \cite{NLH}; therefore, derivatives of $f$ are discontinuous at $t_*$, and Caputo initialization effects seem not to be properly taken into account \cite{HL09}. In our case this is not an issue, both because we do not ask the two derivatives to coincide (we simply make one choice and keep it all along the way) and because the context of fractional evolution is quite different. There is no connection between the lower terminal of integration, representing a boundary of space(time), with the notion of performing observations at space(time) points in the bulk. Here, all functionals are defined to have support in the domain of the fractional integrals, so there is no need to extend the physics outside this domain.
  \item[(v)] The choice of measure \Eq{iavr} and \Eq{cross} may seem too restrictive within all the possible choices of calculus, but this is not the case. On one hand, the more general arbitrary Lebesgue--Stieltjes measure \Eq{mear} is technically untractable \cite{fra1,fra3}, and not much progress can be done even for the absolutely continuous measure \Eq{vrv} without specifying the profile $v(x)$ \cite{fra1,fra2}. On the other hand, there is no apparent advantage in choosing specific profiles different from \Eq{iavr} and \Eq{cross}, which is highly anisotropic in the embedding coordinates. For instance, one might object that anisotropic measures do not have the symmetries we observe in Nature, such as rotation invariance, and one might advocate isotropic measures such as a power-law profile $v(x)\sim |{\bm x}|^{-D(1-\a)}$. Integro-differential operators generating this type of measure constitute the so-called Riesz calculus \cite[Section 2.10]{KST}, which is another version of fractional calculus. In principle, it would be possible to define fractional spaces also with this calculus. However, the factorized measure \Eq{cross} allows us to define a natural set of geometric coordinates (Section \ref{gn}) and a simple differential structure (Section \ref{dis}), while mixing coordinate components would result in something more difficult to deal with. Furthermore, in fact, we do not need the measure to have any of the ordinary Poincar\'e symmetries, because these turn out to be accidental symmetries of fractional spacetimes in the infrared \cite{fra4,frc2}.  
  \end{enumerate}
\end{itemize}
Different choices of calculus are not excluded. In principle, they will lead to different quantitative details (that is, in the mapping between embedding and geometric coordinates and in the definition of the geometric integro-differential operators). Occasionally, we will compare the left theory with one with mixed operators. We saw that these operators do not share many of the simple properties enjoyed by each individual sector, and because of this we will not pursue a full description of a mixed theory. Nevertheless, an example of volume calculation will show how geometry changes from a single-sector to a mixed-sector formulation.

It is important to stress that inequivalent fractional spaces are expected to share the same physical features. The reason is that the scaling property \Eq{scavr2} of the fractional measure is unaffected by the choice of differential operators; see Sections \ref{volume} and \ref{dim}.


\subsection{Geometric notation}\label{gn}

The geometry of fractional space is characterized by a set of integro-differ-ential operators, specifying both the measure on the space and its boundary, and the notion of distance (shortest path) between two points. The latter can be and has been derived within fractional calculus, but these results are more transparent after introducing a novel notation, which we call ``geometric'' and also carries part of the physical interpretation of the model.

The central idea is to regard a \emph{fractional manifold} as embedded in an ordinary $D$-dimensional manifold constituting a mathematical ambient space. A fractional manifold is defined by the multiplet of objects \Eq{fres} but with more general embedding. By embedding we mean the ordinary manifold one would obtain by setting $\a_\mu=1$ in the measure for all $\mu$. A general embedding is a smooth manifold endowed with a metric $g_{\mu\nu}$. For instance, in \cite{frc2} the embedding defining fractional Minkowski spacetime $\cM^D_\a$ is ordinary Minkowski spacetime $M^D$, with metric $g_{\mu\nu}=\eta_{\mu\nu}$. In the case of fractional Euclidean space $\mathcal{E}^D_\a$, the embedding is $\mathbb{R}^D$ and $g_{\mu\nu}=\delta_{\mu\nu}$.

While $x^\mu$, $\mu=1,\dots,D$, are embedding coordinates, on the fractional manifold a natural coordinate system is provided by the ``fractional'' (or ``geometric'') coordinates $\x^\mu$, \Eq{fraco}. The symbol $\x$ leaves implicit both the coordinate dimensionality and the dependence on the embedding coordinate domain, but this does not differ substantially from what one does in ordinary geometry. In fact, hiding a fixed $\a$ in $\x$ is tantamount to hiding the scaling $[x]=-1$ in the symbol $x$. Similarly, one typically specifies a coordinate system $\{x\}$ and its domain ${\rm Dom}(x)$ separately, and not as a joint symbol $x_{{\rm Dom}(x)}$. The coordinates \Eq{fraco}, sometimes called ``generalized'', were introduced in the special case $x_0=0$ (and without the $\Gamma$ factor) in \cite{Tar1,Tar2,Tar3b,Tar4,Rie1,Rie2} in the context of dissipative mechanics.

Equations \Eq{218} and \Eq{218r} state that there is no natural geometric coordinate for Liouville (and also Weyl) calculus, i.e., no function $\x(x)$ such that ${}_{\infty}\p^\a\x=1$. In fact, the right-hand side of \Eq{fraco} diverges for $x_0=-\infty$ (e.g., for global Cartesian coordinates). This fact should not be of concern, since this is just a mapping stating how embedding and fractional coordinate systems are related to each other. A singular mapping does not imply a pathology in the embedding coordinate system, and in fact the final expressions of geometric integrals are perfectly well defined in the language of fractional calculus, even for $|x_{0,1}|=\infty$. In other words, even if there are no geometric coordinates in pure Liouville and Weyl calculi, fractional spaces equipped with these operators are still meaningful.

The ``geometric'' differential associated with the fractional coordinates $\x$ is just $\rmd^\a$:
\be
\bd:=\rmd^\a\,,\qquad [\bd]=0\,.
\ee
We change notation to avoid confusion between the label $\a$ and space indices $\mu,\nu,\dots$. From \Eq{diff2},
\be\label{rmda}
\bd = \bd\x\, \p^\a_\x\,,\qquad \bd\x = (\rmd x)^\a\,,
\ee
where the ``geometric'' derivative is the Caputo fractional derivative regarded as a function of $\x$, 
\be
\p^\a_\x := \frac{\bd}{\bd\x}=\p^\a_x;
\ee
in particular, $\p^\a_\x\x=1$. Note that \Eq{rmda} can be taken as the definition of $\x$ via $\rmd^\a\x$ also in Liouville calculus ($x_0=-\infty$; compare (2.9) in \cite{Svo87}). Also, $\p^\a_\x\neq \p_\x = (\p \x/\p x)^{-1}\p_x$: comparing with \Eq{leru2},
\ba
(\p^\a f)(x) &=&\frac{\a}{\Gamma(\a)\Gamma(2-\a)}\frac{\p f}{\p \x}+\frac{1}{\x}\frac{f(x)-f(x_0)}{\Gamma(1-\a)\Gamma(1+\a)}\nonumber\\
&&+\sum_{j=2}^{+\infty} \frac{\sin[\pi(j-\a)]}{\pi(j-\a)}\frac{\Gamma(1+\a)}{\Gamma(1+j)}(x-x_0)^{j-\a} (\p^j f)(x)\,.\label{leru3}
\ea
In many dimensions,
\be\label{diffmu}
\boxd{\bd: = \bd\x^\mu\, \p^\a_\mu\,,\qquad \p^\a_\mu:= \frac{\p^\a}{\p^\a\x^\mu}\,.}
\ee
By virtue of the unique property $\p^\a_\mu 1=0$ typical of the Caputo fractional derivative, $\p^\a_\mu\x^\nu=\delta_\mu^\nu$. The symbol $\p^\a_\mu$ will indicate both the partial fractional derivative with respect to $x^\mu$ and the one with respect to $\x^\mu$; the context should make the distinction clear.

Finally, we define the ``geometric'' integral
\be\label{geomi}
\fint_0^\x := \frac{1}{\Gamma(\a)}\int_{x_0}^x \left(\frac{\rmd x'}{x-x'}\right)^{1-\a}\,,\qquad [\fint]=0\,.
\ee
The symbol in the left-hand side of the first equation is borrowed from the standard notation for mean integrals, which are, of course, out of the present context. The right-hand side was first introduced by Tarasov \cite{Tar12}. A definite integral over the whole interval is simply obtained by setting $x=x_1$. In one embedding dimension,
\ba
\fint_0^\x \bd\x' &=& \frac{1}{\Gamma(\a)}\int_{x_0}^x \frac{(\rmd x')^{1-\a}}{(x-x')^{1-\a}}(\rmd x')^\a\nonumber\\
&=& I^\a\left\{1\right\}\nonumber\\
&=& \frac{(x-x_0)^\a}{\Gamma(1+\a)}\nonumber\\
&=& \x\,.\nonumber
\ea
Therefore, posed ${\rm f}(\x)=f(x)$, \Eq{nlf} is equivalent to
\be
\fint_0^\x \bd {\rm f}(\x') = {\rm f}(\x)-{\rm f}(0)\,,
\ee
stating that geometric integration in fractional coordinates has formally the same properties as the ordinary integral. By formally, we mean that there will be a difference in the functional space over which the integral operators act.

Geometric integrals with different lower extrema are obtained either by a initialization prescription \cite{LH00} or by the composition law 
\be
\fint_{\x_*}^{\x_1} := \fint_{0}^{\x_1} - \fint_{0}^{\x_*}\,,\qquad \x_{*,1}\equiv \frac{(x_{*,1}-x_0)^\a}{\Gamma(1+\a)}\,,\qquad 0\leq \x_*\leq \x_1\,.\nonumber
\ee
Multiple integrals in the coordinate system $\{\x^\mu|\,\mu=1,\dots D\}$ follow through,
\be
\fint \bd^{D}\x :=\fint\bd \x^1\cdots\fint\bd \x^D\,.
\ee
Each fractional coordinate $\x^\mu$ is mapped into an embedding coordinate $x^\mu$ with a given domain. Examples of nested integral will be seen in Section \ref{volume}.

We summarize the three notation systems employed so far in table \ref{tab1}.
\begin{table}
\begin{tabular}{lccc}\hline
       			    & Geometric        & Fractional calculus & Stieltjes measure  \\ 
       			    & formalism        & formalism           & formalism          \\\hline
Coordinates     & $\x$             & $(x-x_0)^\a/\Gamma(1+\a)$        & $x$		\\
Measure         & $\bd\x$          & $(\rmd x)^\a$       & $\rmd\vr(x)$       \\
Integration     & $\fint$       	 & $I^\a$    & $\int$                       \\
Differentiation & $\p^\a_\x$       & $\p^\a$   & $\p$              					  \\\hline      			 
\end{tabular}
\caption{Equivalent formalisms describing calculus on a fractional manifold. The Lebesgue--Stieltjes measure formalism, with generic measure $\vr$, is the most general but it is often impractical to perform calculations.}
\label{tab1}
\end{table}


\subsection{Metric and distance}\label{dis}

In the language of first-order general relativity, arbitrary coordinate transformations define frames which, in turn, determine the metric. Therefore, the notion of line element naturally emerges. First-order formalism is somewhat of an overkill when dealing with Euclidean space, but one can foresee obvious applications in more general scenarios.

Consider two coordinate systems $\{x^I\}$ and $\{y^\mu\}$, the first (denoted with capital Roman indices) being the Cartesian system and the second a generic curvilinear one. The exterior derivative of order $\a$ can be written in both systems as
\be\nonumber
(\rmd x^I)^\a \p^\a_I=\rmd^\a=(\rmd y^\mu)^\a \p^\a_{\mu}\,.
\ee
Applying \Eq{usef}, we get
\be
(\rmd x^J)^\a =  (\rmd y^\mu)^\a \p^\a_{\mu}\frac{[x^J(y)-x_0^J]^\a}{\Gamma(1+\a)}=: (\rmd y^\mu)^\a e_\mu^J\,,\label{xye}
\ee
where $e_\mu^J$, a $D\times D$ matrix, is the fractional generalization of the vielbein. To avoid confusion with space indices, we shall omit labels $\a$ for fractional vielbein and metric, using the same symbols as in ordinary space. 

Some example of coordinate transformations are given in \cite{CSN1} for Riemann--Liouville fractional calculus. We can easily give the general form of the fractional Jacobian $\cJ_\a$ in a calculus of order $\a$. Let $\cJ(y)=|\p x(y)/\p y|$ be the Jacobian of a coordinate transformation from the system of coordinates $x$ to $y$. The fractional Jacobian $\cJ_\a$ is simply $\cJ$ times the ratio of measure weight factors:
\be
\cJ_\a(y^1,\dots,y^D)=\frac{v_\a[x^1(y)]}{v_\a(y^1)}\cdots\frac{v_\a[x^D(y)]}{v_\a(y^D)}\, \cJ(y^1,\dots,y^D)\,.
\ee
For instance, for left integrals with upper terminal $x_*$
\be\nonumber
\frac{v_\a[x^1(y)]}{v_\a(y^1)}\cdots\frac{v_\a[x^D(y)]}{v_\a(y^D)}=\left[\frac{y_*^1-y^1}{x_*^1(y_*)-x^1(y)}\cdots\frac{y_*^D-y^D}{x_*^D(y_*)-x^D(y)}\right]^{1-\a}\,.
\ee
In compact notation, the integral measure transforms as
\be
\rmd\vr_\a(x)=\rmd\vr_\a(y)\, \cJ_\a(y)\,,\qquad \cJ_\a(y)=\frac{v_\a[x(y)]}{v_\a(y)}\,\cJ(y)\,.
\ee
As in ordinary calculus, the measure in the new coordinates does not factorize over the directions due to the non-trivial coordinate dependence of the fractional Jacobian.

Expressing the 1-form \Eq{1form} in $y$ coordinates and reversing the transformation,
\be
\omega = (\rmd x^J)^\a f_J(x)= (\rmd y^\mu)^\a e_\mu^J f_J[x(y)]= (\rmd x^I)^\a e_I^\mu e_\mu^J f_J[x(y)]\,,\nonumber
\ee
from which it follows the relation
\be
e_I^\mu e_\mu^J = \delta_I^J\,.
\ee
One can also define the \emph{fractional metric}
\be
g_{\mu\nu}:= \eta_{IJ} e_\mu^I e_\nu^J \,,
\ee
where $\eta_{IJ}=\delta_{IJ}$ is the Kronecker delta in Euclidean space. In turn, the fractional metric gives the fractional line element
\be\label{linel}
\rmd s^\a := \left[g_{\mu\nu} (\rmd x^\mu)^\a\otimes (\rmd x^\nu)^\a\right]^{\frac12}\,,
\ee
or, in geometric notation,
\be\label{gele}
\bd {\rm s}^2 = g_{\mu\nu} \bd\x^\mu\otimes \bd\x^\nu\,.
\ee
This result for fractional two-forms suggests a natural definition of the distance between two points. The metric of fractional Euclidean space $\cE_\a^D$ is $g_{\mu\nu}=\delta_{\mu\nu}$, so in geometric coordinates it is expressed as
\be
\Delta_\a(\x,\y):= \sqrt{\Delta_\a\x^\mu \Delta_\a \x_\mu} = \sqrt{[\Delta_\a(\x^1,\y^1)]^2+\dots+ [\Delta_\a(\x^D,\y^D)]^2}\,.
\ee
This only resembles an ordinary Euclidean distance, since $\Delta_\a(\x,\y)\neq |\x-\y|$. From \Eq{linel}, the coordinate distance in (length)$^1$ units is the $2\a$-norm
\be\label{dista}
\Delta_\a(x,y) := \left\{[\Delta(x^\mu,y^\mu)]^\a [\Delta(x_\mu,y_\mu)]^\a\right\}^{\frac{1}{2\a}} :=\left(\sum_{\mu=1}^D|x^\mu-y^\mu|^{2\a}\right)^{\frac{1}{2\a}},
\ee
where $\Delta(x^\mu,y^\mu)=|x^\mu-y^\mu|$. This is a norm only if $\a\geq 1/2$, i.e., when the triangle inequality holds. Therefore, we can further restrict $\a$ to lie in the range
\be\label{newra}
\boxd{\frac12\leq\a\leq1\,.}
\ee 
From the perspective of differential forms, the $2\a$-norm is \emph{the} natural distance in fractional space, which is a metric space. In fact, one should not confuse \Eq{dista} with the choice of a $p$-norm (all topologically equivalent) in a given space: as $\a$ changes, so does the geometry of space.

\begin{figure}
\includegraphics[width=9.6cm]{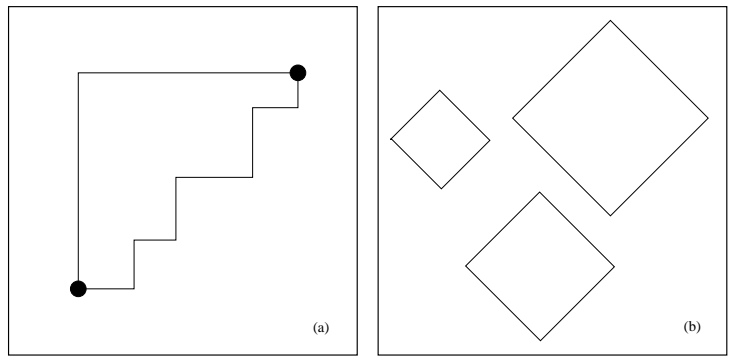}
\caption{1-norm and taxicab geometry in two dimensions. (a) Left panel: the shortest path between two points is not unique. (b) Right panel: circles of radius $R$ are diamonds with edges at $45^\circ$ with respect to the coordinate axes, $|x|+|y|=R$.}
\label{fig3}
\end{figure}

\begin{figure}
\includegraphics[width=5.5cm]{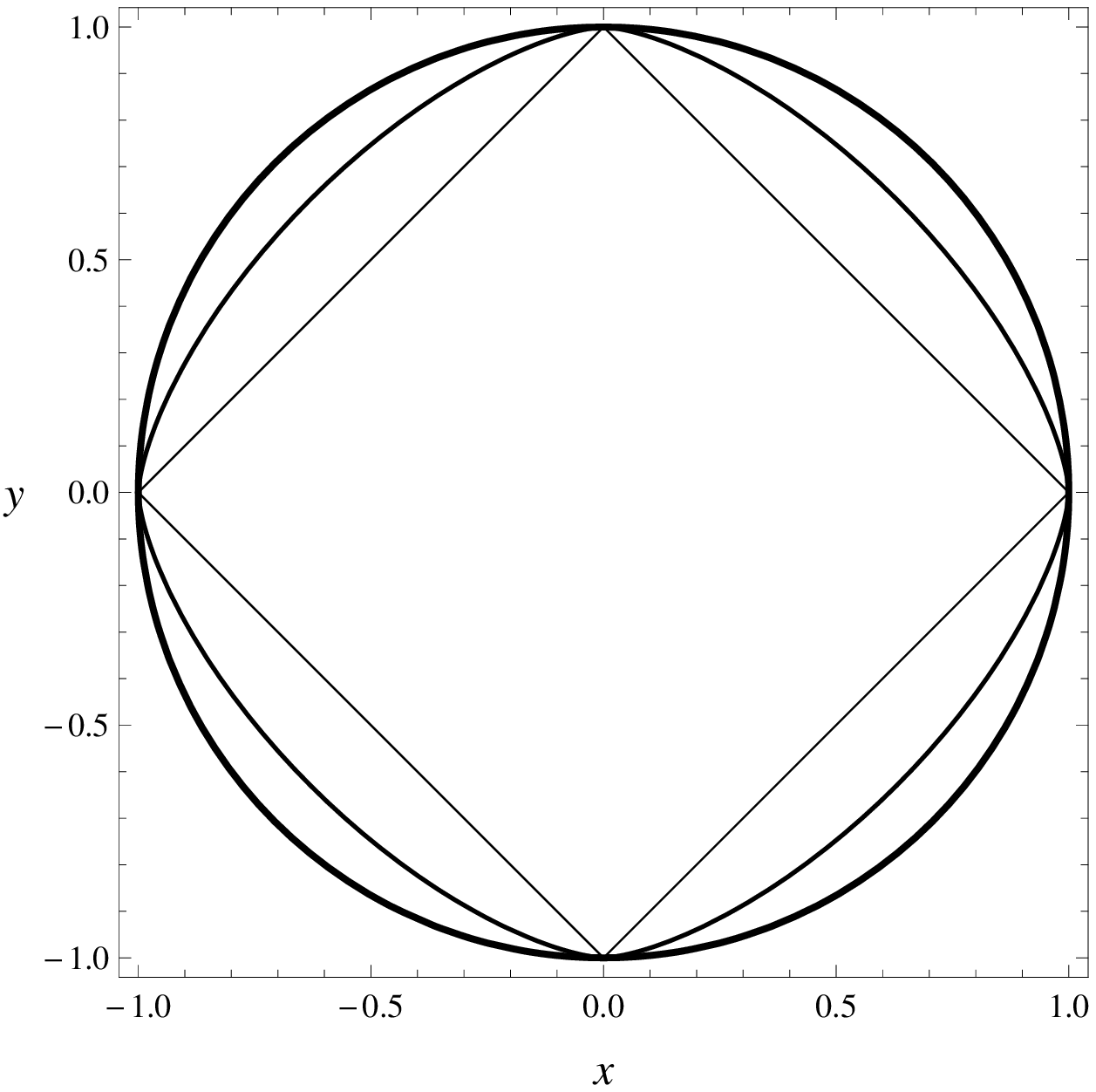} \hspace{1cm}
\includegraphics[width=5.5cm]{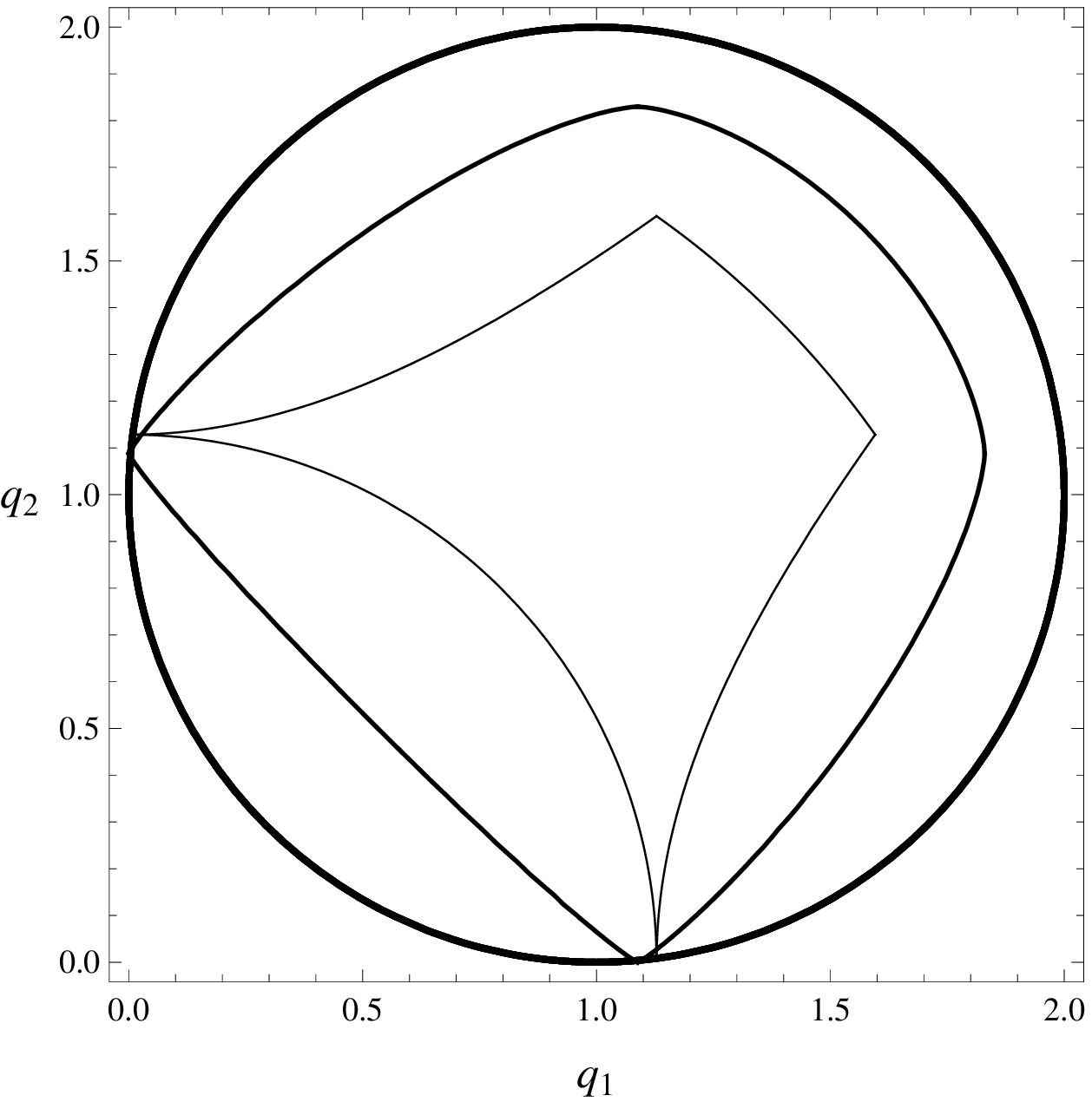}
\caption{Left: circles in two dimensions with unit radius in $2\a$-norm, $|x|^{2\a}+|y|^{2\a}=1$; increasing thickness corresponds to $\a=1/2,\,3/4,\,1$. Right: the same circles in left geometric coordinates $\x_1=(x-x_0)^\a/\Gamma(\a+1)$ and $\x_2=(y-y_0)^\a/\Gamma(\a+1)$, with $x_0=-1=y_0$.}
\label{fig4}
\end{figure}

\begin{figure}
\includegraphics[width=5cm]{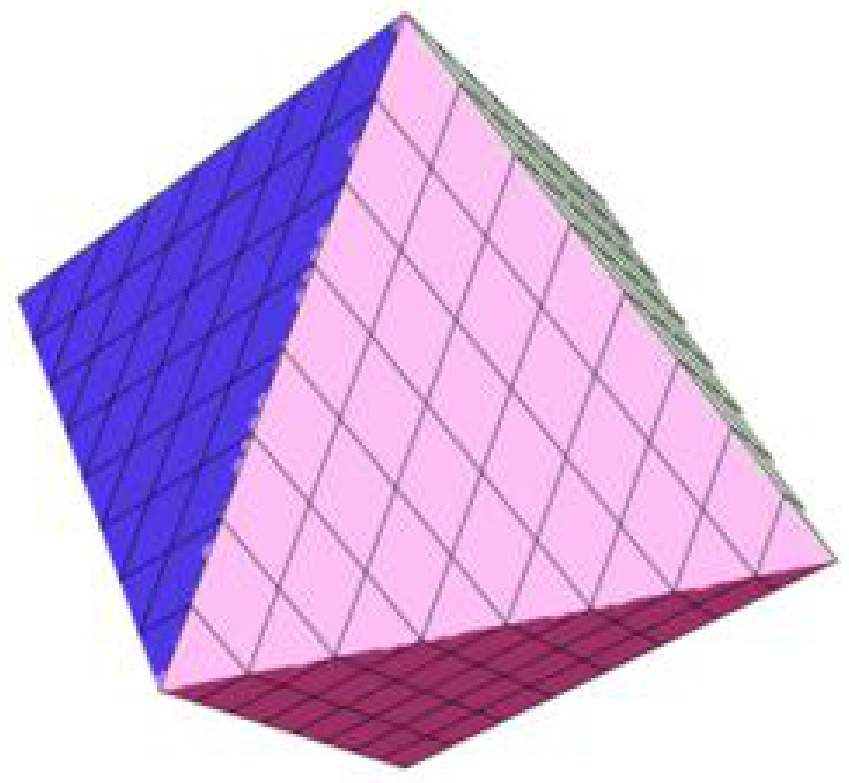}
\caption{Spheres in taxicab geometry are octahedra, $|x|+|y|+|z|=R$.}
\label{fig5}
\end{figure}

In a generic fractional geometry with $\a\neq 1$, the Pythagorean theorem is not valid and the shortest path between two points is neither a straight line (Euclidean distance) nor unique. 
The case $\a=1/2$ corresponds to the so-called ``taxicab'' or ``Manhattan'' distance, given by the rectilinear distance along the axes. In $D=2$, circles in this geometry are diamonds with edges at $45^\circ$; the inclination of the edges is fixed, taxicab distance not being rotation invariant (figure \ref{fig3}).
As $\a$ increases from $1/2$ to 1, the faces of the diamond become convex until they merge into an ordinary circle. These figures are called supercircles (a particular case of superellipse, or Lam\'e curve). See figure \ref{fig4}.
In $D=3$, taxicab spheres are octahedra (figure \ref{fig5}).


\subsection{Volume}\label{volume}

We have already described fractional operators in one dimension, and their replication to $D$ copies is straightforward. As one might expect, there is a fractional analogue of areas and volumes, but their scaling properties and values will differ from their ordinary Euclidean counterparts.

Let $\cM\subset \mathbb{R}^D$ be an arbitrary region in space. The fractional volume of $\cM$ is (e.g., \cite{Tar12})
\be\label{volu}
\cV_\cM^{(D)} :=\prod_{\mu=1}^D I^\a[x^\mu]\,,
\ee
where each fractional integral is defined on an interval $[x_0^\mu,x_1^\mu]\ni x^\mu$. These intervals can be always chosen so that they cover $\cM$. If $\cM$ is a rectangular region, then all the coordinates $x_0^\mu$ and $x_1^\mu$ are constant. Otherwise, one can compute the volume $\cV_\cM$ as a nested integration over elementary domains, as in ordinary calculus.

We give an example of multiple integral in the calculation, in $D=2$, of the volume $\cV^{(2)}(R)$ of a 2-ball with radius $R$ and centred at the origin.\footnote{Clearly, the final result will not depend on the location of the disc in the coordinate plane.} With the $\a$-norm distance \Eq{dista}, this is the area enclosed by a supercircle $\cB_2$, i.e., the set
\be
\cB_2=\left\{(x,y)~:~ |x|^{2\a}+|y|^{2\a}\leq R^{2\a}\right\}\,.
\ee
When $\a=1$, this is a disc with radius $R$; when $\a=1/2$, it is a diamond with vertices $(\pm R,0)$ and $(0,\pm R)$. We integrate first $y$ from $-r(x)$ to $r(x)$, where 
\be\label{rx}
r(x)=(R^{2\a}-|x|^{2\a})^{1/(2\a)}\,.
\ee
Then, we integrate in $x$ from $-R$ to $R$:
\ba
\cV^{(2)}(R) &=& I_{-R,R}^\a\left\{I_{-r(x),r(x)}^\a\{1\}\right\}\nonumber\\
               &=& I_{-R,R}^\a\left\{\frac{[2r(x)]^\a}{\Gamma(1+\a)}\right\}\nonumber\\
               &=& \frac{2^\a f_{2,\a}}{\Gamma(1+\a)}\,R^{2\a}\,,\label{2dball}
\ea
where
\ba
f_{2,\a} &=& \frac{1}{\Gamma(\a)}\int_{-1}^1\rmd x\,(1-x)^{\a-1} \sqrt{1-|x|^{2\a}}\nonumber\\
         &=& \frac{1}{\Gamma(\a)}\int_{0}^1\rmd x\, \sqrt{1-x^{2\a}}[(1-x)^{\a-1}+(1+x)^{\a-1}]\,.\label{f2a}
\ea
We will reconsider this prefactor later for general $D$. Other examples of double integrals can be found in \cite{Tar12}. 
Working in geometric notation, the two fractional coordinates are
\ba
\x_1&=&\frac{(x+R)^\a}{\Gamma(1+\a)}\in [0,\cR]\,,\qquad \cR:= \frac{(2R)^\a}{\Gamma(1+\a)}\,,\label{frax}\\
\x_2&=&\frac{[y+r(x)]^\a}{\Gamma(1+\a)}\in [0,\tilde r]\,,\qquad \tilde r:= \frac{[2r(x)]^\a}{\Gamma(1+\a)}\,.\label{fray}
\ea
Therefore,
\be
\cV^{(2)}(\cR) = \fint_0^\cR \bd\x_1\,\fint_0^{\tilde{r}(\x_1)}\bd\x_2= \fint_0^\cR \bd\x_1\,\tilde{r}(\x_1)= 2^{-\a}\Gamma(1+\a) f_{2,\a} \cR^2\,,\nonumber
\ee
coinciding with \Eq{2dball}. 

\begin{figure}
\includegraphics[width=4.8cm]{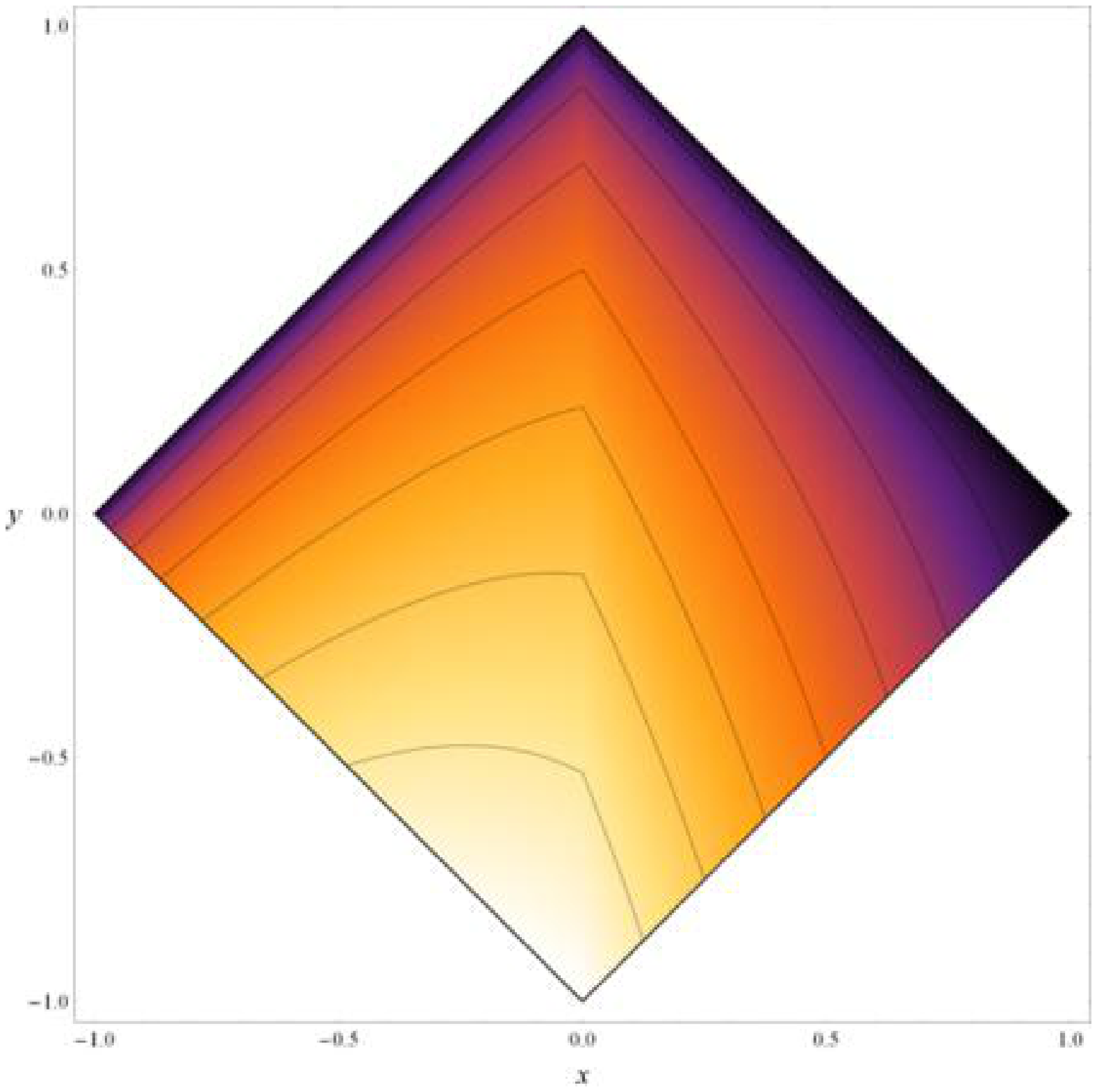}
\includegraphics[width=4.8cm]{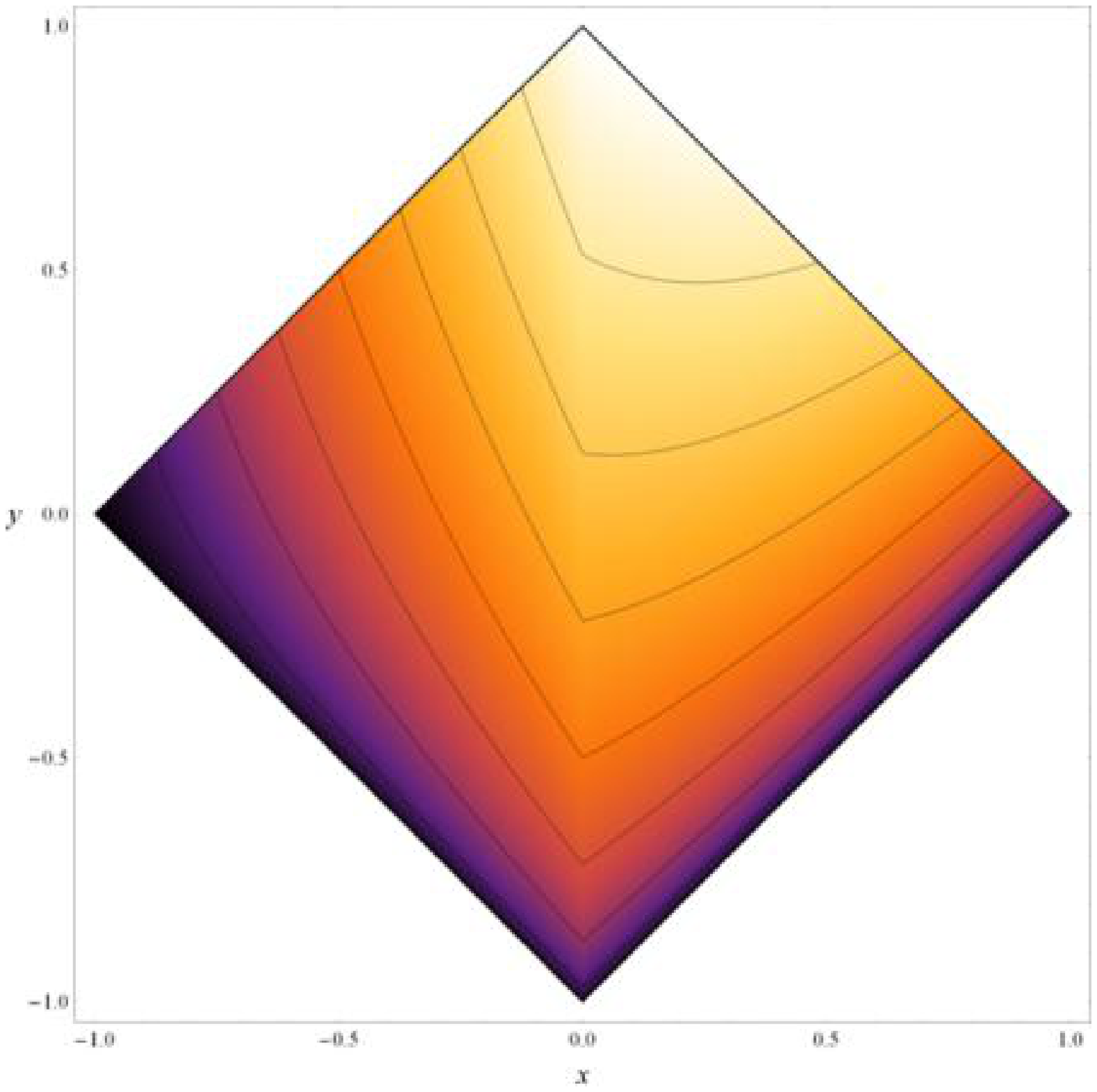}
\includegraphics[width=4.8cm]{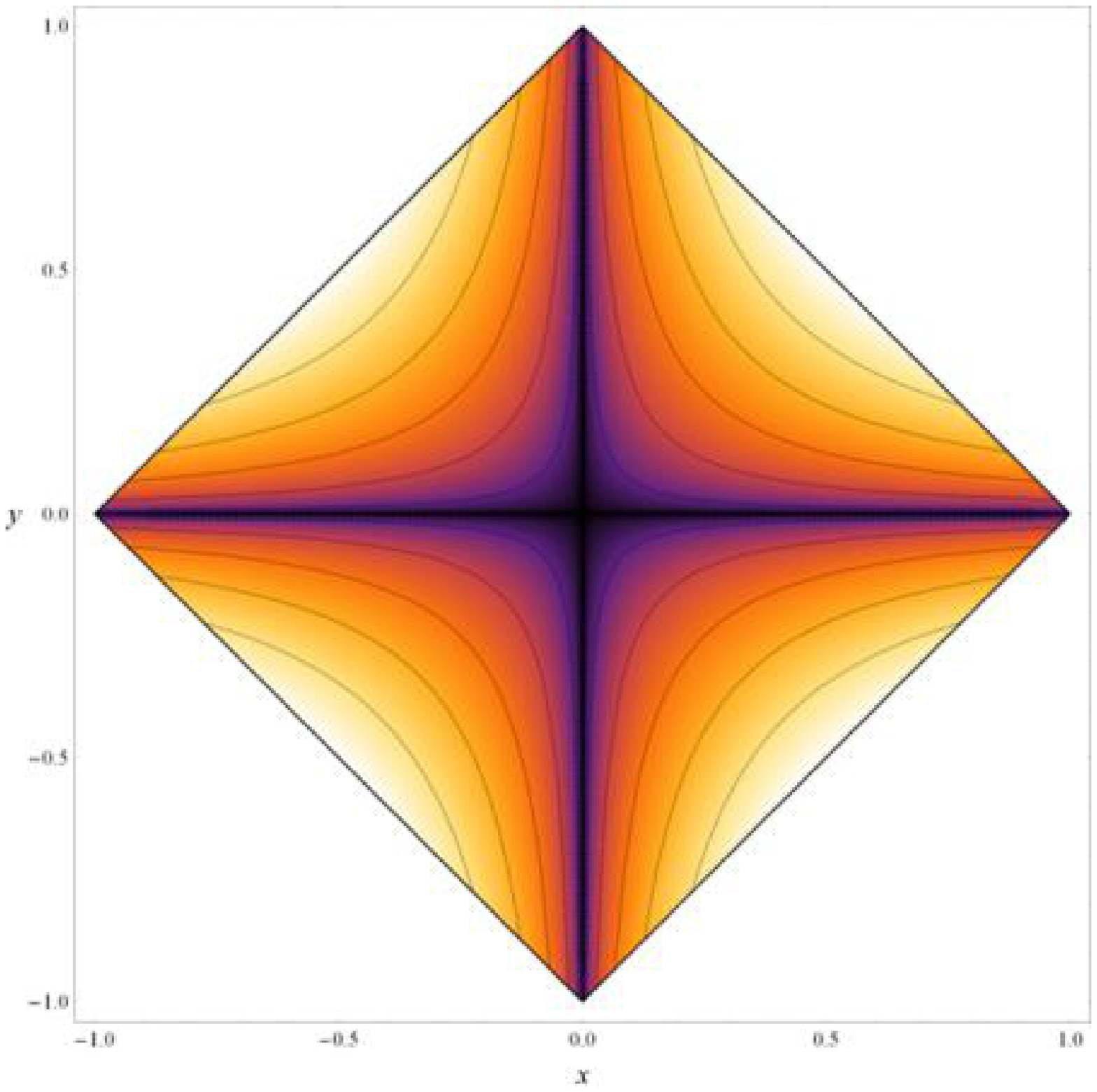}
\caption{Density plots of the area measure of a $1$-norm disc. The integration measure weight is represented in light to dark shade, darkest shade being points where it diverges. From left to right to bottom: left integration, right integration, mixed integration.}
\label{fig6}
\end{figure}

As we said, different presentations of the fractional operators lead to inequivalent fractional spaces, where areas and volumes are weighed according to the measure in the integral. For example, the measure weight of the $\a=1/2$ unit disc is $v(x,y)=[(1-x) (1-|x|-y)]^{-1/2}$, which is heavier for points $(x,y)\sim (1,y)$ and $(x,y)\sim (x,1-|x|)$, corresponding to the right vertex and to the upper edges of the diamond. This is depicted in the left panel of figure \ref{fig6}. Using right integration $\bar{I}^\a[x]\bar{I}^\a[y]$, the measure is $\bar{v}(x,y)=[(1+x) (1-|x|+y)]^{-1/2}$. However, due to the symmetry of the object (\Eq{f2a} is invariant under $x\to -x$), the value of the area is the same; this is not true in general. On the other hand, for mixed integration $\tI^\a[x]\tI^\a[y]$ with $(x_*,y_*)=(0,0)$, we have
\ba
\cV^{(2)}(R) &=& \tI_{-R,0,R}^\a\left\{\tI_{-r(x),0,r(x)}^\a\{1\}\right\}\nonumber\\
             &=& \tI_{-R,0,R}^\a\left\{2\bar{I}^\a_{0,r(x)}\{1\}\right\}\nonumber\\
             &=& \tI_{-R,0,R}^\a\left\{\frac{2[r(x)]^\a}{\Gamma(1+\a)}\right\}\nonumber\\
             &=& \frac{4}{\Gamma(1+\a)}\bar{I}_{0,R}^\a\left\{\sqrt{R^{2\a}-|x|^{2\a}}\right\}\nonumber\\
             &=& \frac{2\tilde f_{\a,2}}{\Gamma(1+\a)}\,R^{2\a}\,,\label{2dballt}
\ea
where we used the fact that the integrand is even and
\be
\tilde f_{2,\a} = \frac{2}{\Gamma(\a)}\int_0^1\rmd x\,x^{\a-1} \sqrt{1-x^{2\a}}\,.\label{tf2a}
\ee
Therefore, in this case the measure weight for $\a=1/2$ is $\tilde v(x,y)=|xy|^{-1/2}$, which diverges along the axes $x=0$ and $y=0$; see the right panel of figure \ref{fig6}.

The angle factors $f_{\a,2}$ and $\tilde f_{2,\a}$ represent the ratio of the area of the disc and a certain power of its radius. Measurements of both would then determine the angle factor and, in principle, provide an experimental discrimination between left/right and mixed theories. Let us recall the overall physical picture outlined in Section \ref{strate}. The fractional space constructed in this paper has fixed dimensionality and there is no dimensional flow. So, the dimension is always different (possibly very different) from $D$ at all scales. To get some physics, one must generalize to spaces which are multi-fractional and whose dimension changes with the probed scale. This is overviewed in \cite{fra4} and done in \cite{frc2}. In the full theory of \cite{frc2}, spacetime has dimension close to $D$ at large scales, and local geometry can be tested against deviations from the ordinary one. With this setting in mind, one can envisage local measurements of geometry in vacuum, made at scales and in conditions where gravitational effects (such as tidal forces) are negligible, to check if geometry in a local inertial frame is Euclidean. For instance, one could take experiments on the equivalence principle and Lorentz violation and, by reverse engineering, place experimental bounds on the angle factor. In standard theories, at human scales (say, below 1 km) Euclidean geometry holds and the theoretical value of the angle factor is $\pi$. However, if a fractional theory with dimensional flow was a correct description of Nature, we would expect the parameter $\a$ not to be exactly equal to 1 at sufficiently small scales. In other words, an expansion in $\a=1-\e/D$, $0<\e\ll 1$, would yield a correction to the Euclidean angle factor, which can be constrained from above by experiments  \cite{frc2}.

We can find the form of this correction in arbitrary dimension. To begin, we prove by induction that the volume of a closed $D$-ball 
\be
\cB_D=\left\{x^\mu~:~ \sum_{\mu=1}^D |x^\mu|^{2\a}\leq R^{2\a}\right\}
\ee
is
\be\label{ballVa}
\cV^{(D)}(R)=\Om_{D,\a} R^{D\a}\,,
\ee
where $\Om_{D,\a}$ is the volume of a unit ball. We first work with fractional calculus and then in geometric notation, to show how the latter is more transparent. Suppose \Eq{ballVa} true in $D-1$ dimensions for a ball of radius $r(x)=(R^{2\a}-|x|^{2\a})^{1/(2\a)}$. Then, integrating also in the direction $x$,
\ba
\cV^{(D)}(R) &=& I_{-R,R}^\a\left\{\cV^{(D-1)}[r(x')]\right\}\nonumber\\
               &=& \Om_{D-1,\a}\, I_{-R,R}^\a\left\{r^{(D-1)\a}(x')\right\}\nonumber\\
               &\ \stackrel{x=x'/R}{=}\ & \Om_{D-1,\a}f_{D,\a}\, R^{D\a}\,,\label{vol1}
\ea
where
\ba
f_{D,\a} &=&\frac{1}{\Gamma(\a)}\int_{-1}^1\rmd x\,(1-x)^{\a-1} (1-|x|^{2\a})^{\frac{D-1}{2}}\nonumber\\
&=&\frac{1}{\Gamma(\a)}\int_0^1\rmd x\,(1-x^{2\a})^{\frac{D-1}{2}} [(1-x)^{\a-1}+(1+x)^{\a-1}]\,.\label{fDa}
\ea
In geometric notation, the fractional coordinate over which one integrates is given by \Eq{frax}. In terms of the fractional radius $\cR$, the volume \Eq{ballVa} scales as 
\be\label{ballV}
\cV^{(D)}(\cR)=\om_{D,\a} \cR^D\,,\qquad \om_{D,\a}=\Om_{D,\a} \left[\frac{\Gamma(1+\a)}{2^\a}\right]^D\,.
\ee
Then, given a $(D-1)$-ball of radius $\tilde{r}=(2r)^\a/\Gamma(1+\a)$, one has
\ba
\cV^{(D)}(\cR) &=& \fint_{0}^{\cR} \bd\x\, \cV^{(D-1)}[\tilde{r}(\x)]\nonumber\\
               &=& \om_{D-1,\a}\fint_{0}^{\cR} \bd\x\, \tilde{r}^{D-1}(\x)\nonumber\\
               &=& 2^{-\a}\Gamma(1+\a)\om_{D-1,\a}f_{D,\a}\, \cR^D\,,\label{crd}
\ea
in agreement with \Eq{vol1}. Formally, this is the same calculation as for (one orthant of) a $D$-ball in Euclidean space, the difference being in the angular factor. Thus, from the point of view of the observer in the fractional manifold, the ball scales ordinarily. However, fractional coordinates/distances have anomalous scaling ($x\to \la x$ implies $\x\to \la^\a\x$), so the embedding volume of the ball scales as $R^{D\a}$. When $D=2$, one recovers the explicit result for the superellipse.

The angular factor $\Om_{D,\a}$ is obtained by solving the recursive equation $\Om_{D,\a}=\Om_{D-1,\a} f_{D,\a}$ with initial condition $\Om_{1,\a}=f_{1,\a}=2^\a/\Gamma(1+\a)$:
\be
\Om_{D,\a}=\prod_{n=1}^D f_{n,\a}\,.
\ee
We were unable to compute $\Om_{D,\a}$ explicitly in finite form for general $\a$, but one can do so for several special cases. For instance, when $\a=1/2$, one obtains \cite[3.197.3]{GR} 
\be
\Om_{D,\frac12}=\sqrt{2}\left(\frac{4}{\pi}\right)^{\frac{D}{2}}\prod_{n=2}^D\left[\frac{1}{n}+\frac{1}{n+1}\,{}_2F_1\left(\frac12,1;\frac{3+n}{2};-1\right)\right]\,.
\ee
When $\a=1-\e/D$, $0<\e\ll 1$, we get the standard result plus corrections. In fact,
\ba
f_{D,1-\frac{\e}{D}} &=& 2\int_0^1\rmd x\,(1-x^2)^{\frac{D-1}{2}}\nonumber\\
           &&-\frac{\e}{D}\int_0^1\rmd x\,(1-x^2)^{\frac{D-1}{2}}\left[2\gamma+\ln(1-x^2)-(D-1)\frac{x^2 \ln x^2}{1-x^2}\right]\nonumber\\
           &&+O(\e^2)\,,\nonumber
\ea
where $\gamma=-\psi(1)\approx 0.577$ is Euler constant and $\psi$ is the digamma function. Using formul\ae\ 3.251.1 and 4.253.1 of \cite{GR},
\be
f_{D,1-\frac{\e}{D}} = \frac{\sqrt{\pi}\Gamma\left(\frac{D+1}{2}\right)}{\Gamma\left(\frac{D}{2}+1\right)}\left\{1-\frac{\e}{2D}\left[2\gamma +\psi\left(\frac{D+1}{2}\right)-\psi\left(\frac{3}{2}\right)\right]\right\}+O(\e^2)\,.
\ee
Then,
\ba
\Om_{D,1-\frac{\e}{D}} &=& \prod_{n=1}^D f_{n,1-\frac{\e}{D}}\nonumber\\
           &=& \Om_{D,1} \left\{1-\frac{\e}{2} \left[2\gamma-\psi\left(\frac{3}{2}\right) +\frac{1}{D}\sum_{n=1}^D\psi\left(\frac{n+1}{2}\right)\right]\right\}+O(\e^2)\,,\nonumber\\
\ea
where
\be
\Om_{D,1}=\prod_{n=1}^D \frac{\sqrt{\pi}\Gamma\left(\frac{D+1}{2}\right)}{\Gamma\left(\frac{D}{2}+1\right)}=\frac{\pi^{D/2}}{\Gamma\left(\frac{D}{2}+1\right)}\label{som}
\ee
is the standard unit volume.

In tables \ref{tab2} and \ref{tab3} we compare these unit volumes with the standard ones, and with the result which would come from a traditional dimensional regularization procedure, where the topological dimension is formally expanded as $D-\e$:
\be
\Om_{D-\e,1}=\Om_{D,1}\left\{1-\frac{\e}{2}\left[\ln\pi-\psi\left(\frac{D}{2}+1\right)\right]\right\}+O(\e^2)\,.\label{some}
\ee
\begin{table}
\begin{tabular}{ccc}\hline
$D$ & $\Om_{D,1/2}$ & $\tilde \Om_{D,1/2}$ \\\hline
2   & $\sqrt{2}$ & $\frac{4}{\pi}\pi=4$ \\
3   & $(\sqrt{2}-1)\frac83\sqrt{\frac{2}{\pi}}\approx 0.88$ & $\frac{8}{\pi^{3/2}}\frac{4\pi}{3}\approx6$ \\
4   & $2\sqrt{2}(\sqrt{2}-1)\left(1-\frac{2}{\pi}\right)\approx 0.43$ & $\frac{16}{\pi^2}\frac{\pi^2}{2}=8$ \\\hline     
\end{tabular} 
\caption{Volume of unit $D$-balls in various dimensions for $\a=1/2$, in left/right and mixed theories (with $x^\mu_*=0$).}
\label{tab2}
\end{table}

Left fractional expressions are not as neatly symmetric as those of integer calculus because fractional integral operators are not even in the embedding coordinates. In the mixed theory, the volume of a $D$-ball scales as in \Eq{ballVa}, but with angle factor given by
\be
\tilde\Om_{D,\a}=\prod_{n=1}^D \tilde f_{n,\a}\,,
\ee
where
\be
\tilde f_{D,\a} = \frac{2}{\Gamma(\a)}\int_0^1\rmd x\,x^{\a-1} (1-x^{2\a})^{\frac{D-1}{2}}\,,\label{tfDa}
\ee
and $\tilde\Om_{1,\a}=\tilde f_{1,\a}=2/\Gamma(1+\a)$. This integral can be expressed in terms of $\Gamma$ functions and one has
\be\label{tom}
\tilde\Om_{D,\a}=\frac{\Om_{D,1}}{[\Gamma(1+\a)]^D}\,.
\ee
In particular, for $\a=1/2$ the volume is enhanced by a factor $(2/\sqrt{\pi})^D$, and in general it is considerably greater than in the left theory (table \ref{tab2}). Expanding in $\a=1-\e/D$, one has
\be
\tilde\Om_{D,1-\frac{\e}{D}}=\Om_{D,1}[1+\e (1-\gamma)]+O(\e^2)\,.
\ee
The coefficients in the $\e$ corrections have opposite sign with respect to the left/right and dimensional-regularization cases (table \ref{tab3}), and are one and the same for any $D$. As a consequence of \Eq{tom}, which define the effective constant $\tilde\pi:= \pi[1+\e(1-\gamma)/n]$, these corrections can be written relative to $\pi$ and are the same for $D=2n$ and $D=2n+1$ dimensions in the mixed theory. Anyway, $\e$ is expected to be constrained to extremely small values by experiments, so the actual coefficients in front of it are not important except for highlighting this concrete comparison of inequivalent theories.
\begin{table}
\begin{tabular}{cccc}\hline
$D$ & $\Om_{D,1-\e/D}$            & $\tilde \Om_{D,1-\e/D}$     & $\Om_{D-\e,1}$ \\\hline
2   & $\pi(1-0.42\e)$             & $\pi(1+0.42\e)$             & $\pi(1-0.36\e)$ \\
3   & $\frac{4\pi}{3}(1-0.54\e)$  & $\frac{4\pi}{3}(1+0.42\e)$  & $\frac{4\pi}{3}(1-0.22\e)$\\
4   & $\frac{\pi^2}{2}(1-0.63\e)$ & $\frac{\pi^2}{2}(1+0.42\e)$ & $\frac{\pi^2}{2}(1-0.11\e)$ \\\hline      			 
\end{tabular}
\caption{Volume of unit $D$-balls in various dimensions, for $\a\sim 1$, in left/right and mixed theories (with $x^\mu_*=0$). The corrections in traditional dimensional regularization are shown in the last column.}
\label{tab3}
\end{table}


\subsection{Hausdorff dimension of space}\label{dim}

When dealing with exotic sets, it is important to define a sensible notion of dimension. Sometimes, the imprecise name ``fractal dimension'' is used to indicate one or more among the many possible (and inequivalent) definitions of dimension, which may create much confusion. Here we specialize to one such definition, the Hausdorff dimension. Before doing so, we make general remarks on dimension counting \cite{Fal03}.

Let $\cF$ be an object living in a $D$-dimensional space. To measure its volume, one can take the minimum number $N(\delta)$ of $n$-balls, $n\leq D$ with radius $\delta$ centred at points in $\cF$ and such that they cover $\cF$ (i.e., each point in $\cF$ lies in at least one ball). The number $N(\delta)$ increases as $\delta$ decreases, approaching the behaviour $N(\delta)\sim \delta^{-d_{\rm B}}$ as $\delta\to 0$. Then, the number
\be\label{bocodi}
d_{\rm B}:=-\lim_{\delta\to 0} \frac{\ln N(\delta)}{\ln \delta}
\ee
is called the \emph{box-counting dimension} of $\cF$. (Strictly speaking, there exist a lower and an upper box-counting dimension, given by the $\lim {\rm inf}$ and $\lim {\rm sup}$, respectively; when they coincide, they reduce to \Eq{bocodi}.) For instance, if $\cF$ is a square or a disc, the covering of $2$-balls will show that $d_{\rm B}=2$, if it is a cube or a $3$-ball, a $3$-ball covering will give $d_{\rm B}=3$, and so on. Very irregular or fractal sets will not be smooth and their dimension will be, in general, non-integer (although there exist also fractals with integer dimension). Intuitively, a set with many irregularities will require more balls for being covered, and their number will increase faster than expected; a typical example is an irregular porous surface, for which $d_{\rm B}>2$ \cite{PA}. On the other hand, a surface with ``too many holes'' may require less balls than a smooth one. For a given $n$, the shape of the covering sets is not important, and one could use, for instance, $n$-cubes of edge length $\delta$ instead of $n$-balls; what matters is how the volume of the probe scales with its size.

The box-counting dimension is only a particular definition of dimension, and it often proves to have a number of inconvenient properties. It is desirable to have a different (not just more general) notion of dimension as follows. The idea is similar, namely, to define a ``minimal'' covering for $\cF\subset\mathbb{R}^D$, but now taking covering sets of different size. Let $|U|={\rm sup}\{\Delta(x,y):x,y\in U\}$ be the diameter of a set $U\subset\mathbb{R}^D$, i.e., the greatest distance $\Delta(x,y)$ between two points in $U$. A $\delta$-cover of $\cF$ is a countable or finite collection of sets $\{U_i\}$ of diameter at most $\delta$ that cover $\cF$: $\cF\subset \bigcup_i U_i$, with $0\leq |U_i|\leq\delta$ for all $i$. If $s\geq0$ is a real non-negative parameter, one can define
\be
\vr^s_{\rm H}(\cF) := \lim_{\delta\to 0}{\rm inf}\left\{\sum_i|U_i|^s~:~\{U_i\}~\textrm{is a $\delta$-cover of $\cF$}\right\}\,.
\ee
This limit exists (it can be also 0 and $+\infty$) and is a measure, the $s$-dimensional Hausdorff measure of $\cF$. One can check that it is proportional to the $n$-dimensional Lebesgue measure for integer $s=n\leq D$ (length, area, volume, and so on). The Hausdorff measure obeys the scaling property (in coordinate notation)
\be\label{scal}
\vr^s_{\rm H}(\la x)=\la^{s}\vr^s_{\rm H}(x)\,,
\ee
where $\la>0$ is the scale factor of a dilation $x\to \la x$. One can show that $\vr_{\rm H}^s$ is non-increasing with $s$ and there exists a critical value of $s$ at which the measure jumps from $+\infty$ to 0. This is the \emph{Hausdorff dimension} (or Hausdorff--Besicovitch dimension) of $\cF$ \cite{Hau18}:
\be\label{defhau}
\dh(\cF) := {\rm inf}\{s~:~\vr^s(\cF)=0\}={\rm sup}\{s~:~\vr^s(\cF)=+\infty\}\,.
\ee
This definition allows one to calculate $\dh$ via the behaviour of the Hausdorff measure. The latter diverges for $s<\dh$, is zero for $s>\dh$, and may be 0, $+\infty$ or finite at $s=\dh$. In general, $d_{\rm B}\neq \dh$.

Taking balls as the covering sets $U_i$, one defines a measure which jumps at the same critical value $\dh$ of the Hausdorff measure \cite{Fal03}. This determines a local, operational definition of the Hausdorff dimension $\dh$ of a smooth set such as an integer or a fractional manifold of topological dimension $D$: $\dh$ is given by the scaling law for the volume $\cV^{(D)}$ of a $D$-ball of radius $R$:
\be\label{ballVh}
\cV^{(D)}(R)\propto R^{\dh}\,.
\ee
Therefore,
\be\label{opdeH}
\dh = \lim_{\de\to 0}\frac{\ln \cV^{(D)}(\de)}{\ln \de} = \lim_{\de\to 0}\frac{\ln \vr[\cB_D(\de)]}{\ln \de}\,,
\ee
where in the last step we formally expressed the volume as the measure of a $D$-ball.

Thus, in Section \ref{volume} we implicitly proved that the Hausdorff dimension of isotropic fractional Euclidean space is
\be\label{dhD}
\boxd{\dh =D\a\,.}
\ee
Also, we calculated volume corrections for a nearly integer dimension, when $\a\sim 1$ and
\be
\dh=D-\e\,.
\ee
The scaling of $\cV^{(D)}$ reproduces the estimates of fractal distributions in the fractional continuum approximation \cite{Tar1,Tar2,Tar3b,Tar4,Tar3}--
\cite{Tar8} as well as the heuristic scaling of general Lebesgue--Stieltjes measures in certain regimes \cite{fra1,fra2,fra3}.

Since fractional space is smooth, one expects most of the inequivalent definitions of fractal dimension to collapse one into the other. For instance, the Hausdorff dimension of the product $\cF_1\times\cF_2$ of two fractals is greater than or equal to the sum of the Hausdorff dimensions of the two sets, but it is strictly equal if the upper box-counting dimension coincides with the Hausdorff dimension for either $\cF_1$ or $\cF_2$ \cite[Corollary 7.4]{Fal03}. The box-counting dimension of a fractional ball is the same as its Hausdorff dimension, 
\be
d_{\rm B}=\dh\,,
\ee
hence the result \Eq{dhD}. In Section \ref{sesi}, we will reobtain \Eq{dhD} two more times by symmetry arguments of fractal geometry.

The class of fractional spaces $\cE_\a^D$ is bounded by two limiting cases. When $\a=1$ in all directions, one recovers $D$-dimensional Euclidean space, where the measure weight is uniform and all points are on an equal footing. A particle has full memory of its past history and the dynamics is determined by certain initial conditions. On the opposite side, when $\a=0$ the measure weight is peaked at the boundary of space, but integration reduces to the identity operator. The action is defined at a point and the space is zero-dimensional. This ``Pointland'' universe has no memory whatsoever of its history and, because it has no extension, it has no dynamics at all.


\section{Fractional versus fractal}\label{fractals}

As soon as fractals made their appearance in the literature \cite{Man67}, the anomalous scaling of fractional measures induced the perception that certain phenomena with fractal properties might be described by fractional calculus \cite{MaV}. Later, it was argued that the fractional charge $\a$ is related to the Hausdorff dimension of certain fractal objects \cite{KoG,tat95}. Fractional equations can approximate, in some sense to be made precise, self-similar deterministic (also known as nested) fractals such as the Cantor set \cite{Nig92} and von Koch curves \cite{tat95}, and random fractals such as (the trail and graph of) Brownian and fractional Brownian motion \cite{tat95}. Criticism on the results of \cite{Nig92} and on the connections between fractional calculus and fractals \cite{Rut1,Rut2} led to their clarification for the Cantor set \cite{LMNN,NLM} and their progressive generalization to self-similar sets (finite or infinite) generated by linear mappings (random self-similar \cite{LMNN,NLM}, self-similar deterministic \cite{RYS,LMNN,NLM}, and generalized self-similar sets \cite{RYS}), generalized cookie-cutter sets where only the first similarity is linear \cite{YRZ,RYZLN}, generalized net fractals (defined by contractions) where only the first mapping is linear \cite{Yu99} and, finally, generalized net fractals generated by non-linear mappings \cite{QL}--
\cite{RLWQ}. 

By now, it is established that fractional systems are not indiscriminately equivalent to fractal systems. On one hand, there are features of deterministic fractals which are not reproduced by the simplest fractional systems. On the other hand, random fractals are indeed describable by fractional tools (Section \ref{fmaf}). Dynamical systems with fractal properties in certain static regimes have been successfully modeled by chaotic and fractional systems \cite{KST,Das08}. Fractional differential equations (such as the generalization of the Fokker--Planck--Kolmogorov equation) well describe, for instance, self-similar dynamics, L\'evy flights, and anomalous diffusion in chaotic Hamiltonian systems and systems close to thermal equilibrium \cite{Tar9,Tar1,Tar3b,Tar4,OSP}--
\cite{WeZ} (see \cite{MeK,Zas3} for reviews). Fractional Brownian motion \cite{Fal03} is also related to fractional calculus. Fractal domains characterized by a mass distribution or correlation functions with anomalous scaling are, in general, very irregular at small scales; this is the case in many physical systems such as porous materials, colloidal aggregates and branched polymers \cite{PA,AFP}--
\cite{ScK}. These media, however, can be considered as continuous at scales much larger than the characteristic size of the irregularities, such as the pores in porous media \cite{Tar3}. Fractional systems, therefore, can be regarded as continuum approximations where the detailed microscopic structure of these materials is smoothened without loosing anomalous scaling \cite{Tar3,Tar7,Tar8}. This picture holds not only for mass distributions but, for instance, also in the description of the propagation of electromagnetic waves in dielectric media \cite{Tar13}, and in other applications (e.g., \cite{Tar8} and references therein).

To understand in what sense fractional models describe fractals, and whether fractional Euclidean space $\cE^D_\a$ is a fractal, we first fix the rules of the game, analyze what properties a fractal should have, and compare these properties with those of $\cE_\a^D$. This is a natural starting point where to draw a more precise comparison. Perhaps the most universal qualities of fractals are \cite{Fal03}
\begin{enumerate}
\item[1.] A fine structure;
\item[2.] An irregular structure;
\item[3.] Self-similarity.
\end{enumerate}
Despite the fact that there are counterexamples of fractals not possessing one or more of these features, one must rely on descriptive properties rather than on a sharp mathematical definition. As a matter of fact, there does not exist a unique definition of ``fractal'', other than ``I know one when I see one'' \cite{Str03}.


\subsection{Fine and irregular structure}\label{finir}

A fractal $\cF$ has a fine structure if it has detail at every scale. Intuitively this means, first of all, that one can zoom indefinitely into a fractal and always meet points belonging to $\cF$, and, secondly, that in doing so one will always see non-trivial details. A smooth manifold $\cM$ as well as fractional space $\cE^D_\a$ can be zoomed in indefinitely (they are continuum structures) but they lack details at all scales.

Typically, fractals are also too irregular to be described with traditional geometric tools. This means that ordinary calculus does not apply to very discontinuous sets, and one must resort to rather advanced techniques to define measures, Laplacians, spectral theory, and so on. A smooth manifold $\cM$ does not satisfy this property. Fractional space $\cE_\a^D$ does by definition, although this is associated with an asymmetry of measure weights rather than manifest irregularity. At this point it becomes clear why fractional models are regarded as approximations of certain fractals: they do possess properties 1.\ and 2., but in a rather ``dull'' way. So $\cE_\a^D$ is a fractal, technically, albeit of a rather uninteresting type as far as these properties are concerned.

This is true only for fractional Euclidean space, where $\a$ is \emph{real} and \emph{fixed}. When lifting both these assumptions, and allowing $\a$ to be complex-valued and vary with the scale, it turns out that the structure becomes extremely rich, and much closer to that of genuine multi-fractal sets \cite{frc2}.


\subsection{Self-similarity and self-affinity}\label{sesi}

Many fractals are self-similar, either exactly, approximately, or statistically. Roughly speaking, a set $\cF$ is exactly self-similar if it is made of $N(\la)$ copies of itself of scale $\la$. Then, the \emph{similarity dimension} or \emph{capacity} of the set is \cite{Kol58,FOY}
\be\label{simd}
d_{\rm C} := -\frac{\ln N(\la)}{\ln \la}\,.
\ee
For instance, a hypercube in $D$ dimensions can be thought of as the union of $N=2^D$ copies of itself, each of size $\la=1/2$ with respect to the original. Then, $d_{\rm C}=D$. One can take finer subdivisions in $N(\la)= (1/\la)^D$ copies of size $\la$, and obtain the same result. In this sense, ordinary Euclidean space is trivially self-similar: $\la$ can be chosen arbitrarily. Genuine fractals have a more interesting self-similarity: for instance, the von Koch curve is the composition of 4 copies, each $1/3$ of the original, hence $d_{\rm C}=\ln 4/\ln 3$; the Cantor set is made of 2 copies of the original, each scaled $\la=1/3$, hence $d_{\rm C}=\ln 2/\ln 3$; and so on. In most cases, like those just mentioned, the similarity dimension coincides with the box-counting dimension \Eq{bocodi} and the Hausdorff dimension  \cite[Section 9.2]{Fal03}. It is instructive to see this via the rigorous definition of self-similar sets. The ensuing calculations, employing the technique of contractions so often used in fractal analysis, will be more lengthy than those stemming from \Eq{simd}, but they will also allow us to obtain very precise information about the structure of fractional space and its Hausdorff dimension.

We begin with sets in $\mathbb{R}^D$. Consider a set of $N$ maps $\cS_i\,:\,\mathbb{R}^D\to\mathbb{R}^D$, $i=1,\dots,N\geq 2$, such that
\be\label{simi}
\Delta[\cS_i(x),\cS_i(y)]\leq \la_i\Delta(x,y)\,,\qquad x,y\in\mathbb{R}^D\,,\qquad 0<\la_i<1\,,
\ee
where the distance $\Delta$ between two points is $\Delta(x,y)=|x-y|$ in ordinary integer geometry. Any such map is called \emph{contraction} and the number $\la_i$ is its \emph{ratio}. If equality holds, $\cS_i$ is a \emph{contracting similarity} or simply a similarity; if, moreover, $\la_i=1$, it is an \emph{isometry}. Thus, a similarity transforms a subset of $\mathbb{R}^D$ into another set with similar geometry. Many fractals are invariant under contraction maps and can be expressed as the union
\be\label{sss}
\cF= \bigcup_{i=1}^N \cS_i(\cF)\,.
\ee
Given $N$ contraction maps $\cS_i$, $\cF$ exists, is unique, non-empty and compact. The writing \Eq{sss} is an ``embedding'' definition of a fractal, which is thought of as a subset of Euclidean space. There is also a topological presentation which requires no embedding. For any non-empty compact set $U$ one can define the transformation
\be
\cS(U) :=\bigcup_{i=1}^N \cS_i(U)
\ee
and its $k$-th iterate $\cS^k:=\cS\circ \cdots \circ \cS$. If $\cS_i(U)\subset U \supset\cF$ for all $i$, then one can show that
\be\label{prefr}
\cF= \bigcap_{k=1}^\infty \cS^k(U)\,.
\ee
In practice, this means that a fractal can be constructed by iterations of contractions. The $k$-th iteration may be regarded as a pre-fractal, an approximation of $\cF$. The Hausdorff dimension of these fractals is bounded from above \cite{Fal03}:
\be\label{ssm0}
\dh(\cF)\leq s\,,\qquad {\rm where}\qquad \sum_{i=1}^N \la_i^s=1\,.
\ee

When the $\cS_i$ in \Eq{sss} are similarities, the attractor $\cF$ is called an exactly (or strictly) \emph{self-similar set}, and it is a union of smaller copies of itself \cite{Hut81}. In general, one further requires that the similarities obey the open set condition: namely, there exists a non-empty open bounded set $U\supset\cF$ containing a disjoint union of its copies,
\be
U \supset \bigcup_{i=1}^N \cS_i(U)\,.
\ee
If this condition holds, then one can prove that
\be\label{ssm}
\boxd{\sum_{i=1}^N \la_i^{\dh(\cF)}=1\,.}
\ee
A heuristic quick proof of this formula makes use of the scaling property \Eq{scal} of the Hausdorff dimension. For a self-similar set \Eq{sss},
\be
\vr^s_{\rm H}(\cF)=\sum_{i=1}^N\vr^s_{\rm H}[\cS_i(\cF)]=\sum_{i=1}^N \la_i^s\vr^s_{\rm H}(\cF)\,,
\ee
and assuming that $\vr^s_{\rm H}(\cF)$ is finite at the critical value $s=\dh(\cF)$, one can divide by $\vr^s_{\rm H}(\cF)$ to obtain \Eq{ssm}.

Thus, if one can define a set $\cF$ via similarities, \Eq{ssm} gives the Hausdorff dimension of $\cF$. Let us reconsider some of the examples mentioned at the beginning of the section. The middle-third Cantor set is defined by
\be\label{mtcs}
\cS_1(x)=\frac13 x\,,\qquad \cS_2(x)=\frac13 x+\frac23\,;
\ee
then, $1=2(1/3)^{\dh}$ implies $\dh =\ln 2/\ln 3$. Another self-similar fractal is the Sierpi\'nski triangle or gasket, the attractor of three similarities of ratio $1/2$; there, $1=3(1/2)^{\dh}$ implies $\dh =\ln 3/\ln 2$. Other fractals, such as the non-linear Cantor set, can be defined by non-linear contractions or similarities. A more trivial but, for our purpose, instructive case is $\cF=\mathbb{R}^D$. In fact, one can work instead with  a compact subset, the unit hypercube $\tilde\cF=[0,1]^D=[0,1]\times\dots\times[0,1]$, where the extrema are chosen without loss of generality; then, the natural unbounded extension of the set will have the same Hausdorff dimension. Take first the case $D=1$ and the interval $\cF=[0,1]$. This can be expressed as the union $[0,1]=[0,\la]\cup[\la,1]$, where $0<\la<1$.\footnote{The argument can be carried out \emph{verbatim} for the semi-open interval $[0,1)$, where $\cS_1(\tilde\cF)\cap\cS_2(\tilde\cF)=\emptyset$.} But this is equivalent to define two similarities such that $\tilde\cF=\cS_1(\tilde\cF)\cup \cS_2(\tilde\cF)$, where
\bs\label{Rdsim}\ba
\cS_1(x) &=& \la x\,,\qquad\qquad\qquad\,\, \la_1=\la\,,\\
\cS_2(x) &=& (1-\la)x+\la\,,\qquad \la_2=1-\la\,.
\ea\es
These maps satisfy the open set condition, as one can verify simply by taking open intervals. Therefore,
\be\nonumber
1=\la^{\dh}+(1-\la)^{\dh}\qquad \Rightarrow \qquad \dh =1\,.
\ee
In more than one dimension, it is convenient to set $\la=1/2$. In two dimensions, one has a square given by four smaller copies, and $4(1/2)^{\dh}=1$ yields $\dh=2$. In general, a hypercube can be expressed by $N=2^D$ similarities (as many as the number of orthants) with equal ratios $\la_i=1/2$, so that
\be
1=2^D\left(\frac12\right)^{\dh}\qquad \Rightarrow \qquad \dh =D\,.
\ee
$\mathbb{R}^D$ is not a genuine self-similar fractal because it is not defined by similarities with a \emph{fixed} $\la$: the similarity ratio $0<\la<1$ is arbitrary. Also, a finite number of iterations $\cS^k$ (actually, just one, $k=1$) is sufficient to obtain the set, which coincides with its pre-fractal approximation. In this last sense, Euclidean space is not a non-trivial self-similar fractal. Furthermore, $\mathbb{R}^D$ is not completely characterized by the similarities \Eq{Rdsim}. In the above construction for a unit hypercube, the similarity $\cS_1(x^\mu)=\la x^\mu+a$ is just a contraction and translation equal along all directions, but $\mathbb{R}^D$ enjoys many more symmetries, including inequivalent contractions/dilations along different directions, rotations ($\Delta[\cS(x),x_0]=\Delta(x,x_0)$), translations ($\cS(x^\mu)=x^\mu+a^\mu$, where also $a$ now is a vector) and reflections ($\cS(x^\mu)=a^\mu-x^\mu$). These are examples of \emph{affine transformations}, linear mappings of the form
\be
{x'}^\mu =\cS(x^\mu) = \cA^\mu_\nu x^\nu+a^\mu,
\ee
where $\cA$ is a $D\times D$ matrix. Linear similarities are a particular case of affine transformations. The attractor of a sequence of affine transformations is a \emph{self-affine} set:
\be\label{safs}
\cF= \bigcup_{i=1}^N [\cA_i(\cF)+a_i]\,.
\ee
Typical examples of self-affine fractals are ``fern-like'' and ``tree-like'' sets. It is rather difficult to find general results on the dimension of self-affine sets, and formul\ae\ such as \Eq{ssm0} or \Eq{ssm} are no longer valid \cite{Fal03}. From our perspective, it is sufficient to note that if one can show that a set $\cF$ is self-similar under a certain sequence of similarities, then \Eq{ssm} allows one to find its Hausdorff dimension, while if one only knows that a set $\cF$ is self-affine, the calculation of the dimension may become less clear. 

In the trivial example of $\cF=\mathbb{R}^D$, one knows $\dh(\cF)$ from a direct calculation of volumes. Then, one notices that $\cF$ is self-affine, but in particular it is invariant under certain similarities $\cS_i$. Using this last property, we have recalculated the Hausdorff dimension. This exercise would be pointless were it not for the insight it can give us for the fractional Euclidean space $\cE_\a^D$. In fact, while the symmetries of fractals are used to calculate or estimate their dimension, here we can reverse the logic and ask what the symmetries characterizing fractional space are. We know its Hausdorff dimension from the operational definition (scaling law of $D$-ball volumes). Therefore, because of \Eq{scal} and \Eq{scavr2}, \emph{if} $\cE_\a^D$ is self-similar, then \Eq{ssm} holds with $\dh=d_{\rm C}$. Sometimes, the scaling property of the measure is erroneously taken as the definition of self-similarity, but we have seen that scaling alone is not sufficient to guarantee self-similarity. Thus, we would like to go into some detail in the symmetry structure of fractional spaces. 

Already in one dimension, however, we see a difference with respect to the ordinary Euclidean case. Take, as before, the unit interval $[0,1]$. If we took two similarities 
\be\label{prosi}
\cS_{\a,1}(x):=\la^{\frac1\a}x\,,\qquad \cS_{\a,2}(x):=(1-\la)^{\frac1\a}x+a\,,
\ee
for some $\la$ and $a$, by \Eq{ssm} we would obtain the expected dimension $\dh=\a$, but the resulting set would be a Cantor dust. In its first iteration, it would be the union of the intervals $[0,\la^{1/\a}]$ and $[a,a+(1-\la)^{1/\a}]$, with gap $a-\la^{1/\a}$ in between. In its second iteration, the intervals would be further split, and so on until one obtains a totally disconnected set. However, this is not what we expected, i.e., a continuous space. What went amiss is the requirement, here ignored, of endowing sets of $\mathbb{R}$ with a fractional measure. In other words, the correct procedure is to define similarities on the ``fractional interval'' $[0,1]^\a$ spanned, by definition, by the geometric coordinate $\x$:
\be\label{Rdsimx}
{\rm S}_1(\x) := \la\x\,,\qquad {\rm S}_2(\x) := (1-\la)\x+\la\,.
\ee
These similarities would guarantee that $[0,1]^\a=[0,\la]^\a\cup[\la,1]^\a$ and that $\dh([0,1]^\a)=1$, consistently with \Eq{crd} in one dimension. From this, one infers that $\dh(\cE_\a)=\a$ and, extending to $D$ embedding dimensions, $\dh(\cE_\a^D)=D\a$. Calculations where $\dh=D-\e$ correspond, in the fractal picture, to regimes with ``low lacunarity'', i.e., where fractal space is almost translation invariant \cite{GMBA}.

As a mapping on $x$, and assuming without loss of generality that $x_0=0$ \cite{frc2}, $\{{\rm S}_1[\x(x)]\}^{1/\a}\propto\cS_{\a,1}(x)$ is a linear similarity, but 
\be\nonumber
\{{\rm S}_2[\x(x)]\}^{\frac1\a}\propto\tilde \cS_{\a,2}(x)=[(1-\la)x^\a+\la\Gamma(1+\a)]^{\frac1\a}\neq \cS_{\a,2}(x)
\ee
is neither linear nor a similarity.\footnote{By using the mean value theorem, one can show that $\tilde \cS_{\a,2}$ is a contraction on any compact interval $[x_a,x_b]$ for $x_a>x_0$ (compare \cite[Example 9.8]{Fal03}).} Therefore, an estimate of $\dh(\cE_\a^D)$ via symmetry arguments seems unpractical in $x$ coordinates, while it is straightforward in geometric coordinates. The reason why we dwelt so long on the topic of self-similarity and self-affinity is that it constitutes the starting point wherefrom to attack the important problem of the isometry group of fractional space. The integer case strongly suggests, in fact, that this group be given by the affine transformations
\be\label{potra0}
{\x'}^\mu ={\rm S}(\x^\mu) := {\rm A}^\mu_\nu \x^\nu+{\rm a}^\mu\,,
\ee
for some $D\times D$ matrix ${\rm A}^\mu_\nu$ and some vector ${\rm a}^\mu$. We shall continue the discussion in \cite{frc2}, where it will be extended to spacetimes with Lorentzian signature.

We wish to insist upon the characterization of fractional spaces as self-similar sets, and rederive the result $\dh=\a$ under yet another perspective. The concept of self-similar measure is fundamental not only to this purpose, but also for determining the spectral dimension of spacetime and for generalizing the fixed-$\a$ case to multi-fractional scenarios.

Let $\cS_i$ be $N$ similarities defining a self-similar set $\cF$. Suppose the strong separation condition holds, i.e., there exists a closed set $U$ such that $\cS_i(U)\subset U$ for all $i=1,\dots,N$ and $\cS_i(U)\cap \cS_{j\neq i}(U)=\emptyset$. $\cF\subset U$ is constructed taking sequences of similarities and the intersection of sets  $U_k=\cS_{i_1}\circ\cdots\circ \cS_{i_k}(U)$. If $|U|=1$, then the diameter of the $k$-th iteration set is the product of similarity ratios, $|U_k|=\la_{i_1}\dots \la_{i_k}$. Let $0<g_i<1$ be $N$ probabilities (or mass ratios, or weights), such that $\sum_i g_i=1$. One can imagine to distribute a mass on sets $U_k$ by dividing it repeatedly in $N$ subsets of $U_k$, in the ratios $g_1\,:\cdots:\,g_N$. This defines a \emph{self-similar measure} $\vr$ with support $\cF$, such that $\vr(U_k)=g_{i_1}\dots g_{i_k}$ and, for all sets $A\subseteq\cF$ \cite{Hut81},
\be\label{ssim}
\vr(A)=\sum_{i=1}^N g_i\,\vr[\cS_i^{-1}(A)]\,.
\ee
For Cantor sets where $N=2$, \Eq{ssim} is said to be a binomial measure (e.g., \cite{Rie95}). The case $N=+\infty$ corresponds to \emph{infinite self-similar} measures, describing fractals with an infinite number of similarities \cite{RiM,YRL}. Given a real number $u$, we define the \emph{singularity} (or \emph{correlation}) \emph{exponent} $\theta(u)$ as the real number such that \cite{Rie95}--
\cite{HJKPS}
\be\label{mfssm}
\sum_{i=1}^N g_i^u \la_i^{\theta(u)}=1\,.
\ee
The correlation exponent exists and is unique, since $0<\la_i,g_i<1$. As a function of $u$, $\theta$ is decreasing and $\lim_{u\to \pm\infty}\theta(u)=\mp\infty$. The \emph{generalized dimensions} are defined as
\be
d(u):= \frac{\theta(u)}{1-u}\,,\qquad u\neq 1\,,
\ee
and a non-singular definition, which we do not report here, is employed for $u=1$.

Self-similar measures are associated with multi-fractal sets, where the mass is not equally distributed among the smaller subsets of $\cF$. Fractals characterized by just one dimension at all scales are special cases of multi-fractals. For nested fractals, the probabilities are all equal to $g_i=1/N$. In all fractals with equal contracting ratios $\la_i=\la$, the generalized dimensions all coincide with the capacity \Eq{simd} of the set, which is also the Hausdorff dimension. In fact, from \Eq{mfssm},
\be\label{mfssm2}
N \frac{\la^{\theta(u)}}{N^u} =1\qquad \Rightarrow\qquad d(u)=-\frac{\ln N}{\ln\la}=d_{\rm C}=\dh\,.
\ee
Trivially, for the unit interval $[0,1]$ the scaling is $\la=1/N=g_i$, and $\dh=1$.

Let us apply \Eq{mfssm} to a fractional line of fixed order $\a$. As for the ordinary real line, the mass is equidistributed on all subsets, and the probabilities are still $g_i=1/N$. $N$ is arbitrary but can be fixed to $N=2$, in which case $\vr_\a$ is a binomial measure. In geometric coordinates, the scaling is $\la=1/N=1/2$ and equals the $g_i$, so $1=2^{1-\dh}$ implies $\dh=1$, and $\dh=\a$ for embedding coordinates. From the point of view of the fractional interval spanned by the $x$, the scaling is
\be\label{galp}
\la=g_i^{\frac1\a}=\left(\frac1N\right)^{\frac1\a}=\left(\frac12\right)^{\frac1\a}\,,
\ee
which is smaller than in the integer case. Then, \Eq{mfssm2} yields $1=2^{1-\dh/\a}$, consistently: the fractional charge $\a$ is the Hausdorff dimension. 


\subsection{Other properties}\label{othe}

After 1.-3., one could mention other properties, which are more model-dependent and hence fail in a number of cases. For instance,
one of the early definitions of fractals was that their dimension (defined in some of the above ways: $d_{\rm B}$, $\dh$, $d_{\rm C}$, and so on) is non-integer and greater than its topological dimension $d_{\rm top}$. There are many counterexamples where the Hausdorff dimension is integer, in some cases smaller than or equal to the topological dimension of the set (or of the graph or the trail of the map it is defined by). In all cases, $\dh$ and $d_{\rm top}$ are smaller than the topological dimension $D$ of the ambient space. Fractals with $\dh=2$ are: for $d_{\rm top}=1$ and $D=2$, the dragon curve, the Sierpi\'nski curve, some plane-filling curves (Moore curve, Peano curve), and the boundary of the Mandelbrot set; for $D\geq 2$, Brownian trails (almost surely, i.e., with probability 1, and $d_{\rm B}=\dh$; however, the graph of Brownian motion has $d_{\rm B}=\dh=3/2$ almost surely \cite{Fal03}); for $d_{\rm top}=2$ and $D=2$, the Mandelbrot set, some Julia sets, some diamond fractals, and Pythagoras tree; for $d_{\rm top}=3$ and $D=3$, the Sierpi\'nski tetrahedron. Fractals with $\dh=3$ and $d_{\rm top}=1$ in $D=3$ are box-filling curves such as the Moore, Hilbert, and Lebesgue curves. 

The fractional space $\cE_\a^D$ can have non-integer dimension, but its Hausdorff dimension is never greater than its topological dimension, $\dh\leq D$. Therefore, we can regard fractional space as a space-filling fractal or, more suggestively, as a fractal associated with a diffusion process.

Because of the empirical nature of the definition of fractals, one can conclude that fractional space $\cE_\a^D$ can be correctly characterized as ``fractal'' but, since it is a special case of fractal with a continuous structure, it may be better to use the less catchy but more specific adjective ``fractional.''


\subsection{Fractional measures as approximations of fractals}\label{fmaf}

A conceptually independent point of view was briefly mentioned at the beginning of this section, where traditional fractals were found to be approximated by fractional measures. We can now look into greater detail at the reason why fractional calculus, under certain assumptions, approximates some classes of fractals. 

A rough understanding of these approximations is actually contained in \Eq{prosi}. Depending on the value of $a$ in $\cS_{\a,2}$, we can either make a connected construction ($a=\la^{1/\a}$) or a Cantor-type one ($a=1-(1-\la)^{1/\a}$). In the first case, the first iteration yields the interval $[0,\la^{1/\a}+(1-\la)^{1/\a}]$, smaller than the desired set by a gap $(\la^{1/\a}+(1-\la)^{1/\a},1]$. This remainder is zero only if $\a=1$. Therefore, we could add at least another similarity $\cS_3$ to fill the gap, but it does not take long to convince oneself that the task is impossible unless one takes an infinite number of similarities as in fractals with infinite-type measure. This is equivalent to a continuum approximation, where $\la$ can be taken arbitrarily small and $\la^{1/\a}+(1-\la)^{1/\a}\to 1$. Conversely, the first iteration of a Cantor-like construction gives $[0,\la^{1/\a}]\cup [1-(1-\la)^{1/\a},1]$, with a central gap $(\la^{1/\a},1-(1-\la)^{1/\a})$. Sending $\la$ to zero would give the same limiting set as in the connected construction. In a qualitative sense, we begin to recognize that real-order fractional measures can be regarded as approximations of self-similar fractals in the limit of the similarity ratio approaching zero. While deterministic self-similar fractals pick a countable number of ratios $\la_i$, infinite and random fractals accept any. So, this limit is associated with random fractal structures.

This intuition is confirmed by precise arguments; we review them from the literature, but add new comments linking independent results. Consider a function $f(x)$ in $D=1$ and the convolution
\be\label{convo}
I_\cF(x)=v_\cF*f:=\int_0^x\rmd x'\, v_\cF(x-x')f(x')\,,
\ee
over a set $\cF\subseteq [0,1]$. The kernel $v_\cF$ depends on the set. For an interval, $v_\cF$ is simply a step function, but on a fractal it can be very complicated; for example, in self-similar fractals $v_\cF$ can be determined recursively at any given order of iteration. In general, it is convenient to Laplace transform the convolution \Eq{convo},
\be
\hat I_\cF(p):=\int_0^{+\infty}\rmd x\, \rme^{-p x}I_\cF(x)=\hat v_\cF(p)\hat f(p)\,.
\ee
If $\cF$ is a self-similar set, $\hat I_\cF(p)$ can be expressed iteratively as an infinite intersection of pre-fractals, \Eq{prefr}. The $k$-th iteration has Laplace-transformed kernel $\hat v_\cF^k(p)=\prod_{n=0}^{k-1} g_n(p)$, for some functions $g_n$. Typically, for self-similar and generalized self-similar sets these functions are equal and with argument $g_n(p)=g(p\la^n)$, where $\la$ is the self-similarity ratio of $\cF$. The asymptotics of $g$ is $g(z)\sim 1+O(z)$ for small $z$ and $g(z)\sim g_1+O(z^{-1})$ for large $z$, where the constant $g_1$ is the first probability weight in the self-similar measure \Eq{ssim} \cite{RYS}. Then \cite{LMNN,NLM},
\be\label{kK}
\lim_{k\to+\infty} \hat v^k_\cF(p) =\hat v_\cF(p) = p^{-\a} F_\a(\ln p)\,,
\ee
where 
\be\label{kKa}
\a=\frac{\ln g_1}{\ln\la}\,,
\ee
and $F_\a$ is a log-periodic function \cite{Sor98} of period $\ln\la$:
\be\label{pia}
F_\a(\ln p+n\ln\la)=F_\a(\ln p)= \sum_{l=-\infty}^{+\infty} c_l \exp\left(2\pi l\rmi\frac{\ln p}{\ln\la}\right)\,,
\ee
for some coefficients $c_l$. The period in $p$ is decreasing according to a geometric series. Combining \Eq{kK} with \Eq{pia},
\be\label{kap}
\hat v_\cF(p)=\sum_{l=-\infty}^{+\infty} c_l \exp\left[(\rmi\om_l-\a)\ln p\right]\,,\qquad \om_l:=\frac{2\pi l}{\ln\la}\,.
\ee
Logarithmic oscillations are a curious feature of the spectral theory on fractals. In fact, it is known that the heat kernel trace for a Laplacian on fractals displays log-oscillations in the scale \cite{KiL,Kaj10}. Oscillatory behaviour has been found analytically and numerically for various fractals \cite{DIL}--
\cite{ABS}, and \Eq{kap} illustrates a rather universal phenomenon. This is one of the most crucial points of the physical scenario that will emerge in \cite{frc2}, where it shall be given adequate space. 

Here, we are focussed only on the relation between \Eq{kap} and real-order fractional integrals. The parameter $\a$
defined in \Eq{kK} coincides, indeed, with the fractional order $\a$ determining the capacity \Eq{simd} of fractional space (remember that $g_1=1/N$). Recognizing $p^{-\a}$ as the Laplace transform of the fractional weight $v_\a(x)=x^{\a-1}/\Gamma(\a)$ and comparing \Eq{convo} with \Eq{kK} and \Eq{galp} with \Eq{kKa}, one sees that $I_\cF$ is quite similar to $I^\a$, were it not for the non-constant contribution \Eq{pia}. The main idea, now, is that \emph{a fractional integral of real order represents either the averaging of a smooth function on a deterministic fractal, or a random fractal support}. The average of a log-periodic function $F_\a$ over the period $\ln\la$ is defined by
\be
b_\a:=\langle F_\a(\ln p)\rangle := \int_{-1/2}^{1/2}\rmd z\, F_\a(\ln p+z\ln\la)\,,
\ee
and depends on the details of $g(p\la^n)$. Then,
\be
\langle v_\cF(x)\rangle = b_\a \frac{x^{\a-1}}{\Gamma(\a)}\,,
\ee
and \cite{LMNN,NLM}
\be\label{aver}
\langle I_\cF f\rangle = \int_0^x\rmd x'\,\langle v_\cF(x-x')\rangle f(x') = b_\a I^\a f\,.
\ee
Taking the average in $\la$ is tantamount to dropping all the oscillatory modes in \Eq{pia} except $l=0$. The $\om_l\to 0$ limit in \Eq{pia} can be regarded as a large-Laplace-momentum limit, so that
\be\label{net}
I_\cF \ \stackrel{{\rm Re}(p)\to +\infty}{\sim}\ b_\a I^\a\,.
\ee
This is in complete agreement with \cite{RYS,YRZ}--
\cite{RLWQ}, where integrals on more general net fractals are shown to be approximated by the left fractional integral, with the order $\a$ being the Hausdorff dimension of the set. The approximation \Eq{net} is valid for large Laplace momenta and drops all the contributions of probability weights $g_i$ for $i\geq 2$ out of the Laplace transform of the measure weight. These weights can be included to better describe the full structure of the Borel self-similar measure $\vr$ characterizing the fractal set $\cF$ \cite{frc2}. References \cite{RYS,YRZ}--
\cite{RLWQ} make this result clear by a detailed Laplace analysis, but \cite{LMNN,NLM} give a sharper physical interpretation of what it means to take the large $p$ limit. $p$ is not a Fourier momentum and $|p|\gg 1$ does not correspond to cutting off large scales. Yet, it is equivalent to randomize the fractal structure: the oscillatory structure disappears, the average of the kernel corresponds to the kernel itself, and the only approximation entailed in the derivation above is in the evaluation of the kernel \Eq{kK} (in fact, it is not obvious that $v_\cF^k*f$ is a Cauchy sequence, i.e., that the limit $k\to+\infty$ commutes with the integration). Thus, fractional integrals of real order represent (or, more conservatively, are intimately related to) random fractals.

Picking up again the example \Eq{prosi}, \Eq{kK} states that $g_1=\la^\a$. In the double limit $\la,g_1\to 0^+$, the frequencies $\om_l$ all vanish (the period of $F_\a$ becomes infinitely long), $\a$ remains finite, and $\hat v_\cF(p)\sim b_\a p^{-\a}$. The limit is only formal because $b_\a(g_1)$ vanishes at arbitrarily small $g_1$, but the main point is that the oscillatory structure is suppressed in the limit of arbitrarily small ratio $\la$. In this sense, fractional integrals of real order are continuum approximations of net fractals. This should not be confused with the low-lacunarity limit $\a\to 1$. Here $\a$ is fixed and defines the quantity $\ln g_1/\ln\la$ in the double limit $\la,g_1\to 0$.

It is quite the rule that fractals are nowhere-differentiable objects. In particular, a truly fractal spacetime is not expected to be represented by a differentiable manifold. Fractional calculus strikes quite a rich compromise between standard fractal geometry, where differentiability at large is given up, and a framework where only \emph{ordinary} differentiability is forfeited. An example from function theory is Weierstrass' function: it is nowhere differentiable in the ordinary sense, yet it is differentiable under fractional calculus \cite{KoG}. Because of the operator $\p^n$ in \Eq{pan}, the functional space on which Caputo derivatives act is the space of integer-differentiable functions, which seems at odds with fractal geometry. If concerned by that, one could take another definition of fractional derivative, differing from Caputo in the absence of integer differentiation inside the definition, but such that $\p^\a 1=0$ \cite{Jum06}. Not much would change in our conceptual framework, though.


\section{Spectral dimension of space}\label{spe}

Spectral theory is a tool to answer the famous ``Can one hear the shape of a drum?'' question \cite{Kac66}: the asymptotic spectrum of eigenvalues of Laplacian operators defined on a set provides information on the boundary of the set. This information is incomplete, inasmuch as ``drums'' with different shapes can vibrate in the same way, but it is nevertheless valuable. While the Hausdorff dimension depends on the local structure of a fractal, the spectral dimension $\ds$ is a local probe of its topology \cite{KiL,AO}--
\cite{bH}. A rigorous and fairly general definition of $\ds$ stems from the spectral theory on fractals \cite{Str99}--
\cite{Str06}. We only sketch some aspects of this theory in Section \ref{dsdef}; details can be found in the references.

\subsection{Harmonic structure}\label{dsdef}

Consider the topological presentation \Eq{prefr} where, now, the maps $\cS^k$ are obtained from $N$ continuous injections $f_i$ which are not necessarily the similarities of the set. The subsets $f_i(\cF)$ are weighted in two ways: by the probabilities $g_i$ of the self-similar measure \Eq{ssim} of the fractal, now constructed with the $f_i$, and by \emph{resistance scaling ratios} $r_i$ appearing in the definition of the Laplacian. A Laplacian $\cK$ on a self-similar fractal is part of the so-called \emph{harmonic structure} of the set, which is characterized by $g_i$ and $r_i$. Together, these quantities are assembled into $N$ parameters\footnote{In \cite{KiL} and other papers, a ``renormalization constant'' is introduced explicitly in the definition of the Laplacian and of the $\g_i$. It renders finite the discrete definition of $\cK$ in the limit of infinite iteration. This constant can be reabsorbed, as explained in \cite{Str06}.}
\be
\g_i:=\sqrt{r_i g_i}\,.
\ee
We can give an intuitive coordinate meaning of the $\g_i$. Let $\cK(x)$ be a Laplacian on a given set defined via the injections $f_i(x)=r_i x+{\rm const}$. In any subcopy $i$, the scaling of the Laplacian is determined by $\cK[f_i(x)]=r_i^{-1}\cK(x)$; summing over all the copies with the appropriate weight, one gets
\be\nonumber
\cK(x)=\sum_i g_i r_i\cK[f_i(x)]=\sum_i \g_i^2\cK[f_i(x)]\,.
\ee
Thus, $\g_i^2$ is the scaling of the Laplacian on the subsets.

The harmonic structure is said to be regular if $0<r_i<1$ for all $i$. When $N=2$, this happens if, and only if, $\g_1^{n_1}=\g_2^{n_2}$, where $n_{1,2}$ are integer numbers. Given what one would call \emph{the} standard Laplacian on a fractal, the \emph{spectral dimension} $\ds(\cF)$ is the unique number satisfying the relation
\be\label{speco}
\sum_{i=1}^N \g_i^{\ds(\cF)}=1\,,
\ee
similar to \Eq{ssm} \cite{KiL}. For simplicity, we can identify $f_i$ with the similarities of the embedding picture, and the resistance ratios with the contracting ratios $\la_i$. Deterministic fractals have a regular harmonic structure with $r_i=\la_i=\la$ for all $i$, hence from \Eq{speco}
\be\label{defra}
1= N \left(\frac{\la}{N}\right)^{\frac{\ds}{2}}\qquad\Rightarrow\qquad \ds = \frac{2\ln N}{\ln(N/\la)}=\frac{2\dh}{\dh+1}\,.
\ee
When $\la=1/N$, $\ds=\dh=1$. The unit interval is a trivial example where the spectral and Hausdorff dimension coincide.


\subsection{Diffusion}\label{diffusion}

The spectral and Hausdorff dimensions of a fractal $\cF$ are related to each other by the dimension of a Brownian motion taking place on $\cF$ \cite{HBA,BaB}--
\cite{GrK}. The anomalous diffusion law on fractals is characterized by the \emph{walk dimension} \cite{HBA,bH}
\be
\boxd{d_{\rm W}:= 2\frac{\dh}{\ds}\,.}
\ee
Since $\ds\leq\dh$ for a fractal, $\dw\geq2$. The mean-square displacement of a random walker is a power law in diffusion time, $\langle r^2(\s)\rangle\sim \s^{2/\dw}$. Processes with $\dw>2$ are of sub-diffusion, since the diffusion speed is lower than for normal diffusion (Gaussian process, $\dw=2$). Intuitively, on a fractal sub-diffusion is due to lacunarity (i.e., the presence of ``many holes'') and/or to the very high multiplicities of certain available states \cite{Dun11}. Systems with $\dw<2$ are caller of super-diffusion or jump processes \cite{GrK,BGK,BBCK}, and do \emph{not} correspond to fractals.

Thus, we realize that the information obtained from the Hausdorff dimension was, at best, incomplete, and we cannot decide about the ``fractal'' nature of fractional spaces before looking at their harmonic structure or, in other words, at their topology, or, in yet other terms, at the way diffusion processes take place in them. Conversely, and contrary to popular belief in part of the physicists community, quantum gravity models with a non-integer spectral dimension do not necessarily entail a ``fractal'' spacetime. To establish whether the latter is ``fractal'' or not, it is important to compare $\dh$ with $\ds$. To this purpose, we review the operative definition of spectral dimension employed in quantum gravity. The main steps are well known (e.g., \cite{SVW2}), but a closer and perhaps pedantic contact with the fractal-geometry perspective will be crucial to avoid confusion in the interpretation and construction of the fractional case.

Since we want to probe the local structure of space,\footnote{Dimension of spacetimes is always computed in Euclidean signature, i.e., after coordinate time has been Wick rotated. Thus, here we refer to spaces rather than spacetimes.} we can imagine to place a test particle in it and let it diffuse in a random walk starting at point $x$ (index $\mu$ omitted) and ending at point $x'$. For a metric space of topological dimension $D$ and Riemannian metric $g_{\mu\nu}$, this process is governed by a diffusion equation for the heat kernel $P(x,x',\s)$,
\be\label{dife}
(\p_\s-\N^2_x)P(x,x',\s)=0\,,\qquad P(x,x',0)=\frac{\delta(x-x')}{\sqrt{g}}\,,
\ee
where $\s$ is diffusion time (a parameter not to be confused with physical or coordinate time), $\N^2=g^{-1/2} \p_\mu(g^{1/2}\p^\mu)$ is the Laplacian defined on the ambient space, $g$ is the determinant of the metric, and $\delta$ is the $D$-dimensional Dirac distribution. 

The Laplacian acts on the $x$ dependence of $P$, even in cases where the ambient space is not translation invariant and $P(x,x',\s)\neq P(x-x',\s)$. In fact, the heat kernel is the matrix element in $x$ representation of the operator $\rme^{\s\N^2}$, $P(x,x',\s)=\langle x'|\rme^{\s\N^2}|x\rangle$. This writing means that, given a complete orthonormal set of eigenfunctions of the Laplacian,
\be\label{sonc}
\N^2\vp_k(x)=-\la_k\vp_k(x)\,,\qquad \sum_k\int\rmd^Dx\,\sqrt{g}\,\vp_k(x)\vp_k^*(x')=1\,,
\ee
the heat kernel can be written as
\be\label{sonc2}
P(x,x',\s)=\sum_k \rme^{-\s\la_k}\vp_k(x)\vp_k^*(x')\,.
\ee
In \Eq{sonc} and \Eq{sonc2}, the sum over $k$ is replaced by an integral if the spectrum is continuous.

The relative sign between $\s$ derivative and Laplacian in \Eq{dife} guarantees causal diffusion, where the particle flows away from the point $x$ as diffusion time increases. The initial condition reflects the pointwise nature of the probe. Probes of finite shape are in principle possible, but they would not lead to a sensible definition of spectral dimension in the present context (in fact, we want to investigate the local manifold structure of a smooth space). In general, given the initial condition $\phi(x,0)$ at $\s=0$, the solution of the diffusion equation is
\be\label{gesode}
\phi(x,\s)=\rme^{\s\N^2}\phi(x,0)= \int\rmd^D x' \sqrt{g}\, P(x,x',\s)\phi(x',0)\,.
\ee
In particular, the effect of the non-local operator $\rme^{\s'\N^2}$ is a shift of the auxiliary variable $\s$: $\rme^{\s'\N^2}\phi(x,\s)= \rme^{\s'\,\p_\s} \phi(x,\s)=\phi(x,\s+\s')$.

For a smooth space with $D$ dimensions, the solution of \Eq{dife} can be found via the Fourier transform method. Consider first flat Euclidean space. The direct and inverse Fourier transforms of a function $f(x)$ are
\bs\label{ft}\ba
\tilde f(k) &=& \frac{1}{(2\pi)^{\frac{D}{2}}}\int_{-\infty}^{+\infty}\rmd^D x\, f(x)\,\rme^{-\rmi k\cdot x}=: F[f(x)]\,,\\
f(x) &=& \frac{1}{(2\pi)^{\frac{D}{2}}}\int_{-\infty}^{+\infty}\rmd^D k\, \tilde f(k)\,\rme^{\rmi k\cdot x}\,.
\ea\es
In particular, this definition is compatible with the definition of the Dirac distribution:
\be\label{dirndl}
\delta(x) = \frac{1}{(2\pi)^D}\int\rmd^D k\, \rme^{\rmi k\cdot x}\,,\qquad (2\pi)^{\frac{D}{2}}F[\delta(x)]=\int\rmd^D x\, \delta(x)\,\rme^{-\rmi k\cdot x}=1\,.
\ee
Phases $\vp_k(x)=\rme^{\rmi k\cdot x}$ are eigenfunctions of the ordinary Laplacian $\N^2=\p_\mu\p^\mu=\p^2_1+\dots+\p^2_D$ with a continuum of eigenvalues,
\be
\N^2 \rme^{\rmi k\cdot x} = -\la_k \rme^{\rmi k\cdot x}\,,\qquad \la_k = k^2:= k_\mu k^\mu = k_1^2+\dots +k_D^2\,,
\ee
which allows us to write the solution of the diffusion equation as
\be\label{inteK}
P(x,x',\s)=\frac{1}{(2\pi)^D}\int\rmd^D k\, \tilde P(k,\s)\,\rme^{\rmi k\cdot (x-x')}\,.
\ee
In particular, due to the composition law $\vp_k(x)\vp_k^*(x')=\rme^{\rmi k\cdot (x-x')}$, the heat kernel \Eq{sonc2} only depends on the relative distance of the initial and end point. $\tilde P(k,\s)$ must obey
\be\nonumber
(\p_\s+k^2)\tilde P(k,\s)=0\,,\qquad \lim_{\s\to 0}\tilde P(k,\s)=1\,,
\ee
yielding
\be\label{tike}
\tilde P(k,\s)=\rme^{-\s k^2}\,.
\ee
Equation \Eq{inteK} is simply a Gaussian in $D$ dimensions,
\be
\label{ker}
P(x,x',\s)=\frac{\rme^{-\frac{(x-x')^2}{4\s}}}{(4\pi \s)^{\frac{D}{2}}}\,.
\ee
For a metric space of topological dimension $D$, the solution at small $\s$ is Weyl's expansion
\be\label{genK}
P(x,x',\s)=\frac{\rme^{-\frac{\Delta(x,x')^2}{4\s}}}{(4\pi \s)^{\frac{D}{2}}}\left[1+\sum_{n=1}^{+\infty}A_n\s^n\right]\,,
\ee
where the $O(\s)$ remainder and the distance $\Delta(x,x')$ between the two points depend on the metric. 
The precise form of the coefficients $A_n$ in the expansion (sometimes called Hadamard--Minakshisundaram--DeWitt--Seeley coefficients) can be obtain via the theory of Green's functions in Riemannian manifolds
\cite{Ham23}--
\cite{DF} (for reviews, see \cite{Avr00}--
\cite{Vas03}).

The spatial average of the heat kernel $P(x,x,\s)$ at coincident points $x=x'$ is the \emph{return probability}
\be\label{retpr}
\cP(\s) := \frac{1}{\cV^{(D)}}\int\rmd^Dx\,\sqrt{g}\,P(x,x,\s)\,,\qquad \cV^{(D)}:=\int\rmd^Dx\,\sqrt{g}\,,
\ee
which is the trace per unit volume of the operator $\rme^{\s\N^2}$. Since the heat kernel $P(x,x,\s)= (4\pi \s)^{-D/2}[1+O(\s)]$ associated with \Eq{genK} is constant in $x$ due to translation invariance, to leading order in $\s$ one has $\cP(\s)\sim \s^{-D/2}$ and the topological dimension of space is given by $D=-2\lim_{\s\to 0}\rmd\ln \cP(\s)/\rmd\ln\s$. This formula suggests an operational definition, for any metric space allowing a random walk process, of the spectral dimension:
\be\label{spedi}
\boxd{\ds : = -2\frac{\rmd\ln \cP(\s)}{\rmd\ln\s}\,,}
\ee
which can be shown to coincide with the more abstract prescription \Eq{speco}. Due to curvature effects, the spectral dimension changes with the diffusion parameter $\s$, and one can find the phenomenon of dimensional flow at different scales already in classical gravity \cite{Car09}. In practice, however, the solution to \Eq{dife} is difficult to compute for arbitrary $\s$, and one confines her/his interest to the heat kernel expansion \Eq{genK}. This expansion is valid also for spaces with boundaries or non-trivial topologies, in which case $\s$ must not be taken ``too large'' (lest global boundary or topology effects vitiate the estimate of $\ds$, which must be local). On the other hand, random walks and Laplacians can be defined even on very non-trivial sets such as fractals. The exponent of the first term in Weyl's expansion is the spectral dimension of the fractal and it can be non-integer \cite{KiL}. Information on the various dimensions can be obtained directly from the scaling property of the heat kernel. In fact, under a coordinate and external-time dilation,
\be\label{scak}
P(\la^{2/\dw}x,\la^{2/\dw}x',\la^2\s) = \la^{-\ds} P(x,x',\s)\,.
\ee
Comparison with \Eq{ker} shows that the walk and spectral dimensions of flat space are, respectively, $\dw=2$ and $\ds=D$.

To summarize, the spectral dimension \Eq{spedi} is obtained with the following ingredients:
\begin{enumerate}
\item An invertible transform between position and momentum space. In flat Euclidean space, the Fourier transform \Eq{ft} is a superposition of phases.
\item A Laplacian $\cK$. There is no unique definition of Laplacian, but the most natural one is such that the expansion basis of the invertible transform is made of eigenfunctions of $\cK$. In flat Euclidean space, the natural Laplacian is the second-order operator $\N^2=\p_\mu\p^\mu$.
\item A diffusion equation $(\bD^\b_\s-\cK_x)P(x,x',\s)=0$, which is defined by $\cK$ (acting on $x$) and by the choice of diffusion process (i.e., of the derivative operator $\bD^\b_\s$ and of the relative coefficient between this and $\cK$). These two independent ingredients correspond to the definition of a harmonic structure. In flat Euclidean space, diffusion is normal ($\bD^\b_\s=\p_\s$, \Eq{dife}).
\end{enumerate}


\subsection{The case of fractional space}\label{dsfras}

The determination of the spectral dimension of fractional spaces highlights once again the main difference between fractal geometry and ordinary space constructions in field theory. In the first case, geometry and topology are defined by the symmetry and harmonic structures of the fractal, which are given at the outset. On the other hand, in field theory the harmonic structure stems from physical considerations. More specifically, we can say that the symmetry structure is first dictated by the action measure $\vr$ and then imposed on the Lagrangian density $\cL$, but the harmonic structure is determined both by the symmetries and by the form of the kinetic operator. For instance, in ordinary field theory the natural Lorentz-invariant Laplace--Beltrami operator is $\cK_1=\B=\p_\mu\p^\mu$. However, any other operator of the form $(\cK_1)^n=\B^n$ respects the same symmetry group, but its harmonic structure is different. Physical requirements (in particular, the absence of ghosts) eventually single out $\cK_1$ as \emph{the} kinetic operator, but this extra input is not always readily available. 

In fractional spaces, the possibility to choose a different symmetry for $\cL$ is the first source of ambiguity. A second source is the non-unique way the diffusion equation is \emph{defined} with fractional calculus. After this preamble, let us examine the three ingredients of the previous section.

\subsubsection{Laplacian}

As we shall see in \cite{frc2}, symmetry arguments select two inequivalent classes of fractional field theories:
\begin{itemize}
\item \emph{Fractional symmetry scenario.} In ordinary integer models, the symmetries ${\rm sym}_\a$ and ${\rm sym}_\cL$ of, respectively, the measure and the Lagrangian density are the same. If we impose ${\rm sym}_\a={\rm sym}_\cL$ also in fractional theories, we obtain an action invariant under what will turn out to be the fractional generalization of rotation/Lorentz transformations. In this case, the invariant Laplacians we shall consider are
\be\label{cka}
\cK_\a:=\de^{\mu\nu}\p_\mu^\a\p^\a_\nu\,,\qquad \bar\cK_\a:=\de^{\mu\nu}{}_\infty\bp_\mu^\a\,{}_\infty\bp^\a_\nu\,.
\ee
Under a scaling transformation $x\to\la x$, $\cK_\a\to \la^{-2\a}\cK_\a$ and $\bar\cK_\a\to \la^{-2\a}\bar\cK_\a$.
\item \emph{Integer-symmetry scenario.} While the symmetry of the measure guarantees protection against proliferation of arbitrary measure operators in the renormalization group flow, one can prescribe a constant symmetry for $\cL$ along the flow. Since the Lagrangian should be Lorentz invariant in the infrared, we assume ${\rm sym}_\cL={\rm sym}_{\a=1}$, the integer Lorentz group in $D$ dimensions. Then,
\be\label{ck1}
\cK_1:=\de^{\mu\nu}\p_\mu\p_\nu\,.
\ee
Under a scaling transformation, $\cK_1\to \la^{-2}\cK_1$. In preparation of building the action functional for a scalar field $\phi$ in fractional space(time) \cite{frc2}, we notice that the presence of a non-trivial measure weight factor
\be\label{frames2}
v_\a(x)=\prod_\mu \frac{(x^\mu)^{\a-1}}{\Gamma(\a)}
\ee
makes the kinetic terms $\phi\cK_1\phi$ and $-\p_\mu\phi\p^\mu\phi$ inequivalent. Upon integration by parts, the latter corresponds to $\phi\cK\phi$, where
\be\label{ck2}
\boxd{\cK:=\de^{\mu\nu}\left(\p_\mu\p_\nu+\frac{\p_\mu v_\a}{v_\a}\p_\nu\right) = \de^{\mu\nu}\left(\p_\mu\p_\nu-\frac{1-\a}{x^\mu}\p_\nu\right)\,.}
\ee
\end{itemize}
Several physical considerations \cite{frc2,frc3} will select $\cK$, over the other choices \Eq{cka} and \Eq{ck1}, as a better Laplacian on fractional spaces. Here we do so also for technical reasons, which are going to become apparent to the reader: $\cK$ allows one the simplest and most natural analytic expansion of the fractional heat kernel in the basis of the fractional transform.

\subsubsection{Fractional Bessel transform}

The construction of a field theory on a fractal is subordinate to the possibility to define a transform thereon and to move at will from position to momentum space and vice versa. Shifting point of view on the same issue, the aim is to define a Laplacian operator on a fractal and study its spectrum. Indeed, the spectral theory of Laplacians on fractals is a hot and evolving topic in the mathematical literature. Currently, the state of the art is that no consistent measure-theoretic definition of a Fourier transform on fractals has been found, although some authors have gone so far as to obtain a Plancherel formula \cite{StT}. Yet, the spectral theory has been formulated in a number of special cases, and general results exist for post-critically finite fractals, i.e., fractals which would become totally disconnected after the removal of a finite set of points (an example is the Sierpi\'nski gasket; textbook introductions are \cite{Kig01,Str06}). These fractals admit an actual Fourier transform \cite{St03a,OkS}. A natural conjecture is to identify the dimension of momentum space with the spectral dimension $\ds$, calculated by looking at the spectral properties of the Laplacian. This can be motivated, for instance, in thermodynamical systems, where the spectral dimension comes from the momentum space trace needed to implement thermodynamical equations of state \cite{Akk2}. In general, $\ds$ differs from the Hausdorff dimension $\dh$ of position space.

On a more phenomenological ground, there are some results also in particular models of spaces with fractal-like features. In \cite{Sti77}, an invertible Fourier transform was defined on a metric space with non-integer dimension, only for the class of functions generated by the Gaussian and used in perturbative field theory. On the same class of functions, an invertible transform exists on spaces with translation-invariant Lebesgue--Stieltjes measure \cite{Svo87}. Then, the measure in momentum space has the same dimension as in position space. This agrees with \cite{fra2}, where we defined a transform with Stieltjes measure such that the engineering dimension of the measure is the same in both momentum and position space and $\ds=\dh$. We do not corroborate the same conclusion in this paper. In fractional spaces, transport can be anomalous with respect to the natural differential structure and, by the diffusion equation method, we shall find that $\ds=\dh$ only in special cases, but not generally. In these spaces, the spectral dimension is not the dimension of momentum space.

The derivation of an invertible transform in fractional spaces considerably differs from that of \cite{fra2} in the final result. The discrepancy lies in the fact that a naive replacement of the measure in the Fourier integral does not lead to an invertible transform,\footnote{I am indebted to G.\ Nardelli for pointing it out.} due to the loss of translation invariance in the measure (such an invariance was formally assumed in \cite{Svo87}). We give here the correct answer, which is proven in a separate publication \cite{frc3}. One can show that, given the fractional measure
\be\label{frames}
\rmd\vr_\a(x)=\rmd^Dx\,v_\a(x)
\ee
with weight \Eq{frames2}, the measure in momentum space is $\tau_\a(k)=\vr_\a(k)$ and the fractional transform and anti-transform are
\bs\ba
\tilde f(k) &:=& \int_0^{+\infty}\rmd\vr_\a(x)\, f(x)\,c_\a(k,x)=: F_\a[f(x)]\,,\label{fst1}\\
f(x)        &=& \int_0^{+\infty}\rmd\vr_\a(k)\,\tilde f(k)\,c_\a(k,x)\,,\label{fstau}
\ea\es
where
\be
c_\a(k,x) := \prod_\mu c_{\a,\mu}(kx):= \prod_\mu\Gamma(\a) (k^\mu x^\mu)^{1-\frac{\a}{2}} J_{\frac{\a}{2}-1}(k^\mu x^\mu)
\ee
and $J_\nu$ is the Bessel function of the first kind of order $\nu$. Notice that, consistently, $c_1(k,x)=(2/\pi)^{D/2} \cos(k_1x_1)\cdots\cos(k_Dx_D)$, and one recovers the ordinary Fourier cosine transform.

The check that the fractional Bessel transform thus defined is invertible makes use of the integral representation of the ordinary Dirac distribution in terms of Bessel functions \cite{frc3}. In one dimension,
\be\label{JJd}
\delta(x-x')=x\int_0^{+\infty} \rmd k\,k J_\nu(kx)J_\nu(kx')\,,
\ee
for ${\rm Re}(\nu)>-1$ ($\a>0$ in the above calculation). In particular, in $D$ dimensions and for $\nu=\a/2-1$, \Eq{JJd} yields the correct integral representation of the ``fractional'' Dirac distribution:
\be\label{deaJJ}
\de_\a(x,x'):=v_\a^{-1}(x)\delta(x-x')=\int_0^{+\infty}\rmd\vr_\a(k)\,c_\a(k,x)c_\a(k,x')\,,
\ee
which is not translation invariant. From \Eq{fstau} and \Eq{JJd}, in one dimension
\be\label{frad}
\int_0^{+\infty}\rmd\vr_\a(x)\,\delta_\a(x,x') f(x)= f(x')\,,
\ee
and the generalization to $D$ dimensions is straightforward.

\subsubsection{Diffusion equation}

Depending on the choice of Laplacian and of the operator $\bD^\b_\s$, the diffusion equation will describe different diffusion processes, characterized by different values of the walk dimension. When anomalous dimensions are involved, there is no unique way to determine the diffusion equation, except on heuristic or phenomenological grounds. Anomalous diffusion can be realized also without fractional derivatives (e.g., \cite{MGN}), but here we pick the form
\be\label{difef}
(\bD_\s^\b-\cK)P(x,x',\s)=0\,, \qquad  P(x,x',0)=\de_\a(x,x')\,,\qquad 0<\b\leq 1\,,
\ee
where
\be\label{dsb}
\bD_\s^\b \in\{\p_\s,\,\p^\b_{0,\s},\,{}_\infty\bp^\b_\s\}\,,
\ee
and we chose the fractional delta as the natural initial condition. Generalized to $(\bD_\s^\b-\bar\cK_\g)P(x,x',\s)=0$, \Eq{difef} is often called fractional wave equation.
When $\b=1=\g$, this is ordinary diffusion (Brownian motion). When $\b\neq 1$ and $\g=1$, it corresponds to fractional sub-diffusion. Finally, when $\b=1$ and $\g\neq 1$ the process is a L\'evy flight \cite{Zas3}. In the literature, in all these cases the initial condition is the usual delta.

\subsubsection{Spectral and walk dimensions}

We have all the ingredients to determine the spectral dimension of fractional space with the heat-kernel method. The geometry of a set affects the spectral dimension in two different ways: through the fractal structure and the metric structure. In a fractional theory, what we call fractal structure is actually the calculus associated with the fractional manifold, and the metric structure is given by the fractional metric. Drawing inspiration from the ordinary manifold result, we can ignore curvature effects to obtain the spectral dimension of fractional spaces for small diffusion parameter.

Solutions of the diffusion equation \Eq{difef} and its generalization $(\bD_\s^\b-\bar\cK_\g)P=0$ are known for phenomenological models of statistical mechanics, where an anomalous diffusion process takes place in ordinary space \cite{Pod99,KST,Tar3b,Bar01}--
\cite{Zas3}. Here, the very geometry of space is modified, and we need to start from scratch. 

Given a diffusion equation (integer or fractional), a most practical way to solve it is to render it algebraic. In ordinary space, one writes the heat kernel as a quadratic superposition of eigenstates $\vp_k(x)$ of the Laplacian, \Eq{sonc} and \Eq{sonc2}. By construction, the Laplacian is chosen so that its eigenfunctions (phases, cosines or sines) also constitute the basis of the momentum expansion of the Fourier transform.

In the fractional case, the eigenfunctions of the Laplacian $\cK_1$ are not the functions $c_\a(k,x)$ in the Bessel transform \Eq{fst1}, so one is left with two different options: to expand the heat kernel in the Laplacian eigenstates $\vp_k(x)$ or in the functions $c_\a(k,x)$. The first possibility seems the one guaranteeing an algebraic solution of the diffusion equation, but in fact it is not compatible with the initial condition in \Eq{difef}. We omit the proof that the expansion \Eq{sonc2} does not work properly. The second case turns out to be both analytically exact and compatible with the initial condition, but only for the Laplacian $\cK$. The $c_\a$ are eigenfunctions of the kinetic operator \Eq{ck2},
\be
\cK c_\a(k,x) = -k^2 c_\a(k,x)\,.
\ee
This is a Bessel equation with two independent solutions given, in one dimension, by $(kx)^{-\nu} J_\nu$ and $(kx)^{-\nu} Y_\nu$, where $\nu=\a/2-1$ and $Y_\nu$ is the Bessel function of the second kind, or by their complex linear combination into Hankel functions. Transforms can be defined also via $Y$ and the Hankel functions, but only the Bessel transform with $J$ is invertible and unitary, i.e., it is the only one where the Dirac distribution admits a Bessel integral representation. Therefore, the existence of an invertible unitary transformation between position and momentum space determines the natural Laplacian in position space. Notice from \Eq{deaJJ} that the $c_\a$ already have the correct normalization.

The heat kernel \Eq{sonc2} is generalized to fractional space by identifying $\vp_k(x)$ with $c_\a(k,x)$, by replacing the sum over $k$ with an integral with the momentum-space measure $\vr_\a(k)$, and by replacing the exponential with the eigenfunction $f_k(\s)$ of the operator \Eq{dsb}, with eigenvalue $-k^2$:
\be
(\bD^\b_\s+k^2)f_k(\s)=0\,.
\ee
For different choices of $\bD^\b_\s$, the solution is
\be
f_k(\s)=\left\{\begin{matrix} \rme^{-\s k^2}\,,           &\qquad&  \bD^\b_\s=\p_\s\\
                              \rme^{-\s (-k^2)^{1/\b}}\,, &\qquad&  \bD^\b_\s={}_\infty\bp_\s^\b\\ 
                              E_\b(-k^2\s^\b)\,,          &\qquad&  \bD^\b_\s=\p_\s^\b\end{matrix}\right.\,,\label{kkk}
\ee
all in agreement with \Eq{tike} when $\b\to 1$ (up to an extra sign for the right derivative). The final form of the fractional heat kernel is then
\be\label{soncf}
P(x,x',\s)=\int_0^{+\infty}\rmd\vr_\a(k)\, f_k(\s)\, c_\a(k,x)c_\a(k,x')\,.
\ee
By construction, it obeys the diffusion equation \Eq{difef} and is not translation invariant. The initial condition is also correct, as one can see by comparing $P(x,x',\s)$ with \Eq{deaJJ}. Furthermore, the general solution $\Phi(x,\s)$ of the diffusion equation for the initial condition $\Phi(x,0)$ is always of the form
\be
\Phi(x,\s)=\Big(\prod\nolimits_\mu x^\mu\Big)^{\frac\a2}\phi(x,\s)\,,
\ee
and is naturally given by the fractional generalization of \Eq{gesode},
\be\label{gesodef}
\Phi(x,\s)= \int_0^{+\infty}\rmd\vr_\a(x')\, P(x,x',\s)\,\Phi(x',0)\,,
\ee
in accordance with \Eq{frad}.


The return probability is the fractional counterpart of \Eq{retpr} and reads
\ba
\cP(\s) &=& \frac{1}{\cV_\a}\int_0^{+\infty}\rmd\vr_\a(x)\,P(x,x,\s)\nonumber\\
&=&\frac{1}{\cV_\a}\int_0^{+\infty}\rmd\vr_\a(x)\int_0^{+\infty}\rmd\vr_\a(k)\, f_k(\s)\, c_\a^2(k,x)\,,\label{retprf}
\ea
where $\cV_\a:=\int\rmd\vr_\a(x)$ is a divergent total volume prefactor. For each of the $D$ directions, one has to solve a double integral of the form
\be\nonumber
\int_0^{+\infty}\rmd x\,x\int_0^{+\infty}\rmd k\,k\, f_k(\s) J_\nu^2(kx)\,.
\ee
In the ordinary flat case, the coordinate dependence factorizes and is cancelled out by the volume prefactor, but here the integrals in $x$ and $k$ do not commute. The choice of integration ordering and the regularization of eventual divergent contributions do not affect the spectral dimension, \Eq{spedi}, because from our choices of $f_k(\s)$ one can always rewrite \Eq{retprf} as a function of $\s$ times a pure ($\a$-dependent) number. Let $\b$ include also the value 1. From \Eq{kkk}, upon the rescaling
\be\label{reskx}
\tilde k =\s^{\frac\b2} k\,,\qquad \tilde x = \s^{-\frac\b2}x\,,
\ee
one can always factorize \Eq{retprf} as
\be\label{ksa}
\cP(\s) = \s^{-\frac{D\a\b}{2}} \left[\frac{1}{\cV_\a}\int_0^{+\infty}\rmd^D\tilde x \int_0^{+\infty}\rmd^D\tilde k\, C(\tilde k,\tilde x;\a)\right]\,.
\ee
The $\s$ dependence comes exclusively from the volume prefactor, while the numerator of \Eq{retprf} is rendered dimensionless by the rescaling \Eq{reskx}. Integration of the $\a$-dependent function $C$ entails an ordering and, eventually, a regularization choice.

Combining \Eq{ksa} with \Eq{spedi}, we finally obtain the spectral dimension: $\ds = D\a\b$ for $\bD^\b_\s=\p_\s$, and $\ds = D\a\b$ for $\bD^\b_\s={}_\infty\bp_\s^\b$, $\p_\s^\b$. In general form,
\be
\boxd{\ds=\b\dh\,.}
\ee
Some remarks are in order:
\begin{itemize}
\item For normal diffusion ($\b=1$),
\be\label{spefra}
\ds=\dh=D\a\,,
\ee
realized by integer operators $\p_\s$ and $\cK$. The order of the diffusion operator is half that of the Laplacian.
\item When $\b<1$, diffusion is anomalous but fractional space can be regarded as a fractal.
\item The case $\b>1$ does not correspond to a fractal, since $\ds>\dh$; the operator $\bD^\b_\s$ is higher order from the point of view of the differential structure of fractional space, and it is responsible for super-diffusion. In particular, for $\b=1/\a$ the spectral dimension coincides with the topological dimension of space.
\end{itemize}
In all these cases, the spectral dimension is constant and non-vanishing. There is no contradiction with the findings of \cite{HaK1}, where the spectral dimension does not converge to the embedding dimension when the lacunarity of fractals becomes asymptotically zero. Here, fractional spaces are not low-lacunarity approximations of fractals and we do not take the limit $\la\to 0$.

The results are summarized in table \ref{tab4} for $\ds$ and the walk dimension $\dw$.
\begin{table}
\begin{tabular}{ccc}\hline
$\bD^\b_\s$ 		                  & $\ds$   & $\dw$  \\\hline
$\p_\s$                           & $\dh$   & $2$    \\ 
${}_\infty\bp_\s^\b$, $\p_\s^\b$  & $\b\dh$ & $2/\b$ \\\hline
\end{tabular}\hspace{2.5cm}
\caption{\label{tab4} Spectral dimension $\ds$ and walk dimension $\dw$ of fractional space $\cE_\a^D$ for the natural Laplacian \Eq{ck2} and different diffusion equations. Fractional space is fractal only if $\dw\geq 2$ ($\b\leq 1$).}
\end{table}


\section{Discussion}

The results presented in this paper are the first step towards a field theory on fractional spacetimes. We have detailed the geometric properties of the fractional equivalent of Euclidean space, with no time, no matter, no gravity, and fixed real dimension. The goal, of course, will be to include the physics in a controlled way. Bits of it, such as the generalization to spacetimes with Lorentzian signature, only require minor modifications of the fractional construction. Others, such as the formulation of a multi-fractal scenario, the recovery of four dimensions at large scales, and inclusion of the log-oscillations of fractal geometry, are not difficult but entail a major change of perspective. We endeavor to complete this programme in \cite{frc2}, still in the absence of gravity.

We would like to conclude with a remark at the interface between quantum gravity and mathematics. It has been recognized that effective spacetime emerging from quantum gravity scenarios has a scale-dependent spectral dimension, a feature on which part of the community has grown the belief that quantum spacetime is, somehow, ``fractal.'' However, aside from the spectral dimension, fractal geometry is an arsenal of tools which has been scantly exploited in quantum gravity. As a consequence, a deeper understanding of fractal properties of spacetime has seldom gone beyond qualitative remarks based on quantitative determinations of $\ds$. Taking advantage of this arsenal in a less frugal way would open up a wealth of possibilities, as we shall argue in the companion paper. On the other hand, progress in pure fractal geometry is very much ongoing and the effective insights in the physics literature can suggest mathematicians some interesting directions of research. For instance, to the best of our knowledge there is no systematic formulation of a Fourier transform on fractals. To a physicist, transforming to momentum space is important both for doing field theory and for computing the spectral dimension. Related to that, we are unaware of any good physical transport model where jump processes naturally occur. These are non-local diffusing processes characterized by discrete jumps, rather than continuous movements. While for local diffusion the walk dimension is bounded by $2\leq \dw\leq \dh+1$, in non-local diffusion $0< \dw\leq \dh+1$ \cite{GrK}. The cases where fractional spaces have $\dw=2\dh/\ds<2$ might correspond to continuum models of jump processes, but the physical meaning of this is presently unclear. Yet, transient regimes where $\ds>\dh$ do arise in other approaches to quantum gravity, as in causal dynamical triangulations (CDTs) \cite{BeH} or in non-commutative spaces \cite{AA}. Further study of the subject promises to be stimulating.


\subsection*{Acknowledgements}

The author is grateful to D.~Benedetti, S.~Gielen, J.~Magueijo, L.~Modesto, D.~Oriti, J.~Th\"urigen, S.~Vacaru, and especially to G.~Dunne and G.~Nardelli for useful discussions. G.~Nardelli is also credited for the construction of the fractional transform \Eq{fst1}--\Eq{fstau}.



\end{document}